\documentclass[letterpaper,titlepage,11pt]{article}
\usepackage{amssymb,amsmath,amsfonts}
\usepackage{epsfig}
\usepackage{epstopdf}
\usepackage{times,cite,nicefrac}

\def\AA{{\cal A}}
\def\BB{{\cal B}}
\def\CC{{\cal C}}

\def\II{{\cal I}}
\def\JJ{{\cal J}}
\def\KK{{\cal K}}

\def\MM{{\cal M}}
\def\NN{{\cal N}}

\def\PP{{\cal P}}

\def\TT{{\cal T}}

\def\VV{{\cal V}}

\def\d{{\partial}}

\def\slr{SL(2,$\mathbb{R}$) }
\def\slc{SL(2,$\mathbb{R}$)/U(1) }
\def\d{{\partial}}
\def\di{\text{d}}
\def\hl{\nicefrac{1}{2}}

\long\def\symbolfootnote[#1]#2{\begingroup%
\def\thefootnote{\fnsymbol{footnote}}\footnote[#1]{#2}\endgroup}

\newcommand{\oao}[2]{{#1\atopwithdelims[]#2}}

\newcommand{\beq}{\begin{equation}}
\newcommand{\eeq}{\end{equation}}
\newcommand{\ba}{\begin{array}}
\newcommand{\ea}{\end{array}}
\newcommand{\bea}{\begin{eqnarray}}
\newcommand{\eea}{\end{eqnarray}}

\newcommand{\Z}{\mathbb{Z}}
\newcommand{\C}{\mathbb{C}}
\newcommand{\R}{\mathbb{R}}

\numberwithin{equation}{section}
\begin{document}

\begin{titlepage}

%\rightline{\vbox{\small\hbox{\tt hep-th/0703151}}}
\rightline{\vbox{\small\hbox{CPHT-RR013.0307}}}
\vskip 2.8cm

\centerline{\LARGE  Orientifolds in $\NN=2$ Liouville Theory and its Mirror}

\vskip 1.6cm
\centerline{\bf Dan Isra\"el$^\spadesuit$ and 
Vasilis Niarchos$^\diamond$\symbolfootnote[2]{Email: israel@iap.fr, 
niarchos@cpht.polytechnique.fr}}
\vskip 0.5cm
\centerline{\sl $^\spadesuit$
Racah Institute of Physics, The Hebrew University, Jerusalem 91904, Israel}
\centerline{\sl 
\textsc{greco}, Institut d'Astrophysique de Paris,
98bis Bd Arago, 75014 Paris, France\footnote{Unit\'e mixte de Recherche
7095, CNRS -- Universit\'e Pierre et Marie Curie}
}

\vskip 0.3cm
\centerline{\sl $^\diamond$Centre de Physique Th\'eorique, Ecole Polytechnique,
 91128 Palaiseau, France\footnote{Unit\'e mixte de Recherche
7644, CNRS -- \'Ecole Polytechnique}}

\vskip 1.4cm

\centerline{\bf Abstract} \vskip 0.5cm 

\noindent
We consider unoriented strings in the supersymmetric \slc coset,
which describes the two-dimensional Euclidean black hole, 
and its mirror dual $\mathcal{N}=2$ Liouville theory.
We analyze the orientifolds of these theories from several complementary points 
of view: the parity symmetries of the worldsheet actions, descent from known
AdS$_3$ parities, and the modular bootstrap method
(in some cases we can also check our results against known constraints 
coming from the conformal bootstrap method). 
Our analysis extends previous work on orientifolds in Liouville
theory, the AdS$_3$ and SU(2) WZW models and minimal models. Compared to
these cases, we find that the orientifolds of the two dimensional Euclidean black hole 
exhibit new intriguing features. Our results are relevant for the study of orientifolds 
in the neighborhood of NS5-branes and for the engineering of four-dimensional 
chiral gauge theories and gauge theories with SO and Sp gauge groups with 
suitable configurations of D-branes and orientifolds. As an illustration,
we discuss an example related to a configuration
of D4-branes and O4-planes in the presence of two parallel fivebranes.

\vfill

\end{titlepage}

\tableofcontents

\section{Introduction}
Orientifolds play an important role in string theory  
(for a review see~\cite{Dabholkar:1997zd}). They appear in non-perturbative dualities
and in many applications with clear phenomenological interest, especially
since the advent of flux compactifications~\cite{Giddings:2001yu}. By nature,
orientifolds are perturbative objects associated to the physics of
unoriented strings that can be studied explicitly in perturbative string theory with the
use of standard conformal field theory (\textsc{cft}) techniques. Their properties become richer
in curved backgrounds where one has to face on the level of the worldsheet the complexities
of a non-trivial~\textsc{cft}. Related \textsc{cft} techniques were  successfully applied 
to study Calabi-Yau compactifications at Gepner points in a 
series of  papers~\cite{Govindarajan:2003vp,Aldazabal:2003ub,
Blumenhagen:2003su,Brunner:2004zd,Brunner:2006yi,Hosomichi:2006yj}.

In this paper we want to study the orientifolds of two related theories: the
$\NN=2$ Liouville theory and the supersymmetric \slc coset. These
theories are known to be dual~\cite{fzz,Kazakov:2000pm} and are mapped to each other by mirror
symmetry~\cite{Hori:2001ax}. From the \textsc{cft} point of view they are interesting
as non-trivial (yet integrable) examples of irrational
conformal field theories and provide a useful testing ground for ideas that may generalize
to other irrational \textsc{cft}s. From the point of view of string theory,
it is known that the supersymmetric coset \slc appears naturally as part of the
worldsheet \textsc{cft} that describes string propagation in the
vicinity of Calabi-Yau singularities~\cite{Ooguri:1995wj,Giveon:1999zm} 
and the near-horizon region of fivebranes in a double scaling 
limit~\cite{Sfetsos:1998xd,Giveon:1999px}. String theory in these situations is related 
holographically to a non-local, non-gravitational 
theory known as  Little String Theory~\cite{Seiberg:1997zk,Aharony:1999ks,Kutasov:2001uf} 
and is in general non-critical.

Adding branes to this context gives another interesting application.
It is well known that one can realize gauge theories with varying dimensionality and amount of
supersymmetry in Hanany-Witten setups where one considers appropriate
configurations of D-branes, orientifolds and  NS5-branes 
(see~\cite{Hanany:1996ie,Witten:1997sc}, the review~\cite{Giveon:1998sr} 
and references therein). Various non-trivial properties of
gauge theories can be studied in this way. Certain Hanany-Witten setups
can be studied directly in pertubative string theory by placing D-branes
in the non-critical string theory of the previous paragraph, which
involves, as we said, the \slc coset as part of its definition. D-branes in the $\NN=2$
Liouville theory and the \slc coset have been constructed  with \textsc{cft} methods in
\cite{Ribault:2003ss,Israel:2004jt,Eguchi:2003ik,Ahn:2003tt,Fotopoulos:2004ut,Hosomichi:2004ph} 
and will be summarized in sect.~\ref{secrev}. This formalism
was applied in the context of six-dimensional non-critical strings 
in~\cite{Fotopoulos:2005cn} where it was shown explicitly how 
to realize four-dimensional $\NN=1$ SQCD
(see also~\cite{Ashok:2005py}), and extended to models with supersymmetry 
breaking~\cite{Israel:2005zp}.  Further aspects of this theory (most notably Seiberg-duality) 
were analyzed in this context in~\cite{Murthy:2006xt}.

Orientifolds in $\NN=2$ Liouville theory and the supersymmetric \slc coset
can be studied with similar \textsc{cft} methods. One can obtain important
insights about these orientifolds from the corresponding analysis in
bosonic Liouville theory~\cite{Hikida:2002bt}, AdS$_3$~\cite{Hikida:2002fh} and 
the non-supersymmetric and supersymmetric minimal 
models~\cite{Hikida:2002ws,Brunner:2002em,Brunner:2003zm}. 
Orientifolds in $\NN=2$ Liouville
theory have been discussed previously in \cite{Nakayama:2004at}. 
The results of that paper will be reproduced here with some important 
additions as a special case of our analysis.

In order to set up our notation and to gather certain facts for later use,
we devote section~\ref{secrev} to a brief review of open and closed strings in
AdS$_3$, \slc and $\NN=2$ Liouville theory.  
Sections~\ref{geomorientsection},~\ref{O2section} 
and~\ref{O1sect} discuss different classes of orientifolds
in the $\NN=2$ Liouville theory and the supersymmetric \slc coset
and contain the main results of this paper.

In this work, we use three different approaches to uncover information
about orientifolds: the explicit form of the allowed symmetries that can be
combined with worldsheet parity, descent
of known AdS$_3$ parities to \slc and a direct modular bootstrap approach
(in some cases, we can also check our results against known conformal bootstrap
constraints). Each approach has its merits and its disadvantages, but
comparison of the information obtained in this way yields important checks 
and helps complete the picture.

In sect.~\ref{geomorientsection} we classify a set of consistent worldsheet parities.
This is most straightforward in the $\NN=2$ Liouville theory because of the simplicity of
the worldsheet action. This approach gives naturally ${\rm O2}$- and
${\rm O}1$-planes that extend towards the weak coupling region of
the theory. One of the interesting results of this analysis is a parity
that can be used to construct non-critical, non-tachyonic type $0'$B
string vacua. The explicit construction of these vacua will appear in a companion paper
\cite{noncritzeroprimeB}. We also analyze parities
that descend from AdS$_3$. This point of view gives a natural set of orientifolds with the
geometry of ${\rm O0}$-, ${\rm O1}$- and ${\rm O2}$-planes on \slc.

In sect.~\ref{O2section} we proceed to analyze with exact \textsc{cft} methods
the crosscap wave-functions of two B-type parities on \slc. The exact result 
reproduces the semiclassical asymptotic Klein bottle amplitude
based on the known action of the parities, but also reveals the presence of
an additional localized orientifold contribution. We propose that the latter 
corresponds to one of the ${\rm O0}$-planes that was found in 
sect.~\ref{geomorientsection}. Hence, we find that the \textsc{cft} gives 
naturally not a single ${\rm O0}$- or an ${\rm O2}$-plane, but a specific 
combination of the two.\footnote{This reminds of the D2-branes 
of~\cite{Ribault:2003ss} which exhibit a localized D0-brane charge.}
We provide a physical interpretation of this result in the context of Hanany-Witten 
setups.

In the final section, we discuss the crosscap wave-function of an
A-type orientifold that gives an ${\rm O}1$-plane on \slc.
We obtain this result by descent from an Euclidean AdS$_2$
orientifold in Euclidean AdS$_3$. The AdS$_2$ orientifold can be obtained
from an H$_2$ orientifold with an SL(2,$\C$) rotation. This provides
and independent derivation of the AdS$_2$ crosscap wavefunction 
in~\cite{Hikida:2002fh}.

Three appendices supplement the material of the main text. In the first two
appendices we summarize some of our conventions and list the known
D-brane wave-functions for quick reference. In the third appendix we derive
the $\PP$-modular transformation properties of the identity character which
will be instrumental in the modular bootstrap approach of sect.~\ref{O2section}.
The derivation appearing in appendix~\ref{Pmatapp} is a generalization of the one appearing
in \cite{Nakayama:2004at} but with some important differences.

\vskip0.5cm

\noindent
{\it Note added.} We are aware that Sujay Ashok, Sameer Murthy and Jan Troost
have been exploring independently a related subject.

%%%%%%%%%%%%%%%%%%%%%%%%%%%%%%%%%%%%%%%%%%%%%%%%%%

\boldmath
\section{Strings and branes in \slc \& $\NN=2$ Liouville } %theory}
\unboldmath
\label{secrev}
We start with a brief review of the \slc conformal field theory and its mirror $\NN=2$
Liouville theory. This will help us set up our conventions and gather some 
important facts for later use. For more details on the material reviewed in this section
we refer the reader to the original references cited below.

%%%%%%%%%%%%%%%%%%%%%%%%%%%%

\subsection*{Closed strings in AdS$_3$}
String theory on AdS$_3$~\cite{Balog:1988jb} with an \textsc{ns-ns}  two-form flux is an exact
solution of string theory, whose background fields read, in global coordinates
\begin{equation}
\di s^2 = \alpha ' k \left[ \di \rho^2 + \sinh^2 \rho \, \di \phi^2
-\cosh^2 \rho\,  \di t^2 \right] , \quad
H   =  2 \alpha' k \cosh \rho \sinh \rho \ \di \rho \wedge
\di \phi \wedge \di t \,
\label{globalcord}
\end{equation}
with a constant dilaton. The global SO(2,2) symmetry of this space-time is enhanced to
an affine $\widehat{\mathfrak{sl}} (2,\mathbb{R})_\textsc{l} \times
\widehat{\mathfrak{sl}} (2,\mathbb{R})_\textsc{r}$ since we can
take the worldsheet theory as the \textsc{wzw} model for the
group \slr. To be more precise, AdS$_3$ space-time with a non-compact global
time $t$ corresponds to the {\it universal cover} of \slr.\footnote{
For some applications it is useful to consider the single cover of \slr for which the
time is periodic $t \sim t + 2\pi$.} In order to obtain superstring backgrounds, 
one can define the super-\textsc{wzw} model for \slr by adding three free 
worldsheet fermions of signature $(-,+,+)$. The central charge
of this $\mathcal{N} =1$ superconformal theory is $c = 9/2+6/k$.

Primary states of the model are classified in terms of 
$\widehat{\mathfrak{sl}} (2,\mathbb{R})$
representations, that can be twisted by an outer automorphism
called {\it spectral flow}~\cite{Henningson:1991jc,Maldacena:2000hw}. 
%(for $w\neq 0$). 
Their conformal weights read, in the \textsc{ns-ns} sector:
\begin{equation}
\Delta_\mathfrak{sl} = -\frac{j(j-1)}{k}-wm+\frac{kw^2}{4} \quad , \qquad
\bar{\Delta}_\mathfrak{sl} = -\frac{j(j-1)}{k}-w\bar{m}+\frac{kw^2}{4}~,
\end{equation}
where $(m,\bar m)$ label the primaries of the elliptic sub-algebra
$(J^3,\bar J^3)$ and $w$ is the spectral flow parameter. Space-time energy is given by
$E=m+\bar m$ whereas the angular momentum (conjugate to $\phi$) is 
$n=m-\bar m \in \mathbb{Z}$. The unitary closed string spectrum
is made of {\it continuous representations} with $j \in \hl + i \mathbb{R}_+$
and {\it discrete representations} in the range $\hl < j< \nicefrac{k+1}{2}$. 
We refer the reader to appendix~\ref{appchar} for more details about these representations.

%%%%%%%%%%%%%%%%%%%%%%%%%%%%%%%%%

\subsection*{Closed strings in \slc}
The \slc conformal field
theory~\cite{Elitzur:1991cb,Mandal:1991tz,Witten:1991yr,Dijkgraaf:1991ba}
is obtained from \slr as a gauged \textsc{wzw} model.
One possibility is to perform an {\it axial gauging} of the elliptic subalgebra, 
corresponding to the time-translation symmetry $t \to t + \lambda_\text{a}$. 
This symmetry has no fixed point, hence the background is non-singular
\begin{equation}
\di s^2 = \alpha ' k \left[ \di \rho^2 + \tanh^2 \rho \, \di \phi^2 \right] \ ,  \quad
\Phi = \Phi_0 - 2\ln  \cosh \rho \, ,
\label{cigarback}
\end{equation}
and has the interpretation of a two-dimensional Euclidean black hole, the {\it cigar}.
Using the standard gauging construction, the primary states of the coset
can be obtained from \slr
primaries with $m+\bar m=0$, with conformal weights (for \textsc{ns-ns} primaries)
\begin{subequations}
\begin{align}
\Delta_\mathfrak{cig} &= \Delta_\mathfrak{sl} + \frac{m^2}{k} = -\frac{j(j-1)}{k}
+ \frac{(n+kw)^2}{4k}\\
\bar{\Delta}_\mathfrak{cig} &= \bar{\Delta}_\mathfrak{sl} + \frac{\bar{m}^2}{k}
= -\frac{j(j-1)}{k} +\frac{(n-kw)^2}{4k}
\end{align}
\end{subequations}
The periodicity $\phi \sim \phi + 2\pi$ of AdS$_3$, see eqn.~(\ref{globalcord}), is 
inherited by the coset. At the asymptotic $\rho \rightarrow \infty$ region, $\phi$
becomes a canonically normalized free boson at radius $\sqrt{\alpha'k}$. 
One identifies $n$ as the momentum of this boson, 
and $w$ as its winding number. Correlators of this theory can be computed by descent 
from the corresponding quantities in H$_3^+$~\cite{Teschner:1997ft,Teschner:1999ug}.

The leading order solution of the background fields~(\ref{cigarback})
is exact to all orders in $\nicefrac{1}{k}$ as the superconformal symmetry
is enlarged to $\mathcal{N}=2$~\cite{Bars:1992sr,Tseytlin:1993my}. 
However it receives non-perturbative corrections in the form
of a "winding condensate"~\cite{fzz,Kazakov:2000pm,Hori:2001ax,Giveon:2001up,Tong:2003ik}.
In the asymptotic region $\rho \to \infty$ where the fields $\rho$, $\phi$ 
and their fermionic superparters $\psi^\pm = \psi^\rho \pm i\psi^\phi$ are free
one can write the winding condensate as a worldsheet interaction of the form\footnote{Henceforth
we will denote the right-moving fields with a bar, or with an explicit $L$ or $R$ subindex
to distinguish between left- and right-movers.}
\begin{equation}
\delta S =  \frac{k}{2\pi}\int \di^2 z \,  e^{-k\rho}
\left[i\mu \, \psi^+ \bar \psi^- e^{ik (\phi_L-\phi_R)}
+ i\mu^\dag\,\psi^- \bar \psi^+ e^{-ik (\phi_L-\phi_R)}\right]\, .
\label{liouvintwind}
\end{equation}

Another consistent theory is defined by a {\it vector gauging} that refers to the symmetry
$\phi \to \phi + \lambda_\text{v}$ and gives the constraint $m-\bar m = 0$.
Since $\rho=0$ is a fixed point of this isometry, the leading order metric
$\di s^2 = \alpha ' k [ \di \rho^2 + \text{cotanh}^2\, \rho \, \di \tilde{\phi}^2 ]$ is a singular
geometry known as the {\it trumpet}. As the geometric interpretation breaks
down it is usually more appropriate to view this model as an $\mathcal{N}=2$ Liouville
theory~\cite{Kazakov:2000pm,Hori:2001ax,Giveon:2001up},  defined as a free $\mathcal{N}=2$ linear
dilaton theory perturbed by a momentum condensate T-dual to~(\ref{liouvintwind}), {\it i.e.} 
with $\phi_L - \phi_R$ replaced by $\tilde{\phi}$ (and $\bar \psi^\pm$ by $\bar \psi^\mp$). 
We will discuss this model in more detail below.

Contrary to the axial gauging, the vectorially gauged \slc coset is sensitive to
the cover of \slr \cite{Israel:2003ry}. 
Starting with the universal cover of AdS$_3$ we obtain a non-compact coordinate 
$\tilde \phi$ -- this coordinate is the time coordinate $t$ in disguise.
Starting with the single cover, the field $\tilde \phi$ corresponds in the asymptotic
region $\rho \to \infty$ to a free boson at radius $\sqrt{\alpha ' k}$. This defines a consistent
\textsc{cft} at the non-perturbative level {\it only} if the level $k$ is an integer,
otherwise the momentum condensate dual
to~(\ref{liouvintwind}) is not periodic. For irrational $k$, the only consistent
theory with a momentum condensate and finite radius is obtained
with T-duality from the cigar~(\ref{cigarback}); in that case the radius of the
transverse coordinate is $\sqrt{\alpha' /k}$.\footnote{In the $\mathcal{N}=2$
Liouville terminology, this is the "minimal radius" solution.} This model {\it cannot} be 
obtained as a gauging of AdS$_3$; however, for integer $k$ it is the $\mathbb{Z}_k$ 
orbifold of the trumpet at radius $\sqrt{\alpha ' k}$. We summarize
the various possibilities in table~\ref{tabmodels}.\footnote{For rational $k$ there
are other possibilities that will not be quoted here, see {\it e.g.}~\cite{Eguchi:2003ik}.}
The last two rows correspond to the same models,
provided the radii are equal.
\begin{table}[!ht]
\centering
\begin{tabular}{|l|l|l|}
\hline
&k integer& k arbitrary \\
\hline
Axial gauging & $R= \sqrt{\alpha' k}$ & $R= \sqrt{\alpha' k}$\\
Vector gauging & $R= \infty$ or $R = \sqrt{\alpha' k}$ & $R= \infty$\\
$\mathcal{N}=2$ Liouville & $R= \infty$, $R = \sqrt{\alpha' k}$ or $R = \sqrt{\alpha'/ k}$
& $R= \infty$, $R = \sqrt{\alpha' /k}$\\
\hline
\end{tabular}
\caption{Various \slc theories}
\label{tabmodels}
\end{table}

%%%%%%%%%%%%%%%%%%%%%%%%%%%%%%

\subsection*{D-branes and boundary \textsc{cft}}
Various D-branes  have been constructed, using the tools of boundary conformal field theory, 
in H$_3^+$~\cite{Lee:2001gh,Ponsot:2001gt} and later in
Lorentzian AdS$_3$~\cite{Israel:2005ek}. They 
are classified by the gluing conditions imposed on
the $\widehat{\mathfrak{sl}}(2,\mathbb{R})$
currents~\cite{Bachas:2000fr}. Extending the coset construction to 
\textsc{bcft}, corresponding branes 
have been obtained in \slc \cite{Ribault:2003ss,Israel:2004jt}. These results are mainly 
in agreement with other approaches, such as 
modular bootstrap~\cite{Eguchi:2003ik,Ahn:2003tt,Fotopoulos:2004ut} and
conformal bootstrap based on the $\mathcal{N}=2$
superconformal algebra~\cite{Hosomichi:2004ph}. 
In what follows we summarize the branes that will be most 
interesting for the analysis below. The corresponding 
wave-functions, {\it i.e.} the coefficient of the one-point functions 
on the disc, are summarized in app.~\ref{appwave-functions}. 

\subsubsection*{D0-branes}
 
The D(-1)-brane (i.e. point-like in space-time) of AdS$_3$ can be obtained with the current algebra gluing conditions 
\mbox{$J^3 = -\bar{J}^3 |_{z=\bar z}$, $J^\pm = -\bar{J}^{\mp} |_{z=\bar z}$} using the 
conventions of~\cite{Ponsot:2001gt}. Its main property is that the spectrum of 
open strings attached to it contains only the {\it identity representation}
of $\widehat{\mathfrak{sl}}(2,\mathbb{R})$. 
It is located at $\rho=0$ and $t=0,\pi$ on the single cover, {\it i.e.} the brane is made of 
two copies. D0-branes in the coset \slc can be obtained from the D(-1)-brane of AdS$_3$ by descend.
Corresponding D0-branes in $\NN=2$ Liouville theory exist by mirror symmetry.

Let us consider first the D0-brane of
non-compact $\mathcal{N}=2$ Liouville theory, or non-compact "trumpet" background.
In the $\mathcal{N}=2$ terminology this is an A-type brane. The quantization of the
brane position at  $\tilde \phi = 2\pi  r/k$
with $r \in \mathbb{Z}$ has its origin in the $\mathcal{N}=2$
Liouville potential that breaks the translation symmetry along
$\tilde{\phi}$ to a $\mathbb{Z}$  subgroup generated by
$\tilde \phi \to \tilde \phi + 2\pi/k$. It is quite analogous
to the "special points" at the boundary of the disc in
SU(2)/U(1)~\cite{Maldacena:2001ky}.\footnote{While the SU(2)/U(1) geometry
is conformal to the interior of the unit disc, the trumpet is conformal to the
exterior of the disc.} 
Using the conventions of app.~\ref{appchar}, 
we can write the annulus amplitude in this theory (in the \textsc{ns} sector), for open strings 
stretched between  D0-branes sitting at  $\phi=2\pi r/k$ and $2\pi r'/k$, as: 
\begin{equation}
\mathcal{A}_{r \, r ' } (t) =  ch_\mathbb{I} ( r' -r;it)\oao{0}{0}\, .
\label{amplD0noncomp}
\end{equation}
It contains only one identity character of the $\mathcal{N}=2$ superconformal
algebra.

The cigar \textsc{cft} (T-dual to the minimal radius
$\mathcal{N}=2$ Liouville theory) is obtained by modding out the
$\mathbb{Z}$ subgroup of the translation symmetry that is not broken 
non-perturbatively. Since this symmetry has no fixed point 
one can obtain the boundary state by summing over the images under 
the orbifold action. As a result, the brane carries no label (apart from the usual
labels characterizing the fermionic boundary conditions). It is  a D0-brane localized
at the tip $\rho=0$ with B-type boundary conditions. One obtains the
annulus amplitude by summing~(\ref{amplD0noncomp}) over $r \in \Z$.
The closed string one-point functions on the disc in all fermionic sectors
are summarized in app.~\ref{appwave-functions}.

Finally, for integer $k$ one can consider the trumpet at radius $\sqrt{\alpha' k}$,
the vector gauging of the single cover. In this case, one mods out the
non-compact model by the subgroup $\tilde \phi \to \tilde \phi + 2\pi$. 
The annulus amplitude for the D0-brane is
obtained from a partial  summation of~(\ref{amplD0noncomp}) as
$r = \hat r + k \mathbb{Z}$. One can repackage the result using the
{\it extended characters} defined in app.~\ref{appchar}. The result in the \textsc{ns-ns}
sector is
\begin{equation}
\mathcal{A}^\mathrm{vect.\,}_{\hat r \, \hat r ' } (t) = 
Ch_\mathbb{I} (\hat r' -\hat r;it) \oao{0}{0}\, .
\end{equation}
Summing over $\hat r \in \Z_k$  gives the annulus amplitude
for the cigar written with extended characters.

\subsubsection*{D1-branes}
The D1-branes of the two-dimensional black hole descend from the AdS$_2$ branes of
AdS$_3$~\cite{Bachas:2000fr}. They are characterized by A-type boundary
conditions of the $\mathcal{N}=2$ superconformal algebra (\textsc{sca}). 
Their embedding equation is
\beq
\label{embed}
\sinh \rho \, \sinh (\phi-\hat \phi) = \sinh \hat \rho
\eeq
with two continuous parameters $(\hat \rho, \hat \phi)$. 
In the asymptotic cylinder region, these branes have the shape of two antipodal 
D1-branes at $\phi=\hat \phi,\ \hat \phi + \pi$. 
The open string spectrum comprises only of continuous representations,
with a non-trivial density of states~\cite{Ribault:2003ss,Israel:2004jt} associated with the 
\begin{figure}[!ht]
\centering
\vskip-.5cm
\epsfig{file=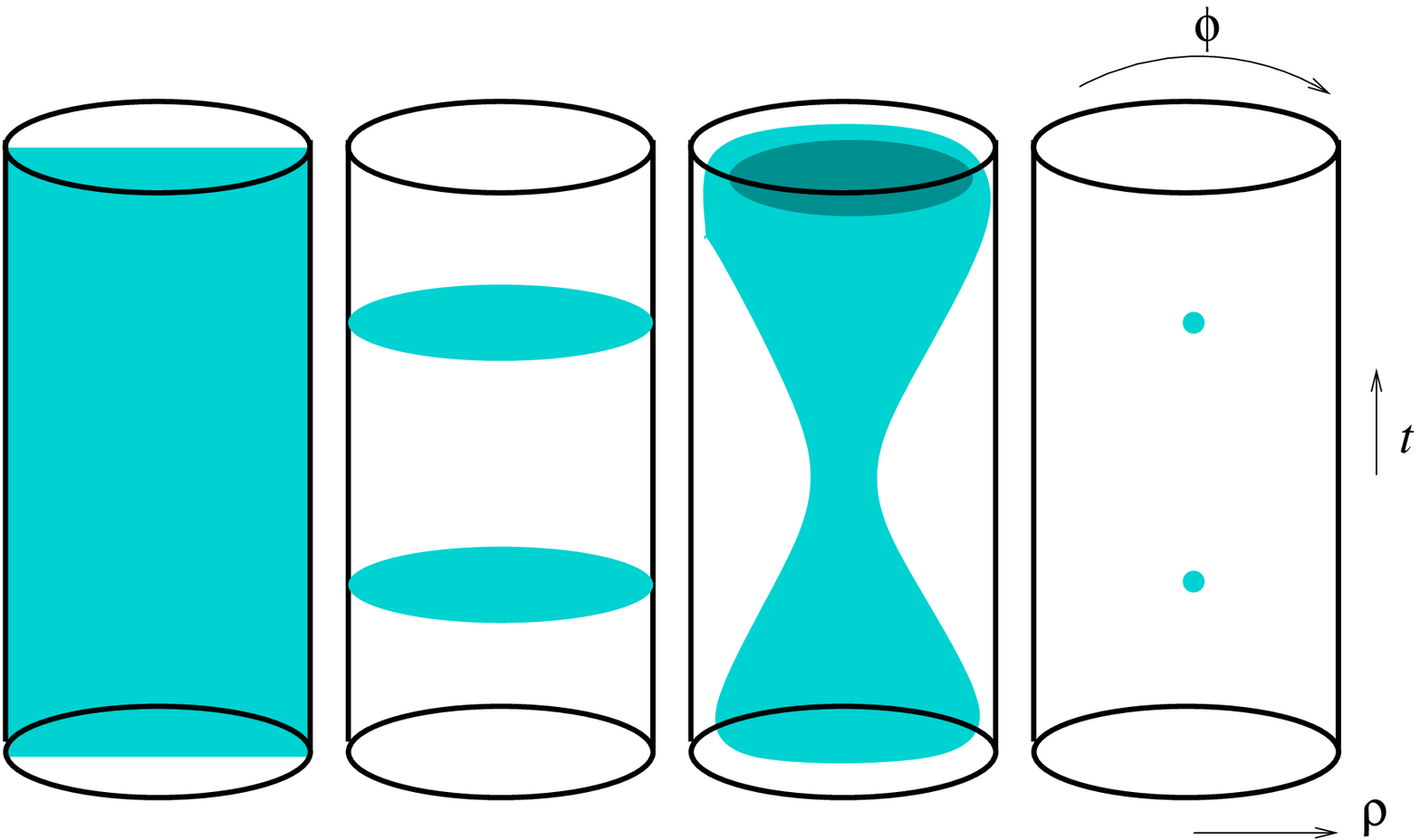,width=80mm}
\vskip3mm \hskip-5mm
\epsfig{file=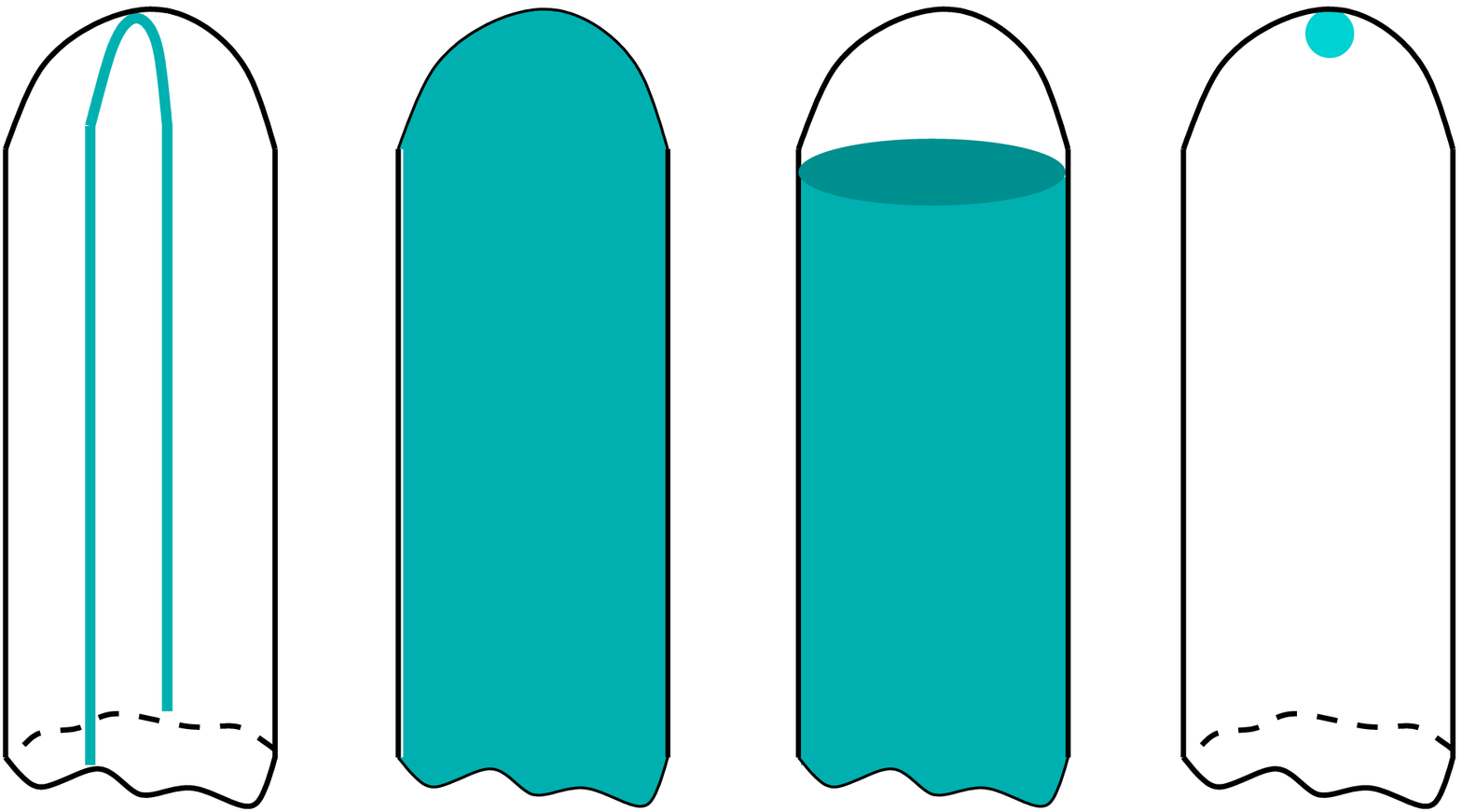,width=80mm}
\caption{\it Geometries of various D-branes on the single cover of AdS$_3$ (upper picture) and
in the cigar (lower picture). From left to right: AdS$_2$ brane with $\hat \rho=0$,
H$_2$ branes with $\hat \sigma = 0$, dS$_2$ brane and D(-1)-branes.
Roughly speaking, the cigar coset is obtained by projecting onto an
"horizontal slice" of the cylinder and the trumpet onto a "vertical slice".
}
\label{branesgeom}
\end{figure}
boundary two-point function. The relevant one-point function on the disc 
is given by eqn.~(\ref{appbae}). In the non-compact trumpet /
$\mathcal{N}=2$ Liouville theory, one obtains a D2-brane with 
a worldvolume $\rho>\hat \rho$ endowed with a magnetic field.

\subsubsection*{D2-branes}
Finally, one can define space-like branes in AdS$_3$ with
H$_2$ geometry. In H$_3^+$ these branes are equivalent to
AdS$_2$ branes by SL(2,$\mathbb{C}$) rotation. By descent they give 
D2-branes on the cigar, with B-type boundary conditions, carrying a magnetic
field~\cite{Ribault:2003ss}. The latter is quantized because the brane 
carries a D0-brane charge near the tip of the cigar. In the non-compact 
trumpet / $\mathcal{N}=2$ Liouville theory, the AdS$_2$ branes give D1-branes with 
embedding equation $\sinh \rho \sin (\tilde \phi - \hat{\varphi}) = \sin \hat \sigma$. In this case, 
the quantization of $\hat \sigma$ is interpreted as the requirement 
that the brane ends on one of the special points at $\rho=0$~\cite{Israel:2005fn} 
(the parameter $\hat{\varphi}$ is also quantized). However, these branes seem to be
inconsistent for irrational $k$ since their open string spectrum contains 
negative multilplicities~\cite{Israel:2004jt}. A different class of D2-branes, related 
to dS$_2$ geometries in Lorentzian AdS$_3$ (this class of branes cannot descend 
from branes in H$_3^+$), have been constructed using modular bootstrap 
methods in \cite{Fotopoulos:2004ut} and conformal bootstrap methods in $\NN=2$
Liouville theory in \cite{Hosomichi:2004ph} (for a \textsc{dbi} analysis of these 
branes see \cite{Fotopoulos:2003vc}). D2-branes in this class are free of the 
abovementioned problems since their open string spectrum is made of 
continuous representations only. They exhibit a double-sheeted structure
that covers the domain $\rho > \rho_\text{min}$ and are labeled by a continuous 
parameter that characterizes the minimal distance $\rho_\text{min}$ from the tip 
of the cigar and a $\mathbb{Z}_2$-valued Wilson line. Their boundary state
wave-functions are summarized in app.\ \ref{appwave-functions}.
It should be pointed out that these D2-branes are related to D2-branes of the 
first category with an overcritical magnetic field~\cite{Fotopoulos:2003vc}.

We sketch the brane geometries in fig.~\ref{branesgeom}.
Note that other types of branes, whose geometrical and physical
interpretation has not yet been elucidated, were
considered in~\cite{Ribault:2005pq,Adorf:2007ah}.

%%%%%%%%%%%%%%%%%%%%%%%%%%%%%%%%%%%%%%%%%%%%

\section{Parities and the geometry of orientifolds}
\label{geomorientsection}
In this section we discuss the orientifolds of the supersymmetric  
\slc coset and the $\NN=2$ Liouville theory from the perspective of the parity
symmetries. This point of view allows for a first look at the semiclassical
features of the orientifolds and provides a useful guide for the exact
analysis of the next section. First, we classify a set of A- and B-type 
parities in $\mathcal{N} = 2$ Liouville theory. Then we 
repeat the exercise with parities in \slc inherited from AdS$_3$.
The simultaneous analysis of parity symmetries in both theories is useful,
because certain orientifolds are easier to analyze in one
theory than the other. Of course, at the end of the day the orientifolds of 
these theories are related by mirror symmetry. We comment on this correspondence 
at the end of this section.

%%%%%%%%%%%%%%%%%%%%%%%%%%%%%%

\subsection*{Parities in worldsheet theories with $\NN=(2,2)$ supersymmetry}
In a two-dimensional \textsc{qft} with $\NN=(2,2)$ supersymmetry
one can define a natural set of parity symmetries (we recommend
\cite{Brunner:2003zm} for an excellent discussion of the general situation).
Here it will be useful to highlight the main points of these symmetries.
The two bosonic coordinates of the $\NN=(2,2)$ superspace will be denoted as
$(z,\bar z)$ and the four fermionic coordinates as $\theta,\bar \theta$ and
$\theta^\dagger, \bar \theta^\dagger$. As above, 
we denote the right-movers with a bar and reserve the dagger for the 
notation of complex conjugate quantities.
The $\NN=2$ superconformal algebra generators will be denoted
as $T(z), J(z), G^{\pm}(z)$ for the left-movers with an analogous notation
for the right-movers.

There are two basic parities in a theory with $\NN=(2,2)$ supersymmetry
that reverse the worldsheet chirality (exchanging the left- and right-movers)
while preserving the holomorphy of the $\NN=2$ supersymmetry. They are
known as A- and B-type\footnote{These parities are analogous to
A- and B-type boundary conditions as we will see more explicitly below.}
and are defined by the worldsheet action
\begin{subequations}
\begin{align}
\label{parityaa}
\Omega_A~: & \quad (z,\bar z,\theta, \bar \theta,\theta^\dagger,\bar \theta^\dagger)
\rightarrow
(\bar z,z, -\bar \theta^\dagger, -\theta^\dagger,-\bar \theta,- \theta)
~, \\
\label{parityab}
\Omega_B~: & \quad (z,\bar z,\theta, \bar \theta,\theta^\dagger,\bar \theta^\dagger)
\rightarrow
(\bar z,z, \bar \theta, \theta,\bar \theta^\dagger,\theta^\dagger)
~.
\end{align}
\end{subequations}
They act on the supercurrents as
\begin{subequations}
\begin{align}
\label{parityac}
\Omega_A ~: &\quad G^{\pm}(z) \rightarrow \bar G^{\mp}(\bar z)~, ~ ~
\qquad \qquad \Omega_A ~: \quad \bar G^{\pm}(\bar z) \rightarrow G^{\mp}(z)
~,\\
\label{parityad}
\Omega_B ~: &\quad G^\pm(z) \rightarrow \bar G^\pm (\bar z)~, ~ ~
\qquad \qquad \Omega_B ~: \quad \bar G^\pm (\bar z) \rightarrow G^\pm (z)
\end{align}
\end{subequations}
and on the R-symmetry currents as
\begin{subequations}
\begin{align}
\label{parityae}
\Omega_A ~: &\quad J(z) \rightarrow -\bar J(\bar z)~, ~ ~
\qquad \qquad\Omega_A ~: \quad \bar J(\bar z) \rightarrow - J(z)
~, \\
\label{parityaf}
\Omega_B ~: &\quad  J(z) \rightarrow \phantom{-}\bar J(\bar z)~, ~ ~
\qquad \qquad \Omega_B ~: \quad \bar J(\bar z) \rightarrow  J(z)
~.
\end{align}
\end{subequations}
The two parities are exchanged by mirror symmetry.
The same thing happens with boundary conditions, where
mirror symmetry exchanges A- and B-type branes.

One can generalize the above parities by combining them
with internal discrete symmetries $\tau$ of the theory (we
will present explicit examples of such symmetries in
$\NN=2$ Liouville theory and its mirror dual below). In this
way one can formulate more general A- and B-type parities
of the form
\beq
\label{parityag}
\PP_A= \tau_A \cdot \Omega_A~, ~ ~
\PP_B =\tau_B \cdot \Omega_B
\eeq
which are still acting on the supercurrents as in~\mbox{(\ref{parityac}, \ref{parityad})} 
and on the R-symmetry currents 
as in~\mbox(\ref{parityae}, \ref{parityaf}) provided, of course, that the internal
symmetries preserve the $R$-symmetry currents.
A general example of such A- and B-type parities are the $(\alpha,\beta)$
parities
\beq
\label{parityai}
\PP_{A_{\alpha,\beta}}=e^{-i \alpha J-i \beta \bar J} \cdot \Omega_A~, ~~
\PP_{B_{\alpha,\beta}}=e^{-i \alpha J-i \beta \bar J} \cdot \Omega_B
~,
\eeq
where one combines the basic worldsheet parities $\Omega_A$, $\Omega_B$
with $U(1)_R$ rotations. It should be pointed out that for general values of 
$\alpha$ and $\beta$ these parities are not involutive ({\it i.e.} $\PP^2 \neq 1$). 
Also they are non-geometric, because they treat the left- and right-movers 
asymmetrically. The latter can have interesting consequences for the resulting theory;
we will mention an interesting example below.

%%%%%%%%%%%%%%%%%%%%%%%%%%%%%

\subsection*{Parities in $\NN=2$ Liouville theory}
We are now in position to examine the A- and B-type
parity symmetries of the $\NN=2$ Liouville action\footnote{A related discussion
of Landau-Ginzburg models can be found in \cite{Brunner:2003zm}.}
(we set $\alpha'=2$)
\begin{equation}
S = \frac{1}{8\pi} \int \di^2 z\, \di^4 \theta \, \Phi\bar{\Phi} + 
\frac{\mu}{2\pi} \int \di^2 z \, \di \theta \di \bar \theta  \, e^{-\sqrt{\frac{k}{2}} \Phi}
+ \frac{\mu^\dag}{2\pi} \int \di^2 z \, \di \theta^\dag \di \bar{\theta}^\dag \, 
e^{-\sqrt{\frac{k}{2}} \bar{\Phi}}
\label{Liouvact}
\end{equation}
written in terms of a chiral $\mathcal{N}=2$ superfield 
\begin{equation}
\Phi = r + i \varphi  + i\sqrt{2}\theta \psi^+  + i\sqrt{2} \bar{\theta} 
\bar{\psi}^+ + 2\theta \bar{\theta} F + \cdots
\end{equation}
$r$ denotes the radial direction and $\varphi$ the angular direction.
This theory is superconformal provided the background charge satisfies $Q = \sqrt{2/k}$. 
In the asymptotic $r \to \infty$ weakly coupled region, the 
left and right $U(1)_R$ currents read, in terms of the component fields:
\begin{subequations}
\begin{align}
J &= \psi^+ \psi^- + i Q  \partial \varphi \\
\bar J &= \bar \psi^+ \bar \psi^- + i Q  \bar \partial \varphi
\end{align}
\end{subequations}

The  potential coming from the superfield action~(\ref{Liouvact}) is similar to the
interaction term~(\ref{liouvintwind}) after the change of normalization of the fields 
$(\rho,\phi) = (2k)^{-\hl} (r,\varphi)$ and T-duality. In this alternate description of the \slc theory, 
the first term in the asymptotic expansion of the cigar geometry, eqn.~(\ref{cigarback}), comes 
as a  a correction to the K\"ahler potential~(\ref{Liouvact})
\begin{equation}
\delta S = \mu_\text{cig} \int \di^2 z \, \di^4 \theta \ e^{-\frac{\Phi+\Phi^\dag}{\sqrt{2k}}}
~.
\label{cigarcorr}
\end{equation}

The basic A- and B-type parities $\Omega_A$ and $\Omega_B$
leave the fermionic measure $d\theta d\bar \theta d\theta^\dagger
d\bar \theta^\dagger$
invariant and act on the $\NN=2$ Liouville chiral superfield $\Phi$ as\footnote{By definition
$\Omega_B$ takes the fermion bilinear $\psi^\epsilon \bar \psi^{\bar \epsilon} \to
-\psi^{\bar \epsilon} \bar \psi^\epsilon$, $\epsilon,\bar \epsilon =\pm 1$.
The standard $\Omega$ worldsheet parity acts on the fermions as $\psi^\pm \to \bar \psi^\pm$,
$\bar \psi^\pm \to -\psi^\pm$ and leaves the fermion bilinear 
$\psi^\epsilon \bar \psi^{\bar \epsilon}$ invariant. The relation between $\Omega$
and $\Omega_B$ is therefore $\Omega_B=(-)^{\bar F} \Omega$, where $\bar F$ is the right-moving
worldsheet fermion number.}
\beq
\label{parityba}
\Omega_A ~:~ \Phi(z,\bar z,\theta,\bar \theta) \rightarrow
\left\{ \Phi(\Omega_A(z,\bar z, \theta,\bar \theta)) \right\}^\dagger
~,~~
\Omega_B ~:~ \Phi(z,\bar z, \theta, \bar \theta) \rightarrow
\Phi(\Omega_B(z,\bar z,\theta,\bar \theta))
~.
\eeq
Hence, if we write the $\NN=2$ Liouville action \eqref{Liouvact} as
\beq
\label{paritybb}
S=S_\text{K\"ahler}+S_\text{Liouville}(\mu)+S^\dagger_\text{Liouville}(\mu^\dagger)
\eeq
we can easily verify that the K\"ahler (kinetic) part of the action is invariant
under both $\Omega_A$ and $\Omega_B$, but the superpotential parts
are transforming as
\beq
\label{paritybc}
\Omega_A~:~ S_\text{Liouville}(\mu) \rightarrow S_\text{Liouville}(\mu)^\dagger
~,
\eeq
\beq
\label{paritybd}
\Omega_B~:~S_\text{Liouville}(\mu) \rightarrow S_\text{Liouville}(-\mu)
~.
\eeq
Consequently, $\Omega_A$ is a true symmetry of the $\NN=2$ Liouville
theory only when $\mu \in \R$.\footnote{This reduction of the closed
string moduli space to a real subspace is a usual feature of $A$-type
orientifolds in vacua with $\NN=(2,2)$ 
worldsheet supersymmetry \cite{Brunner:2003zm,Brunner:2004zd}.}
On the other hand, $\Omega_B$ cannot be a true symmetry
unless we take $\mu=0$, {\it i.e.} unless we drop the $\NN=2$ Liouville
interaction term to be left with a free linear dilaton theory.

The $\NN=2$ Liouville theory has two obvious involutive parities that can be
used to define B-type orientifolds. These are a parity $s$
that shifts the angular coordinate $\varphi$ by half a period, {\it i.e.} 
\beq
\label{paritybe}
s~: \quad \varphi  \rightarrow \varphi+\pi Q
~
\eeq
and the fermionic parity $(-)^{\bar F}$ where $\bar F$ is the right-moving
worldsheet fermion number. Under the parities
$\PP_B=s \cdot \Omega_B$ and
$\PP'_B=(-)^{\bar F} \cdot \Omega_B$ the full $\NN=2$ Liouville action
is invariant.\footnote{It is worthwile mentioning that the parity 
$\PP_B=s \cdot \Omega_B$ has no analogue in sine-Liouville theory (the bosonic 
cousin of $\mathcal{N}=2$ Liouville theory) since the potential is 
odd under $s$.} $(-)^{\bar F}$ can also be combined with $\Omega_A$
to give a consistent A-type parity.

Given the above symmetries of the classical $\mathcal{N}=2$ Liouville theory 
\beq
\label{paritybf}
\PP_A=\Omega_A~, ~ ~ \PP'_A=(-)^{\bar F} \cdot \Omega_A~, ~ ~ 
\PP_B=s \cdot \Omega_B~, ~ ~
\PP'_B=(-)^{\bar F} \cdot \Omega_B
\eeq
we can define the corresponding $(\alpha,\beta)$ $U(1)_R$ rotated
versions as
\beq
\label{paritybg}
\PP_{A_{\alpha,\beta}}=\Omega_{A_{\alpha,\beta}}~, ~ ~
\PP_{A_{\alpha,\beta}}=(-)^{\bar F} \cdot \Omega_{A_{\alpha,\beta}}~, ~ ~ 
\PP_{B_{\alpha,\beta}}=s \cdot \Omega_{B_{\alpha,\beta}}~, ~ ~
\PP_{B_{\alpha,\beta}}=(-)^{\bar F} \cdot \Omega_{B_{\alpha,\beta}}
~.
\eeq
For general $\alpha$ and $\beta$ these parities are non-involutive.

The non-perturbative consistency of these parities requires that
they leave invariant the cigar interaction~\eqref{cigarcorr}. One can check
that this requirement is trivially satisfied by all of the above parities.

In the context of type 0 non-critical strings  the parity 
$\PP_{B_{0,0}}$, as well as $\PP'_{B_{0,\pi}}$, 
leads to an interesting theory of non-oriented type 0 strings without
closed string tachyons, no fermions and no massless tadpoles,
which is a cousin of the type $0'$B theory in ten dimensions.
In the context of two dimensional type 0 strings based on $\NN=1$
Liouville theory it was pointed out in \cite{Gomis:2003vi, Bergman:2003yp} 
that the type $0'$B projection
is not allowed, because it projects out the $\NN=1$ Liouville interaction.
In $\NN=2$ Liouville theory we see, however, that this is no longer the case and
the type $0'$B projection is indeed possible. A detailed analysis of this theory 
will appear elsewere~\cite{noncritzeroprimeB}.

%%%%%%%%%%%%%%%%%%%%%%%%%%%%%%%%%

\subsection*{Geometric parities in AdS$_3$ and its cosets} 

In this subsection we take an orthogonal route to look at the possible geometric
parities in the axial \slc coset, i.e. the cigar geometry given by eqn.~(\ref{cigarback}). For the moment, let us
forget about the details of the worldsheet fermions and supersymmetry and  
look first at the parity symmetries of the bosonic \slr WZW model, 
following~\cite{Bachas:2001id}. Possible
orientifold projections combine the worldsheet orientation
symmetry and a $\mathbb{Z}_2$ isometry. The isometries of the manifold 
are most easily described by embedding AdS$_3$ in $\mathbb{R}^{2,2}$ 
with the equation:  
\begin{equation}
(X^0)^2+(X^3)^2-(X^1)^2-(X^2)^2 = \alpha ' k \, . 
\end{equation}
We will consider geometric $\mathbb{Z}_2$ symmetries that are combinations
of the parities $X^i \to - X^i$ for $i=1,\ldots 4$. 
The global coordinates on the group manifold, 
see the metric~(\ref{globalcord}), are defined as
\begin{equation}
X^0 \pm i X^3 = \sqrt{\alpha' k} \cosh \rho\ e^{\pm i t} \quad , \qquad
X^1 \pm i X^2 = \sqrt{\alpha' k} \sinh \rho\ e^{\pm i \phi}\, .
\end{equation}
In the \textsc{wzw} model the parity symmetry has to reverse the
orientation of the target space manifold in order to preserve
the Wess-Zumino term $\int (g^{-1}\di g)^{\wedge 3}$, {\it i.e.} the coupling 
to the \textsc{ns-ns} two-form. In view of the applications to the
coset we don't restrict ourselves to parities with an invariant 
timelike hypersurface. We give the various inequivalent choices ($i.e.$ which 
are not related one the the other by the isometries of the manifold) for the orientifold
geometry in table~\ref{paritgeom}.
\begin{table}[!ht]
\begin{equation*}
\begin{array}{|c|l|l|l|}
\hline
\text{} & \text{Action in } \mathbb{R}^{2,2} &
\text{AdS$_3$ global coordinates} & \text{Fixed submanifold}\\
\hline
\tau_1 & X^2 \to - X^2 & \phi \to -\phi &
\phi=0,\pi \ (\text{AdS}_2)\\
\tau_2 & X^3 \to - X^3 & t \to -t  & 
t=0 \ (\text{H}_2)\\
\tau_3 & X^1 \to - X^1 \, ,  \ X^2 \to - X^2 \, ,  \
X^3 \to - X^3
& t \to -t \, ,  \ \phi \to \phi + \pi  &
\rho=t=0 \ (\text{point})\\
\tau_4 & X^0 \to - X^0 \, , \ X^2 \to - X^2 \, , \
X^3 \to - X^3
& t \to t+\pi  \ ,  \ \phi \to - \phi  &
\text{none}\\
\hline
\end{array}
\end{equation*}
\caption{Geometric parities in the AdS$_3$ \textsc{wzw} model}
\label{paritgeom}
\end{table}
In the last case the orientifold action
has no fixed submanifold. It is an involution only
on the single cover of the group manifold, for which
$t\sim t + 2\pi$.

Let us also define other parities that do not show up in the above analysis
since they are not strictly speaking geometric.  
It is well known in a free U(1) theory parametrized by a boson $t$ that
the parity $t\to -t$ can be performed together with 
a winding shift, {\it i.e.} a one-half translation of the coordinate T-dual to $t$~\cite{Brunner:2002em}. 
We can consider a similar modification of the $\tau_2$ parity, provided we 
start with the single cover of AdS$_3$.\footnote{Indeed in this model the winding 
around the time direction, which corresponds to the difference between 
left- and right-movers spectral flows, is conserved.}  
It defines a parity $\tilde{\tau}_2$. Geometrically, instead of 
a pair of H$_2$ orientifold planes at $t=0$ and $t=\pi$ with the same tension, we get a pair of 
orientifolds with opposite  tension. Similarly, one can define a $\tilde{\tau}_3$ parity 
corresponding to a pair of O(-1) planes of opposite tension.

Let us consider now the axial coset \slc, {\it i.e.} the cigar, and analyze how the above-mentioned 
parities are realized. The six AdS$_3$ parities give orientifold planes with the following geometries:
\begin{itemize}
\item For $\tau_1$ the geometry is similar to that of a straight D1-brane with $\hat \rho=0$, 
which is localized at $\phi=0,\pi$, see lower-left picture in fig.~\ref{branesgeom}. 
\item For $\tau_2$ the ${\rm O}$-plane covers all the cigar, similar to a 
D2-brane with $\hat \sigma=0$. 
\item For $\tau_3$ the geometry is similar to that of the D0-brane of the cigar, 
{\it i.e.} it is localized at $\rho=0$.
\item For $\tau_4$ we obtain something similar to the O1-plane above, but with an extra 
winding shift  $\phi_L - \phi_R \to \phi_L -\phi_R + \pi$.
\item For $\tilde{\tau}_2$ we obtain again a geometry 
that resembles that of a D2-brane with $\hat \sigma=0$, however 
the parity acts with an extra one-half rotation along the transverse direction $\phi$. 
\item For $\tilde \tau_3$ the geometry is similar to that of $\tau_3$. 
\end{itemize}

The parity $\tau_4$ is identified in the cigar with 
the combination of the inversion $\phi \to -\phi$ and  a winding shift by realizing that the translation symmetry 
along $t$, which amounts to the translation symmetry in the vector coset ({\it i.e.} the trumpet) becomes the winding symmetry in the axial coset. 
However, we know that this symmetry is broken at the
non-perturbative level by the winding condensate~(\ref{liouvintwind}). Hence, 
we conclude that this parity is not consistent in \slc. All other parities leave the 
winding condensate invariant, in agreement with the analysis done in $\NN=2$ Liouville 
as we will see in the next paragraph.

In the vector coset, or $\NN=2$ Liouville theory, the parities
$\tau_1$ and $\tau_3$ give respectively an O2-plane  
and a localized A-type orientifold whose geometrical nature is not well-defined. The parity 
$\tau_2$ gives a pair of antipodal O1-planes of the same tension. On the single cover of the trumpet, 
one can define a parity $\tilde{\tau}_2$ as we saw above, which includes a winding shift. It gives 
a pair of antipodal O1-planes of opposite tension. On the universal cover there is of course 
no such parity, or better saying it cannot be distinguished from the parity $\tau_2$. 

The case of the trumpet/$\mathcal{N}=2$ Liouville at minimal radius, which is well-defined for any $k$, 
cannot be obtained directly from AdS$_3$ by gauging; however it is T-dual to the cigar.  Since in this model the 
winding is conserved one can define parities similar to $\tilde{\tau}_2$ and 
$\tilde{\tau}_3$, that include a winding shift.

%%%%%%%%%%%%%%%%%%%%%%%%%%%%%%%%%%

\subsection*{The supersymmetric \slc and its parities}
As we saw previously, the supersymmetric coset \slc at level $k$ can
be realized by a suitable gauging of the supersymmetric \slr \textsc{wzw} model. To make
the fermionic action of the above parities more transparent, we recall the basic
features of the gauged action. It has the well-known form
\beq
\label{sucosetaa}
S_{coset}=S_{\textsc{wzw},k} (A,g)+\frac{i}{2\pi} \int d^2z ~ {\rm Tr}
\left( \bar \Psi {\rm D} \bar \Psi + \Psi \bar {\rm D} \Psi \right) \, ,
\eeq
and depends on the gauge field $A$, the \slr elements $g$ and $\Psi$,
a Dirac fermion which can be conveniently arranged in a Hermitian $2\times 2$
matrix\footnote{A similar expression holds for the right-movers.}
\beq
\label{sucosetab}
\Psi=\left( { 0 \atop \psi^+} {\psi^- \atop 0} \right)\, .
\eeq
$S_{\textsc{wzw},k}$ is the bosonic \textsc{wzw} action at level $k+2$,
whose explicit form will not be needed here (see {\it e.g.}~\cite{Tseytlin:1993my}), and the
covariant derivative ${\rm D}_{\mu}\Psi =\d_\mu \Psi+\left[ A_\mu,\Psi \right]$.
In terms of the global coordinates $(t, \rho, \phi)$ the generic \slr element
$g$ is written as\footnote{In this parametrization $g$ is actually written as an 
SU(1,1) element.}
\beq
\label{sucosetac}
g=e^{i(\phi+t)\sigma^3/2} e^{\rho \sigma^1} e^{i(t-\phi)\sigma^3/2}
=\left( {e^{it} \cosh \rho \atop e^{-i\phi} \sinh \rho} ~{e^{i\phi}
\sinh\rho \atop
e^{-it} \cosh \rho} \right) \, ,
\eeq
where $\sigma_i$ ($i=1,2,3$) are the usual Pauli matrices
\beq
\label{sucosetad}
\sigma_1=\left( {0 \atop 1} ~{1\atop 0} \right)~, ~~
\sigma_2=\left( {0 \atop i} ~{-i \atop 0}\right)~, ~ ~
\sigma_3=\left( {1\atop 0}~{0 \atop -1}\right)
~.
\eeq
The axial $U(1)$ gauge transformation of interest under 
which~\eqref{sucosetaa} is invariant has the form
\beq
\label{sucosetae}
A \to hAh+hdh~, ~ ~ g \to hgh~, ~ ~ \Psi \to h\Psi h~, ~~ \bar \Psi
\to h\bar \Psi h \, ,
\eeq
with $h=e^{it \sigma^3/2}$.
It turns out that the gauged theory has $\NN=(2,2)$ supersymmetry.

Now one can easily check that the fermionic completion of the parities
that appear in tab.~\ref{paritgeom} is
\begin{subequations}
\begin{align}
\label{sparity1}
&\tau_1~: ~ (A,g,\Psi) \to (\sigma_2 A\sigma_2,\sigma_2 g^{-1}
\sigma_2, \sigma_2 \Psi \sigma_2)
~,\\
\label{sparity2}
&\tau_2~: ~(A,g,\Psi) \to (\sigma_3 A\sigma_3,\sigma_3 g^{-1}
\sigma_3, \sigma_3 \Psi \sigma_3)
~,\\
\label{sparity3}
&\tau_3~: ~(A,g,\Psi) \to (A,g^{-1},\Psi)
~,\\
\label{sparity4}
&\tau_4~: ~ (A,g,\Psi) \to (-\sigma_1 A\sigma_1,-\sigma_1 g^{-1}
\sigma_1,- \sigma_1 \Psi \sigma_1)
\end{align}
\end{subequations}

To obtain orientifolds of the supersymmetric coset we need to combine the above
symmetries with the worldsheet parity $\Omega_B$. Immediate candidates are the
parities $\tau_i \Omega_B$ ($i=1,2,3$) ($\PP_4$ will not be considered, because,
as explained above, it is non-pertrubatively inconsistent). However, not all of these
symmetries are automatically well-defined. 
As explained in subsect.\ (4.2.1) of~\cite{Brunner:2003zm} for the
$SU(2)/U(1)$ case, there are possible anomalies from the fermionic sector.
In our case, one can check that  the parities 
$\PP_2=\tau_2 \Omega_B$, $\PP_3=\tau_3 \Omega_B$ and
$\tilde \PP_2=\tilde \tau_2 \Omega_B$, $\tilde \PP_3=\tilde \tau_3 \Omega_B$,
which are  B-type parities, are anomaly free, but the 
parity $\PP_1=\tau_1 \Omega_B$, which is A-type, has an anomaly. 
The anomaly can be cancelled by combining $\PP_1$ with $(-)^{\bar F}$. 
Additional parities can be obtained 
as $U(1)_R$ rotated versions of the above anomaly free
parities (see eqn.~\eqref{parityai}).

As a final comment, notice that it is possible to define 
another set of consistent orientifold projections as 
$\PP_i (-)^{\bar F}=\tau_i \Omega$, $\tilde \PP_i (-)^{\bar F}=\tilde \tau_i \Omega$
($i=2,3$). These parities are such that the fermion bilinears 
$\psi^\epsilon \bar \psi^{\bar \epsilon}$ ($\epsilon,\bar \epsilon=\pm$) 
are invariant (see comments in the footnote around eqn.\ \eqref{parityba}). 
For concreteness we will discuss in the following section
mostly the parities with $\Omega_B$,
but will indicate what changes for the parities with $\Omega$.

%%%%%%%%%%%%%%%%%%%%%%%%%%%%%%%%%%

\subsection*{Comparison of \slc and $\NN=2$ Liouville parities}

In the previous subsection we analyzed the parities of the 
$\NN=2$ Liouville theory / \slc  coset from two different 
point of views. First, following a general discussion of $\NN=(2,2)$ field theories, 
secondly as geometric parities inherited from AdS$_3$. The asymptotic analysis
in $\NN=2$ Liouville theory gives a nice and simple picture of the action of
orientifolds that extend to the asymptotic semiclassical region. The
discussion of orientifolds in AdS$_3$ and its cosets gives, on the other
hand, an intuitive geometric picture and also points towards the existence
of localized ${\rm O}0$-planes on the cigar (those associated with the
parities $\PP_3$, $\tilde \PP_3$). In the next section, we will see how
the exact CFT analysis blends the above information in a picture of mixed
${\rm O}2$/${\rm O}0$-planes. 

In general, we expect that for each A(B)-type orientifold presented in 
$\NN=2$ Liouville theory there is a corresponding B(A)-type orientifold 
in the supersymmetric \slc coset related to it by mirror symmetry and vice versa.
For instance, one can associate the $\NN=2$ Liouville A-type parity 
$\PP_A$ with the cigar B-type parity $\PP_2$. However, it is not always 
straightforward to match parities one-to-one, since we determined
the parities on each side with different methods and some of these methods
capture only the features of the asymptotic region where the worldsheet theory 
is weakly coupled.

%%%%%%%%%%%%%%%%%%%%%%%%%%%%%%%%%%%%%%%%%%%

\section{B-type orientifolds on the cigar: O2/O0-planes}
\label{O2section}
In this section we present a detailed analysis of the properties of
the orientifold planes arising from the B-type cigar parities $\PP_2$,
$\tilde \PP_2$ and $\PP_3$, $\tilde \PP_3$  which appeared
above. For simplicity, we will concentrate only on parities of the
supersymmetric \slc coset, but it should be kept in mind that
for each of the orientifolds presented here there is a mirror orientifold 
in $\NN=2$ Liouville theory whose properties can be deduced in a 
very similar manner. The properties of the B-type orientifolds
will be analyzed from several complementary points of view. Using the
explicit knowledge of the parity symmetries we compute directly the
volume diverging asymptotic Klein
bottle amplitude. The result of this semiclassical calculation gives a
non-trivial check for the exact crosscap wave-functions that we derive
in the ensuing by modular bootstrap
from the M\"obius strip amplitude of the D0-brane. Another check comes
by comparison with the known conformal bootstrap constraints
of~\cite{Nakayama:2004at}. The geometry of the orientifolds presented
here exhibits some intriguing
features which can be read off the crosscap wave-functions. In
particular, we will see that the
M\"obius strip amplitudes lead naturally to an intricate combination  of O2- and
O0-planes, which incorporate a similtaneous action of $\PP_2$ and
$\PP_3$ parities or $\tilde \PP_2$ and $\tilde \PP_3$ parities. We
expect these features to have
an interesting relation to the physics of orientifolds in the presence
of NS5-branes in the context of
Hanany-Witten setups. At the end of each subsection, we present for
completeness the M\"obius strip amplitudes of open strings on
D2-branes and comment on the action of the parities on the open string
densities.

%%%%%%%%%%%%%%%%%%%%%%%%%%%%%%%%%

\boldmath
\subsection{An ${\rm O}2/{\rm O}$0-plane}
\unboldmath
\label{OtwoOzero}

We begin with the analysis of the orientifold associated to the B-type parity
$\PP_2$. We will call this orientifold ${\rm O}_B$.
In the previous section, we argued by descent from AdS$_3$ that
$\PP_2$ gives an O2-plane which is spacefilling on the cigar. We will  soon
see that the full story is more involved.

In order to familiarize ourselves
with the properties of this orientifold we will first analyze the
Klein bottle amplitude in the asymptotic linear dilaton region of the  cigar
where the worldsheet theory becomes the free theory of two bosons and
two fermions.

\subsubsection*{The asymptotic Klein bottle amplitude}
The torus partition function of the supersymmetric cigar \textsc{cft}
has been discussed in a series of
papers~\cite{Hanany:2002ev,Eguchi:2004yi,Israel:2004ir,Fotopoulos: 2004ut}.
It receives several contributions: a piece which involves
the continuous representations with a non-trivial density of states
and a piece with the contributions of the discrete states that are  exponentially
supported near the tip of the cigar. The details of the fermion
contribution depend on
the \textsc{gso} projection; in this paper we will focus for
concreteness on the simplest type
0B diagonal torus partition sum. Furthermore, for the purposes of the  present
exercise we will be interested only on the contribution of the  continuous
representations. The density of these states has an \textsc{ir}
divergence at zero radial momentum,
which is associated with the infinite volume of the asymptotic
cylinder region of the
cigar. This will be the contribution of interest here. It is captured
by the asymptotic free
linear dilaton theory and takes the simple form (for the explanation
of our conventions and the
definitions of the relevant \slc characters see app.~\ref{appchar})
\begin{multline}
\label{p2aa}
\TT (\tau ) =V 
\sum_{a,b\in \Z_2} \sum_{n,w \in \Z} \int_0^\infty \di P \
ch_c\left(P,\frac{n+kw}{2} ; \tau \right)\left[ {a \atop b} \right] \
\bar{ch}_c\left(P,\frac{n-kw}{2}; \bar{\tau}\right)\left[ {a \atop b}
\right]\, ,
\end{multline}
where $V$ is the regularized volume of the asymptotic cylinder and
$ch_c(P,m;\tau)\left[ {a \atop b}\right]$ the continuous character
with $a,b \in \Z_2$ the standard fermionic indices labeling the spin  structures
on the torus.

The $\PP_2$ parity acts on the bosonic part simply as $\Omega$
and therefore on the \textsc{ns-ns} coset primaries with momentum $n$ and winding $w $ as
\beq
\label{p2ab}
\PP_2~:~ |P,n,w\rangle \to |P,n,-w\rangle \, .
\eeq
On the worldsheet fermions it acts as:\footnote{When $\PP_2$ acts on
the product of two fermions it takes
$\psi^\epsilon \bar \psi^{\bar \epsilon} \to -\psi^{\bar \epsilon}
\bar{\psi}^{\epsilon}$,
$\epsilon, \bar \epsilon =\pm 1$.}
\beq
\label{p2ac}
\PP_2~:~ \psi^\pm \to -\bar \psi^\pm~, ~ ~ ~
\bar \psi^\pm \to -\psi^\pm \, .
\eeq
Combining these facts it is straightforward to determine the asymptotic
expression for the Klein bottle amplitude
\beq
\label{p2ad}
\KK_{{\rm O}_B} (t) =V  \sum_{a \in \Z_2}
\sum_{n \in \Z} \int_0^\infty \di P \,
ch_c\left(P,\frac{n}{2};2it \right) \left[ {a \atop 1} \right] \, .
\eeq
As expected from the B-type nature of the parity only
momentum modes contribute in eqn.~\eqref{p2ad}.

For later purposes it will be useful to perform an $S$-modular  transformation
$(\tau \to -\frac{1}{\tau})$ on~\eqref{p2ad} to obtain the Klein
bottle amplitude in the transverse crosscap channel. With the use of
the $S$-modular property of the continuous characters, eqn.~(\ref{p2ae}), 
we deduce the crosscap channel expression
\beq
\label{p2af}
\tilde{\KK}_{{\rm O}_B} (t) =\frac{kV}{4} 
\sum_{a \in \Z_2} e^{\frac{i\pi a}{2}}\sum_{\ell \in \Z}
ch_c\left(0,k\ell; -\nicefrac{1}{2it}\right)
\left[ {1 \atop a} \right] \, .
\eeq
The only contribution comes from the zero radial momentum modes,
which is expected since we perform an asymptotic free field analysis.
Furthermore, we see that the orientifold sources winding modes
in the \textsc{r-r} sector with even winding $w=2\ell$.
In a little while, we will reproduce this result from an exact  modular bootstrap
analysis that is not restricted to the asymptotic linear dilaton
region of the theory. 

Repeating the above exercise with the parity $\PP_2 (-)^{\bar F}$
would give similar relations, the important difference being that 
in~\eqref{p2ad} states in the \textsc{ns} and \textsc{r}
sector would appear. Hence, we would obtain an orientifold that
sources winding modes in the \textsc{ns-ns} sector.

\subsubsection*{M\"obius strip amplitude for the D0-brane}
The diagonal modular invariant theory that we are considering
here has four different D0-branes characterized by two fermionic
labels $a,b \in \Z_2$. The open string spectrum between two D0-branes
with labels $\left[ {a_1 \atop b_1} \right]$ and $\left[ {a_2 \atop
b_2} \right]$ can be derived easily from the annulus amplitude
\beq
\label{mobaa}
\AA_{\left[ {a_1 \atop b_1} \right];\left[ {a_2 \atop b_2} \right]} (t) =
\delta_{b_1,b_2}^{(2)} \, 
\sum_{r\in \Z} ch_{\mathbb{I}} (r;it) \left [{a_1-a_2 \atop b_1} \right]\, ,
\eeq
where only the identity representation appears. For the precise  definition
of the identity representation character see
app.~\ref{appchar}.\footnote{To compare
with another  terminology used in the literature, one may identify the
brane $\oao{0}{0}$ with the boundary state $|NS,+\rangle$, the brane
$\oao{1}{0}$ with the boundary state $|NS,-\rangle$, $\oao{0}{1}$ with
$|R,+\rangle$ and $\oao{1}{1}$ with $|R,-\rangle$.}
Since the $\tau_2$ symmetry of \slr has no obvious action on the open  strings
attached to the D0-branes of the cigar it is sensible to postulate the
open string channel M\"obius strip amplitude, for an open string sector  
corresponding to a D0-brane of fermionic labels $[a\, b]$: 
\beq
\label{mobab}
\MM_{\left[ {a \atop b} \right]} (t) =
\delta_{b,1}^{(2)} \, 
\sum_{c \in Z_2}\sum_{r\in \Z}
~\widehat {ch_{\mathbb{I}}}(r;it)
\left[ {1 \atop c} \right]
~.
\eeq

As usual with M\"obius strip amplitudes the character that appears on  the
\textsc{rhs} of this equation is a hatted character (see
app.~\ref{Pmatapp}), {\it i.e.} 
it corresponds to $\mathrm{Tr} (\Omega e^{-2\pi t H_\text{o}})$ in the
appropriate open string sector. The overall Kronecker $\delta$ symbol
has been inserted by using the input of the asymptotic Klein bottle amplitude
\eqref{p2af} which shows that the orientifold sources only \textsc{r-r}
fields in the bulk. This will be justified in a minute when we derive
the crosscap state and compare with the asymptotic Klein bottle
amplitude~\eqref{p2af} to see how everything fits nicely together with
the postulate \eqref{mobab}. Also, notice that the amplitude is independent
of the $a$ fermionic label of the D0-brane.

\subsubsection*{Getting the crosscap wave-function}
Given the M\"obius strip amplitude \eqref{mobab} we can  determine
the full crosscap wave-function of the orientifold with modular  bootstrap.
In the transverse channel, the \textsc{lhs} of \eqref{mobab} can be  expressed as
an overlap of the crosscap state --~that we call $|{\rm O}_B\rangle $~-- and
the D0-brane boundary state $|D0; \left[ {a \atop b} \right] \rangle$.
Performing a $\PP$-modular tranformation on the \textsc{rhs} of
eqn.~\eqref{mobab} we find
\begin{multline}
\label{bootaa}
\left \langle D0; \left[ {a \atop b} \right]  \left|
e^{-\frac{2\pi}{t} H_c}\right| {\rm O}_B \right \rangle
\\ = \delta_{b,1}^{(2)} \, \sum_{c\in \Z_2}
\sum_{w\in \Z} \int_0^\infty  \di p~
\PP_{\mathbb{I}; \left[ {1 \atop c} \right]}^{c;(p,kw/2)}
\widehat{ch_c}\left(p,\frac{kw}{2}; -\frac{1}{4it}\right) \left[ {1 \atop c}
\right]
+ {\rm discrete}
~,
\end{multline}
where $\PP_{\star}^{\star}$ are the matrix
elements of the $\PP$-modular transformation for the hatted identity  character
$\sum_{r\in \Z} \widehat{ch_{\mathbb{I}}} (r;\tau)\left[ {1 \atop c}\right]$
in the \textsc{r} sector.\footnote{The $\PP$-matrix implements the  transformation
$\tau \to -\nicefrac{1}{4\tau}$. It allows to transform the M\"obius  amplitude
from the open to the closed channel. We refer the reader {\it e.g.} to the
review~\cite{Angelantonj:2002ct} for more details.} The derivation of  these
elements is given in detail in app.~\ref{Pmatapp}. In the \textsc {rhs} of
eqn.~(\ref{bootaa}), we denote by  ``discrete'' the contribution of
discrete representation characters.  We will not deal explicitly here  with
this contribution because it can be obtained  from the coupling
of the continuous states by analytic continuation. The full explicit modular 
transformation can be found in eqn.~(\ref{Pmatcigfull})

The overlap on the \textsc{lhs} of \eqref{bootaa} can be
re-expressed in terms of the known D0-brane wave-functions
$\Phi_{D0;\left[{a \atop b}\right]}(p,m)$ and the crosscap wave-functions
$\Psi_{{\rm O}_B}\left(p,m; \left[ {a \atop b} \right]\right)$ as
\begin{multline}
\label{bootab}
\left \langle D0; \left[ {a \atop b} \right]  \left|
e^{-\frac{2\pi}{t} H_c}\right| {\rm O}_B \right \rangle  
%=\\
= \, 
\sum_{c \in \Z_2} \sum_{w\in \Z} \int_0^\infty \di p \
\Phi_{D0;\left[{a \atop -b}\right]}\left(-p,-\frac{kw}{2}\right)
\Psi_{{\rm O}_B}\left(p,\frac{kw}{2}; \left[ {b \atop c}
\right]\right) \ \times\\ \times \
\widehat {ch_c}\left( p,\frac{kw}{2};-\nicefrac{1}{4it}\right) \left[
{b \atop c-a}\right]
+{\rm discrete.} 
\end{multline}
From eqns.~(\ref{bootaa}, \ref{bootab}) we deduce the crosscap
wave-function\footnote{Similar results can be obtained for the 
parity $\PP_2 (-)^{\bar F}$. Most notably, in \eqref{bootac} one should 
replace the fermionic index 1 by 0, because in this case the orientifold sources
states in the \textsc{ns} sector. Furthermore, in deriving 
\eqref{bootac} one should use fermionic Ishibashi states in the \textsc{ns-ns} 
sector for which the natural normalization is
$|\CC\rangle_{NSNS\pm}=e^{\pm i \frac{\pi}{4}}e^{i\pi(L_0\mp \frac{1}{4})}
|\BB\rangle_{NSNS\pm}$ \cite{Hosomichi:2006yj}.
$|\CC\rangle$ and $\BB\rangle$ are respectively
crosscap and boundary Ishibashi states.\label{hosofoot}}
\beq
\label{bootac}
\Psi_{{\rm O}_B}\left(p,\frac{kw}{2};\left[ {b \atop c} \right]\right)=
\delta^{(2)}_{b,1} ~
\frac{\PP_{\mathbb{I};\left[{1 \atop c-a}\right]}^{c;(p,kw/2)}}
{\Phi_{D0;\left[{a \atop -1}\right]}\left(-p,-\frac{kw}{2}\right)}
~.
\eeq
Substituting the explicit formulae of apps.~\ref{Pmatapp}
and~\ref{appwave-functions}
we obtain the final expression
\begin{multline}
\Psi_{{\rm O}_B}\left(p,\frac{kw}{2};\left[ {b \atop c} \right]\right)=
2\sqrt{k} \delta^{(2)}_{b,1} e^{\frac{i\pi(c-1)}{2}} \,  \nu^{-ip}
\frac{\Gamma(-2ip) \Gamma(1-\nicefrac{2ip}{k})}
{\Gamma(1-ip+\frac{kw}{2})\Gamma(-ip-\frac{kw}{2})} \times \
\\ \times \
\cosh(\pi p)\
\frac{ \delta^{(2)}_{w,0} \, e^{\frac{i\pi w}{2}} \, \sin \frac{\pi
kw}{2} \cosh \frac{\pi p}{k}
+\delta^{(2)}_{w,1}\, e^{\frac{i\pi}{2} (2c-1+w)} \, \sinh \pi p
\sinh \frac{\pi p}{k}}
{\sinh\pi(p+i\frac{kw}{2})\sinh\pi(p-i\frac{kw}{2})}
\label{bootad}
\end{multline}
which, as expected, is independent of the fermion number $a$, {\it i.e.} independent
of the D0-brane that we use to perform the modular bootstrap.
By definition, the wave-function~\eqref{bootad} gives the one-point  functions
on $\R \mathbb{P}_2$ of all the fields in the continuous  representation that are
sourced by the orientifold ${\rm O}_B$. The discrete couplings can be
determined from the analyticity properties of \eqref{bootad}. Indeed,  taking
the analytic continuation $p = -i(j -\hl)$ in eqn.~(\ref{bootad}) one  finds
poles on the real $j$-axis whose residues correspond to the couplings
to discrete representations. This can be checked explicitly using the
discrete $\PP$-matrix elements~(\ref{discPmat}) computed in app.~\ref {Pmatapp}.

A first non-trivial check of~\eqref{bootad} can be obtained by
comparing with the asymptotic Klein bottle amplitude~\eqref{p2af}.
In the $p \to 0$ limit the wave-function \eqref{bootad} simplifies  considerably.
The contribution of the odd winding numbers drops out completely -- ~this is one
of the first requirements of~\eqref{p2af}~-- and the remaining
expression becomes
\begin{equation}
\label{bootae}
\Psi_{{\rm O}_B}\left(0,\frac{kw}{2};\left[ {b \atop c} \right]\right)=
\frac{2 \sqrt k}{\pi}
e^{\frac{i\pi}{2}(c+1+w)}
\Gamma(0) \delta^{(2)}_{b,1} \delta^{(2)}_{w,0}\, .
\end{equation}
The divergent $\Gamma(0)$ gives the volume divergent factor $V$ in~ \eqref{p2af}.
With a simple calculation one can verify that the Klein bottle  amplitude in
the crosscap channel computed with the wave-function~\eqref{bootae}  reproduces
the independent result~\eqref{p2af}.

Another non-trivial check of the techniques used here is as follows. 
Starting with a M\"obius amplitude for a {\it single} 
hatted identity character 
$\widehat{ch}_{\mathbb{I}} (r)\left[ {1 \atop c}\right]$ we obtain, 
using the arguments around eqn.~(\ref{amplD0noncomp}) and the $\PP$-matrix elements~\eqref{contPmat} 
of app.~\ref{Pmatapp}, the type A crosscap wave-function for the trumpet \textsc{cft} at  infinite
radius (or the $\NN=2$ Liouville theory at infinite radius), of  similar form as 
in~\eqref{bootac}. This  wave-function turns
out to be the same as the one derived by Nakayama in~\cite{Nakayama: 2004at}
where it was shown that it passes the non-trivial check of one of the  conformal
bootstrap equations. In reference to~\cite{Nakayama:2004at},
we would like to point out here that our computation in app.~\ref {Pmatapp}
is similar to the one of \cite{Nakayama:2004at}
in the case $r=0$, which was the only case considered there.
Also, certain important details of the derivation of the $\PP$
matrix elements  are different in our work and help clarify
some unjustified statements in~\cite{Nakayama:2004at}.\footnote{
In particular the choice of 
hatted characters made in~\cite{Nakayama:2004at} (see 
the footnotes p.~6) does not match the usual definition 
of  hatted characters for the chiral algebra of a \textsc{cft}.}

\subsubsection*{Other amplitudes}
For completeness we conclude this subsection with a list
of the M\"obius strip amplitudes on D2-branes and a related discussion
on open string densities.

We will focus on the D2-branes of~\cite{Fotopoulos:2004ut,Hosomichi: 2004ph}
which source states only in the continuous representations.
The corresponding boundary states will be denoted as $|D2;P,M;\left [ {a \atop
b} \right]\rangle$ and the explicit form of their wave-functions can  be found in
app.~\ref{appwave-functions}. We would like to compute the M\"obius
strip amplitude between these branes and the orientifold
$|{\rm O}_B\rangle$. In the crosscap
channel it is straightforward to compute the overlap
\begin{equation}
\label{obootaa}
\mathcal{M}_{P,M,\left[ {a \atop b} \right]} (t) = 
\left \langle D2;P,M; \left[ {a \atop b} \right] \left|
e^{-\frac{\pi}{2t} H_c} \right|
{\rm O}_B \right \rangle
\end{equation}
with the use of the crosscap and boundary state
wave-functions~(\ref{bootad}, \ref{appbab}).
Then, the $\PP$-modular transformation of the continuous characters, 
eqn.~(\ref{obootab}) leads to an open string channel M\"obius strip amplitude that 
is \textsc{ir} divergent as usual because of the infinite volume of the brane. 
The full amplitude reads:
\bea
\label{obootac}
&&\MM_{P,M,\left[ {a \atop b} \right]} (t) =-\frac{2^{5/2}}{k^2}
\delta_{b,1}^{(2)}
\\
&&
\int_0^\infty \di p' \sum_{\ell \in \Z} \sum_{c\in \Z_2}
\Bigg\{ (-)^{a+c} \rho(p';P)~ \widehat{ch_c}\left(p',\ell+2M+\frac{1} {2};it\right)
\left[ {1 \atop c}\right]+
\nonumber\\
&&+\frac{1}{2}\sum_{\epsilon=\pm} \epsilon
\widetilde\rho(p';P)~
\widehat{ch_c}\left(p',\ell+2M+\frac{1}{2}+\frac{\epsilon k}{2};it \right)\left[
{1 \atop c}\right]  \Bigg\}
\, , \nonumber
\eea
where $\rho(p';P)$ and $\widetilde \rho(p';P)$ are the spectral densities
\begin{subequations}
\begin{align}
\label{obootad}
&\rho(p';P)=\frac{1}{4} \int_0^\infty \di p ~
\frac{1}{\cosh(\frac{\pi p}{k})} \cos\left(\frac{4\pi p
P}{k}\right)\cos\left(\frac{2\pi p p'}{k}\right)
~,\\
&\widetilde\rho(p';P)=\int_0^\infty \di p~
\frac{\cosh \pi p\, \cosh\frac{\pi p}{k}}{\sinh 2\pi
p \, \sinh \frac{2\pi p}{k}}
\cos \frac{4\pi p P}{k}\, \cos\left(\frac{2\pi p p'}{k}\right)
\, .
\end{align}
\end{subequations}
This result should be compared to the annulus amplitude for open strings 
stretched between two different D2-branes
\begin{multline}
\label{obootaf}
\AA_{P_1,M_1,\left[ {a_1 \atop b_1}\right];P_2,M_2,\left[ {a_2 \atop
b_2}\right]} (t) = \\
= \frac{4}{3k}\delta^{(2)}_{b_1,b_2} e^{\frac{i\pi  b_1(a_2-a_1)}{2}}
\int_0^\infty \!\!\! \di p' \sum_{\ell \in \Z}
\Bigg\{ \rho(p';P_1|P_2) ch_c(p',\ell+M_1-M_2;it) \left[ {a_2-a_1
\atop b_1} \right]+
\\
+\sum_{\epsilon=\pm}
\frac{(-)^{b_1}}{2}\widetilde \rho(p';P_1|P_2)
ch_c\left(p';\ell+M_1-M_2+\frac{\epsilon k}{2};it\right)\left[
{a_2-a_1 \atop b_1} \right]\Bigg\}
\end{multline}
where the spectral densities now read:
\begin{subequations}
\begin{align}
\label{obootaeeven}
\rho(p';P_1|P_2)&=\int_0^\infty \di p~
\frac{\tanh 2\pi p}{\sinh \frac{2\pi p}{k}}
\cos \frac{4\pi pP_1}{k} \cos \frac{4\pi p P_2}{k}
\cos \frac{4\pi p p'}{k} ~,\\
\widetilde\rho(p';P_1|P_2)&=\int_0^\infty \di p~
\frac{\cos \frac{4\pi pP_1}{k} \cos \frac{4\pi pP_2}{k}
\cos \frac{4\pi p p'}{k}}
{\sinh(2\pi p)\sinh \frac{2\pi p}{k}}
~.
\label{obootaeodd}
\end{align}
\end{subequations}

A few comments are in order here: 
\begin{itemize}
\item[($i$)] Both in the annulus and the M\"obius strip amplitudes
the character that appears with density $\tilde \rho$ exhibits
an intriguing angular momentum shift by $\pm \frac{k}{2}$ which
implies a mild breaking of the open string momentum number quantization
law. In the case of the $\NN=2$ Liouville theory, this shift was also  noticed
for the annulus amplitude in \cite{Hosomichi:2004ph}, where it was  suggested
that it can be understood as due to the boundary interaction terms
on the D2-branes.
\item[($ii$)] The spectral densities $\rho$ and $\tilde \rho$ are  different
in the annulus and M\"obius strip amplitudes. They also appear in  front of
different characters (notice the extra $\frac{1}{2}$ shift in the
argument of the continuous character in \eqref{obootac}). 
The origin of this shift lies in the winding
dependent phases $i^{w}$ and $i^{w-1}$ that appear in the $ {\rm O}_B$
crosscap wave-function \eqref{bootad}. Note that this shift would not  exist for
the orientifold based on the $\tau_2 \Omega$ (versus $\tau_2 \Omega_B$)
parity which sources fields in the NS sector (versus the R sector  above).
The fact that the M\"obius strip spectral densities are different from
the annulus spectral densities and are not related in the obvious way to the open
string reflection amplitudes suggests a subtle property of the action of the parity on
the open string spectrum. This feature doesn't have a clear explanation, 
but has been  noticed previously both in the context of bosonic and 
supersymmetric  Liouville theory 
\cite{Nakayama:2003ep,Nakayama:2004vk, Nakayama:2004at}.
In relation to this point, notice in the present context that both in
the annulus and the M\"obius strip amplitude the density $\rho$ is a  finite quantity.
The densities $\tilde \rho$ have the usual IR divergence at $p=0$ that
needs to be regularized. Incidentally, for integer level $k$ the  contribution that involves
$\tilde \rho$ in the M\"obius strip amplitude cancels out completely.
This cancellation, however, would not occur for the orientifold based on
the $\tau_2 \Omega$ parity.
\end{itemize}

%%%%%%%%%%%%%%%%%%%%%%%%%%%%%%%%%%%%%%%

\boldmath
\subsection{An ${\rm \tilde O}2/{\rm \tilde O}$0-plane}
\unboldmath

In this subsection we discuss the properties of the B-type parity $\tilde \PP_2$. 
We will denote the corresponding orientifold as $\tilde{\rm O}_B$.
In sect.~\ref{geomorientsection} we argued by descent from AdS$_3$
that this parity  gives another type of ${\rm O}2$-plane which is also space-filling in
the cigar geometry. Many of the details of the following analysis are similar  to the
ones of the above subsection, so here we will be brief emphasizing  mostly
the details that are different. Also, it should be noted that, as before, one can repeat
the exercise for the $\tilde \PP_2 (-)^{\bar F}$ parity, but we will not present
this case explicitly here.

\subsubsection*{The asymptotic Klein bottle amplitude}
We are concentrating again on the asymptotic linear dilaton region
of the cigar. The $\tilde P_2$ parity acts on the bosonic part
of the asymptotic \textsc{cft} as an $s\Omega$ parity, {\it i.e.}
\beq
\label{kleinaa}
\tilde \PP_2~:~|p,n,w\rangle \to (-)^n |p,n,-w\rangle
~.
\eeq
On the single complex fermion it acts as in \eqref{p2ac}.
Consequently, the asymptotic expression of the Klein bottle
amplitude is
\beq
\label{kleinab}
\KK_{{\rm \tilde O}_B} (t) = V 
\sum_{a \in \Z_2} \sum_{n \in \Z} \int_0^\infty
\di p~ (-)^n ch_c\left(p,\frac{n}{2};2it\right)\left[ {a \atop 1}  \right]\, .
\eeq
In the transverse crosscap channel it gives:
\beq
\label{kleinac}
\tilde{\KK}_{{\rm \tilde O}_B} (t) = \frac{kV}{2}  \sum_{a \in
\Z_2} \sum_{\ell \in \Z} e^{\frac{i\pi a}{2}} ch_c
\left(0,k\ell+\nicefrac{k}{2};-\nicefrac{1}{2it}\right)\left[ {1 \atop
a}
\right] \, ,
\eeq
which implies that the orientifold
couples in the asymptotic region only to odd winding states
(compare this to the case of ${\rm O}_B$, eqn.\ \eqref{p2af}).

\subsubsection*{M\"obius strip amplitude for the D0-brane}
For the $\tilde \PP_2$ parity we postulate the M\"obius strip amplitude
\beq
\label{tmobaa}
\widetilde \MM_{\left[{a \atop b}\right]} (t) =\delta_{b,1}^{(2)}
~\sum_{c \in \Z_2} \sum_{r\in \Z}
(-)^r~\widehat{ch_{\mathbb{I}}}(r) \left[ {1 \atop c}\right](it)
~,
\eeq
which compared to \eqref{mobab} has an extra $(-)^r$ phase
in front of each character. We will see in a moment that this ansatz
is consistent with the above-mentioned semiclassical properties of the
parity $\tilde P_2$.\footnote{Another way to motivate this ansatz is the following.
In the $\mathcal{N}=2$ Liouville description of the theory, 
the label $r$ of the identity characters that appear in the annulus
amplitude~\eqref{mobaa} can be thought of as the (fractional) winding of 
open strings stretched between two copies of the localized brane. Since we want to implement 
a winding shift as part of the definition of the T-dual of the parity $\tilde {\mathcal P}_2$, 
it is sensible to postulate a M\"obius strip amplitude of the form \eqref{tmobaa}.}

\subsubsection*{Getting the crosscap wave-function}
We can now determine the full crosscap state $|{\rm \tilde O}_B\rangle$
by re-expressing the M\"obius strip amplitude \eqref{tmobaa}
in the transverse channel. The $S$-modular transformation
of the \textsc{rhs} of \eqref{tmobaa} gives
\begin{multline}
\label{tbootaa}
\left \langle D0; \left[ {a \atop b} \right]  \left|
e^{-\frac{2\pi}{t} H_c}\right|
{\rm \tilde O}_B \right \rangle
= \\ = \delta^{(2)}_{b,1} ~\sum_{c\in  \Z_2}
\sum_{w\in \Z} \int_0^\infty  \di p~ 
\widetilde \PP_{\mathbb{I}; \left[ {1 \atop c} \right]}^{c;(p,kw/2)}
\widehat{ch_c}\left(p,\frac{kw}{2};-\frac{1}{4it}\right) \left[ {1  \atop c}
\right]
+ {\rm discrete}\, ,
\end{multline}
where $\widetilde\PP_{\star}^{\star}$
are the matrix elements of the $\PP$-modular transformation of the 
combination of characters $\sum_{r\in \Z} (-)^r \widehat{ch_{\mathbb{I}}} \left[ {1
\atop c}\right]$ in the \textsc{r} sector, which can be found in
app.~\ref{Pmatapp}.
Expressing the \textsc{lhs} of eqn.~\eqref{tbootaa} as in~\eqref{bootab}
(with $\Psi_{{\rm O}_B}$ replaced by $\Psi_{{\rm \tilde O}_B}$)
we deduce an expression analogous to~\eqref{bootac} which gives
\begin{multline}
\label{tbootab}
\Psi_{{\rm \tilde O}_B}\left(p,\frac{kw}{2};\left[ {b \atop c} \right] \right)=-
2 \sqrt k ~\delta^{(2)}_{b,1} e^{\frac{\pi i}{4}} e^{\frac{i\pi c} {2}} \nu^{-ip}
\frac{\Gamma(-2ip) \Gamma(1-\frac{2ip}{k})}
{\Gamma(1-ip+\frac{kw}{2})\Gamma(-ip-\frac{kw}{2})}\ \times
\\ \times \
\frac{\cosh \pi p \left[ \delta_{w,0}^{(2)} e^{i\pi(c+\frac{w}{2})}
\sinh \pi p
\sinh \frac{\pi p}{k}-
\delta_{w,1}^{(2)} e^{i\pi \frac{w-1}{2}} \sin \frac{\pi k
w}{2} \cosh \frac{\pi p}{k}\right]}
{\sinh\pi(p+i\frac{kw}{2})\sinh\pi (p-i\frac{kw}{2})} \, .
\end{multline}
Again, the discrete couplings can be determined from the analyticity
properties of \eqref{tbootab} or by using the results of app.\ \ref {Pmatapp}.

As we send $p\to 0$ the wave-function \eqref{tbootab} becomes
\beq
\label{tbootac}
\Psi_{{\rm \tilde O}_B}\left(0,\frac{kw}{2}; \left[ {b \atop c}  \right]\right)=-
\frac{2\sqrt k~e^{\frac{i \pi}{2}(c+w)}
e^{-\frac{\pi i}{4}}}{\pi}
\Gamma(0) \delta^{(2)}_{b,1} \delta^{(2)}_{w,1}
~.
\eeq
The contribution of even winding numbers drops out and we are left
with an expression which is consistent with the asymptotic Klein bottle
amplitude \eqref{kleinac}.

Repeating the above exercise in the trumpet \textsc{cft} at infinite  radius, or
in $\NN=2$ Liouville theory at infinite radius, we find that the  orientifolds
${\rm O}_B$ and ${\rm \tilde O}_B$ are identical. This is sensible  from the
AdS$_3$ point of view for the following reason. As explained in sect.\
\ref{geomorientsection}, in the single cover of AdS$_3$
the parities $\tau_2$ and $\tilde \tau_2$ give two distinct pairs of
H$_2$ orientifold planes
at $t=0$ and $t=\pi$. For $\tau_2$ the orientifolds have the same  tension,
for $\tilde \tau_2$ they have opposite tension. As we go from the  single cover
to the universal cover, the two H$_2$ planes are separated at  infinite distance
and the two parities $\tau_2$ and $\tilde \tau_2$ become  indistinguishable.
Correspondingly, in the vector coset, or in $\NN=2$ Liouville theory
at infinite radius, the orientifolds ${\rm O}_B$ and ${\rm \tilde O}_B $ become
identical essentially because there is no summation over $r$ in the open string
spectrum on the D0-brane. The result is given by eqns.~(\ref {orientNC}, \ref{orientNC2}) 
in the next section.

\subsubsection*{Other amplitudes}
The M\"obius strip amplitude on D2-branes can be obtained as in
the previous subsection. As in the case of the ${\rm O}_B$-plane,
one finds a non-trivial action of the orientifold on the annulus open
string densities. The explicit form of the M\"obius strip amplitude is
not very illuminating and will not be quoted here, but statements  analogous
to those appearing in the previous subsection for the ${\rm O}_B$-plane
apply in this case as well.

%%%%%%%%%%%%%%%%%%%%%%%%%%%%%%%%%%%%%%%%%%

\subsection{The orientifold geometry and Hanany-Witten setups}
One can obtain a simple intuitive picture of the geometry of the
orientifold planes ${\rm O}_B$ by descent from AdS$_3$.
As explained in section~\ref{geomorientsection}, the parity $\PP_2$
descends from $\tau_2 \Omega$ in AdS$_3$ (see {\it e.g.} tab.~\ref{paritgeom})
and gives naturally an ${\rm O}2$-plane that covers
the cigar. This expectation is borne out nicely by the asymptotic
semiclassical features of the exact result~\eqref{bootad}.
The one-point function reveals, however, additional
features which are not amenable to the semiclassical analysis.
The orientifolds ${\rm O}_B$ have additional couplings
to odd winding modes, which are \textsc{ir} finite, {\it i.e.} the  corresponding
one-point functions do not exhibit a pole at $p=0$.
These couplings indicate the presence of a localized orientifold source
based on a parity, which morally speaking, acts on the $U(1)$ part of
the \slc closed string sector as an $s\Omega$ parity (in the notation of~\cite {Brunner:2002em}).
We pointed out in section~\ref{geomorientsection} that there is an
AdS$_3$ parity $\tau_3 \Omega$ which gives a localized orientifold
in AdS$_3$ and upon descent an orientifold localized in \slc at the tip
of the cigar at $\rho=0$. This parity involves a half-period shift $s$
around the angular direction $\phi$ of the cigar and is a natural candidate for the extra localized
orientifold source that gives rise to the second term on the  numerator of the
second line in~\eqref{bootad}.
Consequently, we would like to propose
that the orientifold planes ${\rm O}_B$ are geometrically
a combination of an ${\rm O}0$-plane, localized near the tip
of the cigar sourcing odd winding modes, and an ${\rm O}2$-plane
which extends to the asymptotic cylinder region, covers the whole cigar
and sources even winding modes.\footnote{Moreover, in the parent \slr theory
we expect that the isometry $\tau_3$ acts trivially on the open string sector
of the D(-1)-brane, which contains only the identity representation of the \slr algebra. 
Hence, upon descent it is natural to expect that 
the $\mathcal P_3$ parity shares the same M\"obius strip amplitude as $\PP_2$.
This is consistent with the interpretation of the properties of the ${\rm O}_B$ plane that we
propose here.} 
We will make this geometric
statement more precise in sect.~\ref{O1sect} by using a different basis
of wave-functions in H$_3^+$.

A similar story holds for the second B-type orientifold that we 
constructed.  By descent from AdS$_3$ and from the asymptotic semiclassical analysis
of the parity $\tilde \PP_2$ on the cigar we learn that the geometry of 
the ${\rm \tilde O}_B$ orientifold is that of an ${\rm O}2$-plane with an 
$s\Omega$ action in the angular direction of the cigar. However, as in the above case of the 
${\rm O}_B$ orientifold, the exact one-point functions \eqref{tbootab} reveal an extra 
localized contribution which suggests the presence of a localized orientifold source 
that couples to even winding states. As explained in sect.~\ref{geomorientsection}, 
there is a localized orientifold in AdS$_3$, based on the parity 
$\tilde \tau_3$, which descends to an ${\rm O}0$-plane on the cigar that couples
to even winding states. This is a natural candidate for the localized
source in eqn.~\eqref{tbootab}. Hence, we propose that ${\rm \tilde O}_B$ is
a combination of an ${\rm \tilde O}2$- and an ${\rm \tilde O}0$-plane based respectively on the
parities $\tilde \PP_2$ and
$\tilde \PP_3$.

\subsubsection*{Hanany-Witten setups}
The above combination of localized and extended B-type orientifolds as 
consistent conformal field theory objects may have a natural
interpretation in Hanany-Witten setups. In these setups one is able to engineer a variety of 
gauge theories with suitable configurations of D-branes, orientifolds and 
fivebranes. For example, in type IIA superstring theory one can 
suspend a stack of D4-branes between two parallel fivebranes 
to engineer super-Yang-Mills theory in four dimensions 
with $\mathcal{N}=2$  supersymmetry and unitary gauge 
groups~\cite{Hanany:1996ie,Witten:1997sc,Giveon:1998sr}.

It is well known~\cite{Giveon:1999px,Israel:2004ir} that the cigar \textsc{cft} appears
naturally as part of the worldsheet theory in the near horizon region of
NS5-branes in a double scaling limit. For example, it can be argued
that string theory in the near-horizon geometry of two parallel fivebranes separated in a
transverse direction (say direction $x^6$) is described in a double scaling limit by type
II non-critical string theory on $\R^{5,1} \times$ \slc, where the coset is at level $k=2$. 
In this context the D4-branes correspond in the \slc space to D0-branes 
at the tip.\footnote{One can easily generalize this construction to a ring of more 
than two NS5-branes, for which the superstring theory is really critical, and involves 
an $\mathcal{N}=2$ minimal model SU(2)/U(1); see~\cite{Israel:2005fn} for more details.}

The non-critical string picture, which can be generalized to include also 
other configurations of fivebranes, {\it e.g.} two orthogonal fivebranes, allows for a 
perturbative string theory analysis of Hanany-Witten configurations that
takes into account the gravitational backreaction of the NS5-branes. In this way, one  
can test explicitly whether some heuristic rules of brane
constructions hold~\cite{Israel:2005fn,Ashok:2005py,Fotopoulos:2005cn,Murthy:2006xt}. 

\begin{figure}[t]
\centering
\epsfig{file=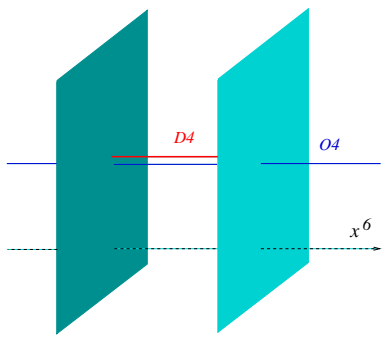,width=50mm}
\caption{\it Hanany-Witten setup for  $\mathcal{N}=2$ super Yang- Mills with
orthogonal and symplectic gauge groups.}
\label{figHW}
\end{figure}

In addition to the NS5-branes and D4-branes, it is possible to include an ${\rm O}4$-plane 
along $x^6$ with the rest  of its directions parallel to the fivebranes (see fig.~\ref{figHW}). 
On the D4-branes this leads to $\mathcal{N}=2$ gauge theories with orthogonal 
and symplectic gauge groups 
(see the review~\cite{Giveon:1998sr} and references therein).
In the 6-direction the ${\rm O}4$-plane breaks into three pieces: two  pieces
extending to infinity from the left and the right of the fivebranes and
a finite piece in between. Based on the known dictionary
between D-branes in the presence of fivebranes and D-branes on the cigar~\cite{Israel: 2005fn},
one would be urged to conjecture a correspondence between
the ${\rm O}4$-plane of fig.~\ref{figHW} and the ${\rm O}_B$ orientifold of this work 
(of course appropriately translated in type II string theory where the \textsc{gso} projection 
involves  an asymmetric orbifold of \slc$\!$).

This correspondence indicates that one can  
match the extensive ${\rm O}2$ and localized ${\rm O}0$ 
contributions to the crosscap states with respectively the left, right semi-infinite 
pieces of the ${\rm O}4$-plane and the finite ${\rm O}4$ piece
in between. From this point of view it is natural to have both the
${\rm O}2$ and ${\rm O}0$ contributions to ${\rm O}_B$, 
because each of them separately would correspond to an ${\rm O}4$-plane
ending on a fivebrane, which is certainly not a consistent configuration. 

Moreover it is known~\cite{Evans:1997hk} that the parts of the O4-plane 
on each side of the NS5-brane carry opposite \textsc{r-r} charge. If 
one starts with two NS5-branes on top of each other and an O4$^+$ plane as in 
fig.\ \ref{figHW} 
and begins to separate the fivebranes in the $x^6$ direction, 
the part of the orientifold that stays between the NS5-branes is negatively charged, 
which requires the addition of a pair of D4-branes to ensure charge conservation 
across the fivebranes.\footnote{Conversely, a configuration with 
(negatively charged) O4$^-$-planes requires the addition of a pair of 
semi-infinite D4-branes on each side of the fivebrane interval.} 
It should be possible to reproduce this feature from the details
of our ${\rm O}_B$ crosscap state. We will see below that the 
couplings to closed string modes of the localized and extended parts of the orientifold   have 
in fact  opposite signs.

In Hanany-Witten setups one can
engineer a wide class of four-dimensional gauge theories
with $SU(N)$, $SO(N)$ or $Sp(N)$ gauge groups and non-chiral or chiral
matter. For instance, one can obtain $\NN=1$ SQCD in this way
with a combination of D4- and D6-branes in type IIA. This configuration
has been analyzed in the dual cigar \textsc{cft} language
in~\cite{Fotopoulos:2005cn,Murthy:2006xt}. In these more general constructions
${\rm O}4$- and ${\rm O}6$-planes play an important role. It would be
very interesting to investigate in general how known properties of these 
constructions translate in the language
of the exact \textsc{cft} description of this paper and vice versa  and what
lessons we can learn in this way about gauge theory dynamics.

\subsubsection*{Some quantitative results}
In order to obtain a better understanding of the properties of orientifolds 
in the context of Hanany-Witten setups and their relation with our 
work, we elaborate a bit further here on a configuration 
including ${\rm O}4$-planes in six-dimensional 
non-critical type II superstrings using  an ${\rm O}_B$-plane 
similar to those constructed above.

We start with type IIA superstrings on $\mathbb{R}^{5,1}\times$ \slc$\!|_{2}$. The
angular coordinate of the cigar for $k=2$ is asymptotically a free $U(1)$ boson at level 2. 
It is well-known that upon a $\mathbb{Z}_2$ orbifold this $U(1)$ is the same as the theory 
of a Dirac fermion.\footnote{This allows the $U(1)_2$ to play the role of the two 
"missing" fermions (compared to ten-dimensional flat space-time) 
in the \textsc{gso} projection.}  
We therefore define special combinations of the \slc characters at level
$k=2$ that appear naturally in the fermionic description; let us consider the example of the identity character
\begin{equation}
Ch^\textsc{nc}_{\mathbb{I}} (\tau) \oao{a}{b}  = e^{\frac{i\pi ab}{2}}
\sum_{r \in \mathbb{Z}} e^{i\pi rb} ch_{\mathbb{I}} (r;\tau) \oao{a}{b}
\label{idNCchar}
\end{equation}
and  the continuous representations
\begin{equation}
Ch^\textsc{nc}_{c} (P;\tau) \oao{a}{b}  = e^{\frac{i\pi ab}{2}}
\sum_{r \in \mathbb{Z}} e^{i\pi rb} ch_{c} (P,r+\nicefrac{a}{2};\tau) \oao{a}{b}
~.
\label{contNCchar}
\end{equation}
As far as the $[a,b]$ labels are concerned,
these characters are such that  their modular transformation is similar to that  
of two left-moving complex fermions. The usual  momentum and winding of the cigar are
\begin{equation}
n=r+\bar r + \frac{a+\bar a}{2} \quad , \qquad w = \frac{r-\bar r}{2} + \frac{a-\bar a}{4}~,
\end{equation}
where $\bar r$ enters into the definition of the right-moving analogue 
of~(\ref{contNCchar}). 
Since we are dealing with an asymmetric orbifold of the cigar, $n$ and $w$ are not necessarily 
integer.\footnote{The standard extended characters of \slc (see app.~\ref{appchar}),  
are closely related to the above characters via a $\mathbb{Z}_2$ orbifold. 
In the type II context considered here this orbifold is asymmetric.} 
In the context of two parallel fivebranes,
the momentum $n$ of the cigar (which is conserved) corresponds 
to the (unbroken) rotational symmetry in the plane $(x^8,x^9)$ while 
the fractional winding symmetry corresponds  
to the (broken to $\mathbb{Z}_2$) rotational symmetry in the plane $(x^6,x^7)$ 
where the fivebranes have been separated~\cite{Murthy:2003es,Israel:2005fn}.\footnote{This is consistent with the 
fact that the winding number can be violated 
by any integral amount by insertion of the winding 
condensate~(\ref{liouvintwind}) in the correlators. In the $\mathbb{Z}_2$ 
orbifold model it remains a conserved $\mathbb{Z}_2$ charge.}
Since we are dealing here with an asymmetric $\mathbb{Z}_2$ 
orbifold of the cigar, acting precisely as  shifts along $\phi$, 
the distinction between  O$_B$ and $\tilde{\rm O}_B$ is 
intertwined with the details of the \textsc{gso} projection. We find 
that the O$_B$ orientifold seems to be the relevant parity here as the $\tilde{\rm O}_B$ 
amounts, in the fermionized picture, to reversing the \textsc{gso} projection for one 
of the complex fermions in the transverse direction to 
the fivebranes.

Using the set of characters defined above one can write the 
torus amplitude of the type IIA non-critical superstring theory of interest as
\begin{multline}
\mathcal{T} =  V\int_\mathcal{F} \frac{\di \tau \di \bar \tau}{4\tau_2^2}
\frac{1}{(8\pi^2 \tau_2)^6 \eta^4 \bar{\eta}^4}\, 
\frac{1}{2}\sum_{a,b \in \mathbb{Z}_2}\frac{1}{2}\sum_{\bar a,\bar b \in \mathbb{Z}_2}
(-)^{a+b+\bar a + \bar b + \bar a \bar b} \frac{\vartheta^2 \oao{a}{b}
\bar{\vartheta}^2 \oao{\bar a}{\bar b}^2}{\eta^2 \bar{\eta}^2}\ \times \\
 \times \ \int_0^\infty \!\!\! \di P \ Ch^\textsc{nc}_{c} (P;\tau) \oao{a}{b}
\bar{Ch}^\textsc{nc}_{c} (P;\bar{\tau}) \oao{\bar a}{\bar b}~,
\end{multline}
where $\mathcal F$ is the PSL(2,$\mathbb{Z}$) fundamental domain. 
Now let us add $N$ D4-branes suspended between the NS5-branes. 
In our exact \textsc{cft} setup each D4 has four Neumann boundary conditions
in the six flat directions of $\R^{5,1}$ and is 
a D0-brane on the cigar part of the worldsheet \textsc{cft}. Using the above modified set of 
coset characters, one requires for these branes the annulus amplitude
\begin{equation}
\mathcal{A} =  N^2 \, V_4 \int \frac{\di t}{2t}
\frac{1}{(16\pi^2 t)^4 \eta^4}\
\frac{1}{2} \sum_{a,b \in \mathbb{Z}_2}
(-)^{a+b} \frac{\vartheta^2 \oao{a}{b}}{\eta^2 } Ch^\textsc{nc}_{\mathbb{I}} (it) \oao{a}{b}\, .
\end{equation}
In addition, we consider an O$_B$ orientifold of the cigar with four 
Neumann dimensions in $\R^{5,1}$. Requiring again
similar modular properties as those of two Dirac fermions we define the hatted
version of the characters~(\ref{idNCchar}) as follows:
\begin{align}
\widehat{Ch}^\textsc{nc}_{\mathbb{I}} (it) \oao{a}{b} &= e^{\frac{i\pi}{4}(1-a^2)}
e^{\frac{i\pi ab}{2}} \sum_{n \in \mathbb{Z}} e^{i\pi n b}
ch_{\mathbb{I}} (n;it+\hl) \oao{a}{b}\nonumber \\ &= 
e^{\frac{i\pi (ab-1)}{2}+ \frac{i\pi a}{4}}\sum_{n \in \mathbb{Z}} e^{i\pi\left( 
n (b+1)+\frac{n^2+an}{2}\right)}
\widehat{ch}_{\mathbb{I}} (n;it) \oao{a}{b}\, .
\label{hattedNSchar}
\end{align}
These hatted characters are such that the orientifold action on the fermionized $U(1)_2$ is 
similar to that on the other worldsheet fermions. They are related non-trivially to a sum of unextended hatted characters, 
as defined in app.~\ref{Pmatapp} and used in sect.~\ref{O2section}. 
Indeed, in the definition~(\ref{idNCchar}, \ref{contNCchar}) one sums over the spectral 
flow orbit of the $\mathcal{N}=2$ algebra, so that the states are reorganized 
in terms of the extended symmetry that appears for $k \in \mathbb{Z}_{>0}$. 
As a consequence,  states with $r \neq 0$ are considered as primaries of the 
unextended symmety, but are 
not primaries of the extended one. This is why the character~(\ref{hattedNSchar}) 
contains the phase factor $\exp{i\pi(r^2/2+ar)}$ if one compares with 
app.~\ref{Pmatapp}.\footnote{The 
situation is exactly the same in $U(1)_k$ vs.\ generic $U(1)$ orientifolds 
as discussed in~\cite{Brunner:2002em}.} The general guideline is to obtain 
modular properties consistent with the generalized \textsc{gso} projection and 
spacetime supersymmetry.

Accordingly, we make the following M\"obius strip amplitude ansatz for an orientifold extended along 
$\mathbb{R}^{3,1}$ and of the O$_B$ type in \slc --~expected to correspond to an O4-plane in the five-branes background~--  
and $N$ D4-branes in the non-critical superstring (for the overall phase in the \textsc{ns} sector, see 
footnote~\ref{hosofoot}):
\begin{multline}
\mathcal{M} =   N \epsilon\, V_4 \int \frac{\di t}{2t}
\frac{1}{(16\pi^2 t)^4 \eta^4 (it+\hl)}
\frac{1}{2} \sum_{a,b \in \mathbb{Z}_2} e^{\frac{i \pi (1-a^2)}{4}}
(-)^{a+b} \frac{\vartheta^2 \oao{a}{b}(it+\hl)}{\eta^2(it+\hl) }
\widehat{Ch}^\textsc{nc}_{\mathbb{I}} (it) \oao{a}{b}
~.
\end{multline}
In this expression $\epsilon = \pm 1$ denotes the usual sign ambiguity of the
M\"obius strip amplitude. 
We can modular transform this result to the closed string channel using the results
of app.~\ref{Pmatapp}. Then we obtain
\begin{multline}
\mathcal{M} =   N \epsilon \, V_4 \int \frac{\di t}{t^3}
\frac{1}{(16\pi^2 )^4 \eta^4 (-\frac{i}{4t}+\frac12)}\
\frac12 \sum_{a,b \in \mathbb{Z}_2} 
(-)^{b+ab+1}
\frac{\vartheta^2 \oao{a}{a-b+1}(-\nicefrac{i}{4t}+\hl)}{\eta^2(-\nicefrac{i}{4t}+\hl) }\ \times\\
\times \ \sqrt{2} 
\int_0^{\infty} \!\!\! \frac{\di P}{\cosh \pi P}
\left[ \cosh \frac{\pi P}{2} - \sinh \frac{\pi P}{2}\, \sinh \pi P \right]
e^{\frac{i\pi(a^2-1)}{4}}\widehat{Ch}^\textsc{nc}_{c}\left( P,-\frac{1}{4it}\right)\oao{a}{a-b+1}
~.
\label{HWM}
\end{multline}
There is also a contribution of $j=1$ discrete characters in this expression which 
we will not write out explicitly.
We observe that the contributions of the O2- and O0-planes, respectively the 
first and second terms inside the square brackets, couple to the same characters 
(in contrast with the un-orbifoldized \slc theory discussed in subsection~\ref{OtwoOzero}). More importantly, 
these two parts of the orientifold wave-function 
come with opposite signs, suggesting as discussed above that, while the O2-part 
(mapped to the semi-infinite parts of the O4-plane outside the five-branes) 
corresponds to an O4$^+$ plane, the O0-part (mapped to the segment of the 
O4-plane between the five-branes) is similar to an O4$^-$ plane. 

Using the explicit expression for the continuous representation characters, see 
app.~\ref{appchar}, one finds, using Jacobi's abstruse identity, 
that the amplitude vanishes as expected from supersymmetry:
\begin{multline}
\mathcal{M} =   N \epsilon \,   V_4 \int \frac{\di t}{t^3}
\frac{1}{(16\pi^2 )^4 \eta^9 (-\frac{i}{4t}+\frac12)}\
\ \times\\
\times \ \frac{1}{\sqrt{2}} 
\int_0^{\infty} \!\!\! \frac{\di P}{\cosh \pi P}
\left( \cosh \frac{\pi P}{2} - \sinh \frac{\pi P}{2}\, \sinh \pi P \right) e^{-\frac{\pi }{4t} P^2}  \  
\Big[ \vartheta_3^4  -\vartheta_4^4  - 
\vartheta_2^4    \Big] (-\tfrac{i}{4t}+\tfrac{1}{2})   = 0 \, .  
\label{HWsusy}
\end{multline}

We leave a more detailed analysis of these results and the
corresponding spacetime physics for future work.

%%%%%%%%%%%%%%%%%%%%%%%%%%%%%%%%%%%%%%%%%%%

\section{A-type orientifolds on the cigar: an O1-plane}
\label{O1sect}
In this last section we construct A-type orientifolds of the \slc coset \textsc{cft}. In the 
axial coset (the cigar) they correspond to O1-planes extending in the asympotic region, 
with a geometry similar to the lower-left picture of fig.~\ref{branesgeom}.  
Algebraically they are related to the parity $\mathcal{P}_1 =  \tau_1 \Omega_B$, 
using the notation of sect.~\ref{geomorientsection}. In terms of the natural 
$\mathcal{N}=2$ parity $\Omega_A$ we can write $\mathcal{P}_1$ 
as $(-)^{\bar F} \Omega_A$. Let us recall that the "twisted" 
version of this orientifold, the $\tau_4 \Omega$ parity of sect.~\ref{geomorientsection}, 
is non-perturbatively inconsistent because it projects out the winding 
condensate~(\ref{liouvintwind}).

Although A-type boundary conditions are usually more straightforward to deal with, a problem
arises when one tries to apply modular boostrap methods to this case. Indeed, the D0-brane 
of the cigar has B-type boundary conditions, therefore its M\"obius strip amplitude with
the ${\rm O}1$-plane would involve mixed boundary conditions and would
either vanish or turn the computation of the corresponding $\mathcal{P}$-matrix 
into a complicated problem. We could try to circumvent this
difficulty by starting with the M\"obius strip amplitude for an A-type brane, {\it i.e.}\ a D1-brane. 
However the latter has a continuous open string spectrum, with a regularized density 
of states that contains most of the information about the boundary state. Previous 
experience teaches us that dealing with this volume divergence 
is quite intricate.

Our strategy to solve this problem will be to study this orientifold in the parent 
\textsc{wzw} model \slr --~or more conveniently in its Euclidean counterpart 
H$_3^+$~-- rather than in its cosets. Indeed, it is rather straightforward to 
"lift" the results of section~\ref{O2section} concerning the ${\rm O}2$-planes 
of the axial coset \slc 
to H$_3^+$. There the problem is simpler since the ${\rm O}2$-plane and the 
O1-plane are related to each other by an SL(2,$\mathbb{C}$) relation, as the corresponding 
D-branes~\cite{Ponsot:2001gt}.\footnote{We essentially apply backwards  the strategy used 
in~\cite{Ribault:2003ss} to find the D2-brane of the cigar \textsc{cft}. Note that 
we use the equivalence \mbox{\slc $\sim$ H$_3^+/\mathbb{R}$}.} This will furthermore 
allow us to compare with the conformal bootstrap results in H$_3^+$ that 
were obtained in~\cite{Hikida:2002fh}.

\subsection*{The asymptotic Klein bottle amplitude}
We can gain intuition about the ${\rm O}1$-plane by looking at the way the orientifold 
projection acts on the closed string spectrum of the cigar and computing the Klein 
bottle amplitude in the direct channel. 
As in sect.~\ref{O2section} we consider first  the extensive part of the 
torus amplitude. The contribution of the finite regularized density of 
states will be computed later using the exact crosscap state.
There we will also see that the ${\rm O}1$-plane couples only to the
continuous representations.

The leading part of the torus amplitude with a type 0B modular invariant 
appears in eqn.\ \eqref{p2aa}.
Following the geometric and algebraic descriptions of sect.~\ref{geomorientsection}, 
we find that the parity $\mathcal{P}_1 = (-)^{\bar F} \tau_1 \Omega$  acts on an  
\textsc{ns-ns} primary state $|P;n,w\rangle \otimes |0\rangle_\textsc{ns-ns}$ as follows:
\begin{equation}
\mathcal{P}_1 \ : \quad |P;n,w\rangle \otimes |0\rangle_\textsc{ns} \ \longrightarrow
\  |P;-n,w\rangle \otimes |0\rangle_\textsc{ns-ns}\, .
\end{equation}
The trace over the bosonic oscillators in unaffected, since $\Omega$ requires the pairing of
left- and right-movers and the geometric involution $\tau_1$ leaves the paired combinations 
invariant. The parity, including the $(-)^{\bar F}$ factor, acts on worldsheet fermions as
\begin{equation}
\mathcal{P}_1 \ : \quad \psi^{\epsilon} \to  \bar{\psi}^{-\bar{\epsilon}} \ , \quad
\quad \bar{\psi}^{\bar \epsilon} \to  -\psi^{-\epsilon}\, ,
\end{equation}
so that each term of the winding $\mathcal{N}=2$ Liouville interaction, 
see eqn.~(\ref{liouvintwind}), is separately invariant. Because of the 
diagonal \textsc{gso} projection only states with $F+\bar F = 0 \mod 2$ 
contribute to the torus amplitude. One can similarly trace the action in 
the \textsc{r-r} sector. The Klein bottle amplitude reads:
\begin{equation}
\mathcal{K}_\text{O1} (t) = V \sum_{a \in \mathbb{Z}_2}
\int_0^\infty \di P \, \sum_{w \in \mathbb{Z}} 
e^{-4\pi t \left(\frac{P^2}{k} + \frac{kw^2}{4}\right)}  
\frac{\vartheta \oao{a}{0} (2it)}{
\eta (2it)^3}
\label{asymptklein}
\end{equation}
Using the standard $S$-modular 
transformation~(\ref{p2ae}), one finds in the transverse channel
\begin{equation}
\tilde{\mathcal{K}}_\text{O1} (t) =  \frac{V}{2} \sum_{a \in \mathbb{Z}_2} 
\sum_{n \in \mathbb{Z}} \delta_{n,0}^{(2)}
e^{-\frac{\pi}{t}  \frac{n^2}{4k}}  \
\frac{\vartheta \oao{0}{a} (-\frac{1}{2it})}{
\eta (-\frac{1}{2it})^3}
\label{asymptkleindual}
\end{equation}
Therefore the orientifold plane sources only even momentum 
states in the \textsc{ns-ns} sector in this context.

\subsection*{The crosscap wave-function by rotation}
In order to obtain the full crosscap state of the ${\rm O}1$-plane in \slc, we  start 
from the ${\rm O}2$ crosscap wave-function in H$_3^+$. The latter is obtained using  the coset construction backwards, 
as explained in the beginning of this section, from our results for the ${\rm O}_B$-plane in the coset model.  
Let us write the one-point functions on $\mathbb{RP}_2$ for the O2/O0-plane, 
for generic $j$ in the \textsc{ns-ns} sector as follows: 
\begin{equation}
\langle V^{\textsc{ns-ns}}_{j,m\, \bar m} (z,\bar z) \rangle_\text{O2} = 
\frac{C^{\text{O2}}_{j\, m \, \bar m}}{
|1+z\bar z|^{2\Delta_{jm \bar m}}\!\!\!\!\!\!\!\!\!\!\!\!\!\!} 
\label{orientNC}
\end{equation}
with
\begin{multline} 
C^{\text{O2}}_{j\, m \, \bar m} = \mathcal{N}_k e^{\frac{i\pi}{4}} \delta_{m,\bar m} 
\nu^{\hl-j} \Gamma \left(1+\frac{1-2j}{k} \right)
 \ \times \\ \times \ 
\left\{ \cos \frac{\pi (j-\hl)}{k}
\Gamma (1-2j) \left(\frac{\Gamma(j-m)}{\Gamma(1-j-m)} +
\frac{\Gamma(j+m)}{\Gamma(1-j+m)} \right)-\right. \\ \left. -
i \sin \frac{\pi (j-\hl)}{k}\, \frac{\Gamma(j+m) \Gamma (j-m)}{\Gamma(2j)} 
\right\}\, .
\label{orientNC2}
\end{multline}
This expression is obtained from the $\mathcal{P}$-matrix element of one unextended 
character, see app.~\ref{Pmatapp}, analytically continued in the complex $j$-plane.
In the axial/vector coset, there will be some condition over $m$ and $\bar m$ 
($i.e.$ $m\pm \bar m=0$), otherwise 
this result applies readily (up to a $k$-dependent normalization factor $\mathcal{N}_k$)
to the parent H$_3^+$ theory.\footnote{Note that in H$_3^+$ there are no spectrally-flowed 
states in the spectrum.}

Since we consider below the H$_3^+$ 
model, for which the Euclidean time is non-compact, there is no room for an 
analogue of the $\tilde{\rm O}_B$ orientifold. It will be more convenient here to 
use the $(x,\bar x)$ basis, related to the $(m,\bar m)$ basis through the Mellin transform 
(see {\it e.g.}~\cite{Ponsot:2001gt}):
\begin{equation}
\hat{f}_{m \, \bar m} = \frac{1}{4\pi^2} \int_\mathbb{C} \di^2 x 
\ x^{j-1+m} \, \bar x^{j-1+\bar m} f(x,\bar x)
\label{Mellin}
\end{equation}
Therefore we can re-express the crosscap couplings as
\begin{multline} 
C^{\text{O2}}_{j\, x \, \bar x} = 4\pi \mathcal{N}_k \,
\nu^{\hl-j} \Gamma \left(1+\frac{1-2j}{k} \right)\ \times \\
\times\ \left\{ \cos \frac{\pi (j-\hl)}{k} |1-x\bar x|^{-2j}
-i \sin \frac{\pi (j-\hl)}{k}\,  |1+x\bar x|^{2j}
\right\}~.
\label{crosscapxbasis}
\end{multline}
Comparing this expression to the one-point functions for the branes 
found in~\cite{Ponsot:2001gt} in H$_3^+$, one confirms the geometrical 
interpretation outlined in sect.~\ref{O2section}. The first 
term corresponds to an H$_2$ orientifold in H$_3^+$, since it has 
the geometry of an H$_2$ brane with no magnetic field, while the second term 
corresponds to a point-like orientifold with the same geometry as a "spherical brane" 
in H$_3^+$ with zero radius.\footnote{In fact, the wave-functions of both terms in 
eqn.~(\ref{crosscapxbasis}) agree with the corresponding wave-functions of 
the Euclidean D1- and D(-1)-branes, up to the quantum corrections 
(the $\cos \pi (j-\hl)/k$ and $\sin \pi (j-\hl)/k$ factors) which are not fixed by the gluing conditions alone,  
and therefore may differ between a brane and an orientifold with the same geometry. 
} As argued previously from several point of views, 
these orientifolds are tied together in the coset  
and cannot make sense separately.

Now, in order to obtain the Euclidean AdS$_2$ 
orientifold in Euclidean AdS$_3$, we consider an SL(2,$\mathbb{C}$) rotation acting 
on the SL(2,$\mathbb{C}$)/SU(2) eigen-functions as follows:
\begin{equation}
U\ : \quad \Phi^{j}_{x \bar x} \ \longrightarrow \  \left|\frac{x+1}{\sqrt{2}}\right|^{-4j} 
\Phi^{j}_{  \frac{x-1}{x+1} ,  \frac{\bar x-1}{ \bar x+1}}
\end{equation}
Under this rotation, the crosscap wave-function~(\ref{crosscapxbasis}) transforms as:
\begin{multline}
C^{\text{O2}}_{j\, x \, \bar x} \ \longrightarrow  4\pi \mathcal{N}_k\,
\nu^{\hl-j} \ \Gamma \left(1+\frac{1-2j}{k} \right)\ \times \\
\times\ \left\{ \cos \frac{\pi (j-\hl)}{k} |x+\bar x|^{-2j}
-i \sin \frac{\pi (j-\hl)}{k}\,  |1+x\bar x|^{-2j}
\right\}\, .
\label{crosscaprotxbasis}
\end{multline}
This result can be interpreted as follows. The first term 
of the crosscap wave-function that exhibits the H$_2$ geometry, 
is rotated to an orientifold with an AdS$_2$ geometry.  The second term 
is invariant as it should, since the O(-1) has a point-like geometry and sits at 
the center of rotation.

We now come back to the axial coset \slc. There we would like to 
argue that by descent from the first term {\it alone}, {\it i.e.}\ the AdS$_2$-plane, we obtain a 
consistent O1-plane. First notice that the two terms of eqn.~(\ref{crosscaprotxbasis}) 
will give rise to different boundary conditions for the $\mathcal{N}=2$ superconformal algebra 
(A-type for the first one and B-type for the second one). In other words, 
the corresponding crosscap states will be constructed out of a different 
set of Ishibashi states. We learned, however, in sect.~\ref{O2section} that
an ${\rm O}0$-plane alone (this would come from the second term of 
eqn.~(\ref{crosscaprotxbasis})), cannot be a consistent
orientifold on its own.
In particular, the couplings to discrete representations will 
not be consistent, see app.~\ref{Pmatapp}. Furthermore, we will 
see below that the first term of~(\ref{crosscaprotxbasis}), after descent to 
the coset theory, does not contain couplings to discrete representations and 
therefore is free of this problem. To summarize, from the first piece we get 
the O1-plane wave-function in the $x$-basis
\begin{equation}
C^{\text{O1}}_{j\, x \, \bar x} =
4\pi \mathcal{N}_k \, 
\nu^{\hl-j} \Gamma \left(1+\frac{1-2j}{k} \right)
\cos \frac{\pi (j-\hl)}{k} |x+\bar x|^{-2j}
~.
\label{crosscapO1xbasis}
\end{equation}
We can now go back to the $(m,\bar m)$ basis using the Mellin transform~(\ref{Mellin}) 
and finally obtain the crosscap wave-function in the cigar for the \textsc{ns-ns} sector:
\begin{equation}
C^{\text{O1}\, \textsc{ns}}_{j\, \frac{n+kw}{2} \, \frac{n-kw}{2}} =
\mathcal{N}'_k  \, \delta_{w,0}\ \delta_{n,0}^{(2)}
\nu^{\hl-j} \cos \frac{\pi (j-\hl)}{k}
\frac{\Gamma \left(1+\frac{1-2j}{k} \right) \Gamma(1-2j)}{\Gamma(1-j+\frac{n}{2})
\Gamma(1-j-\frac{n}{2})}
~.
\label{crosscapO1mbasis}
\end{equation}
Up to a cosine term (which accounts for "quantum" corrections to the semi-classical result) 
we have indeed the same wave-function as for a straight ({\it i.e.}\ with $\hat \rho=0$) D1-brane 
in the cigar, see eqn.\ \eqref{appbae} in app.~\ref{appwave-functions} or 
ref.~\cite{Ribault:2003ss}. 
One can check that the crosscap wave-function is compatible 
with the reflection symmetry~(\ref{reflproper}). Similarly to the D1-brane case, 
this crosscap wave-function does not 
possess couplings to discrete representations of \slc as we advertised above.

It is rather straightforward to obtain the
wave-function in the \textsc{r-r} sector. Indeed, using the coset construction, 
the H$_3^+/\mathbb{R}$ coset superconformal field theory
can be represented as the constrained product of \textsc{cft}s \mbox{ H$_3^+$ 
$\times$ U(1) $\times$ [Dirac fermion] $\times$ [ghosts] $\times$ [superghosts]}. 
The wave-function~(\ref{crosscapO1mbasis}) is written in terms 
of the eigenvalues of the {\it bosonic} \slr algebra 
$(m_b ,\bar{m}_b) = (m-m_f, \bar{m}-\bar{m}_f)$ 
where $(m_f,\bar{m}_f)$ is the same as the left- and
right-fermion number of the  free Dirac fermion. 
Therefore we find the generic wave-function (with a similar notation
as in sect.~\ref{O2section})
\begin{equation}
\Psi_{{\rm  O}1}\left(p,n; \left[ {b \atop c} \right]\right)=
\mathcal{N}'_k  \, e^{i\varphi(b,c)}\ \delta_{n,0}^{(2)}
\nu^{-ip} \cosh \frac{\pi p}{k}\,
\frac{\Gamma \left(1-\frac{2ip}{k} \right) \Gamma(-2ip)}{\Gamma(\frac{1}{2}-ip +\frac{n+b}{2})
\Gamma(\frac{1}{2}-ip-\frac{n+b}{2})}\, .
\label{crosscapO1mbasisallsec}
\end{equation}
We will fix below the normalization of the wave-function~(\ref{crosscapO1mbasis}) and the
phases $\varphi(b,c)$ in the different fermionic sectors by computing different amplitudes.

\paragraph*{Comparison with conformal bootstrap}
Partial results for the conformal bootstrap of Euclidean AdS$_2$ orientifold planes 
in H$_3^+$ were obtained in~\cite{Hikida:2002fh}. The author of this paper 
considered the auxillary two-point function 
$\langle \Phi^j_{x\bar x} \Phi^{-\hl}_{y \bar y}\rangle_{\mathbb{RP}_2}$ with
a degenerate representation $j=-\hl$ in order to constrain the form of the crosscap 
wave-function. In this way he proposed couplings to the continuous representations 
which are identical to our result (\ref{crosscapO1xbasis}) when evaluated 
at $j=\hl+ip$. However we should emphasize that the conformal bootstrap method used there 
was not powerfull enough in order to fully determine the crosscap wave-function. 
Our approach allows to remove this freedom and find the full wave-function, up 
to an overall normalization that is fixed by a Cardy-like condition.

\subsection*{Asymptotic M\"obius strip amplitude for the D1-brane}
We will study here the effect of the parity $\mathcal{P}_1$ on open 
string sectors attached to D1-branes of the cigar. 
As reviewed in sect.~\ref{secrev} these branes, which 
extend to the asymptotic region, are characterized by two 
parameters $(\hat{\rho},\hat{\phi})$. From eqn.~(\ref{embed}) 
we observe that $\hat{\rho}$ parameterizes the position of the turning point
of the brane near the tip of the cigar at $\rho=0$. Since the ${\rm O}1$-plane 
corresponding to $\mathcal{P}_1$ has a similar geometry as the D1-brane 
with $\hat{\rho}=0$, it is clear that only the D1-branes with $\hat{\rho}=0$ are 
invariant.

The second parameter $\hat{\phi}$ gives the position of the
brane on the transverse circle in the asymptotic region $\rho \to \infty$ where the geometry
is approximated by a semi-infinite cylinder. The brane possesses two branches, 
at ($\hat \phi,\hat \phi + \pi)$. Consequently, there are two kinds of open strings, 
one kind with integral winding where both ends of the string are on the same branch, and
another kind with half-integral winding where the open string has one end on 
each branch.\footnote{As with closed strings,
winding number on D1-branes in the full cigar geometry
is not conserved since an open string can slip around the tip. Technically
this effect comes from the boundary interaction associated with the branes 
which breaks the winding symmetry, just like the winding condensate~(\ref{liouvintwind}) 
breaks it in the closed string sector.}
The parity  $\mathcal{P}_1$ as  defined in sect.~\ref{secrev} 
corresponds at infinity to a pair of O1-planes of equal tension located at $\phi =0,\pi$.

In the asymptotic region, one finds that the action of the parity on open string states 
with integral winding $w$ is
\begin{equation}
\mathcal{P}_1  \ : \quad |w\rangle_{\hat \phi , \hat {\phi}} \longrightarrow
|w\rangle_{-\hat {\phi},-\hat{\phi}}\, .
\end{equation}
There are invariant states provided $\hat{\phi}=0 \mod \pi$. 
Since the brane has two branches there is actually only one possibility. 
Let us now consider open strings with half-integral windings. The action
of the parity reads
\begin{equation}
\mathcal{P}_1  \ : \quad |w \rangle_{\hat \phi , \hat {\phi}+\pi} \longrightarrow
|w\rangle_{\pi-\hat {\phi},-\hat{\phi}}\, .
\end{equation}
In this case, invariant open string states exist when $\hat \phi = \pi /2 \mod \pi$.
Accordingly we will distinguish between two different cases: the case of a D1-brane
with $\hat \phi=0$ and the case with $\hat \phi=\pi/2$. In the first case,
integral windings will contribute to the M\"obius strip amplitude, in the 
second half-integral windings will contribute.

Let us start with the first case. The annulus amplitude for a D1-brane,
see~\cite{Ribault:2003ss,Israel:2004jt}, comes with two different regularized 
densities of states for the integral and half-integral winding modes depending
on the parameter $\hat{\rho}$. The extensive
part of the open string partition function, however, is the same in both cases and can 
be written as
\begin{multline}
\mathcal{A}_{\hat{\phi}=0 \,\oao{a_1}{b_1}\ ;\ \hat{\phi}=0\, \oao{a_2}{b_2}} (t)
=\\ V\delta_{b_1,b_2}^{(2)}  \int_0^\infty\!\!\! \di P  \sum_{w \in \mathbb{Z}}
\left\{ ch_c (P,kw;it)\oao{a_1-a_2}{b_1} +  ch_c (P,k(w+\hl);it) \oao{a_1-a_2}{b_1} \right\}
\end{multline}
Acting with the parity $\mathcal{P}_1$ one finds the M\"obius strip amplitude:
\begin{equation}
\mathcal{M}_{\hat{\phi}=0 \,\oao{a}{b}}
= V \delta_{b,0}^{(2)} \,  \int_0^\infty\!\!\! \di P  \sum_{c \in \mathbb{Z}_2}
\sum_{w \in \mathbb{Z}}
\widehat{ch}_c (P,kw;it)\oao{0}{c}
\label{moebintwinding}
\end{equation}
With the help of eqn.~(\ref{obootab}) we perform now a $\PP$-modular 
transformation to the closed string channel to obtain the amplitude
\begin{subequations}
\begin{align}
\tilde{\mathcal{M}}_{\hat{\phi}=0 \,\oao{a}{b}} (t) 
= V \delta_{b,0}^{(2)} e^{\frac{i\pi}{4}}
\,  \sum_{c \in \mathbb{Z}_2}
\sum_{n \in \mathbb{Z}} e^{-\frac{i\pi c}{2}}
\delta_{n,0}^{(2)}\, \widehat{ch}_c \left( 0,\frac{n}{2};-\frac{1}{4it}\right) \oao{0}{1-c}
\label{mobO1dual}=\\
= V
\sum_{c \in \mathbb{Z}_2} \sum_{n \in \mathbb{Z}}
\Phi_{D1;(\hat \rho=0;\hat{\phi}=0) \oao{a}{-b}}^{\text{sing}} (0;-n)
\Psi_{{\rm  O}_A}^{\text{sing}} \left(0,n; \left[ {b \atop c} \right]\right)
\widehat{ch}_c \left( 0,\frac{n}{2};-\frac{1}{4it}\right) \oao{b}{c-a}\, .
\label{singMOB}
\end{align}
\end{subequations}
In~(\ref{singMOB}) only the residues of the poles that the wave-functions 
have for $p \to 0$ appear, since we started with the extensive part of 
the annulus amplitude. Using the brane 
wave-function~(\ref{appbae}) one finds in the limit $p\to 0$
\begin{equation}
\Phi_{D1;(\hat \rho=0;\hat{\phi}=0) \oao{a}{b}} (p;n)
 \stackrel{p\to 0}{\sim} 
 \delta_{n,0}^{(2)}
 \frac{\nu^{-ip}\, \Gamma(0)}{\sqrt{2}\Gamma(\frac{1}{2}-\frac{n+b}{2})
 \Gamma(\frac{1}{2}+\frac{n+b}{2})} =
 \frac{ (-)^{\frac{n}{2}}}{\sqrt{2}}
 \delta_{n,0}^{(2)}
\delta_{b,0}^{(2)} \,  \nu^{-ip}\, \frac{\Gamma (0)}{\pi}\, .
\end{equation}
In this way, we obtain the volume diverging part of the crosscap wave-function as
\begin{align}
\Psi_{{\rm  O}_A}^{\text{sing}} \left(0,n; \left[ {b \atop c} \right]\right)
&= \sqrt{2}\delta_{b,0}^{(2)}\delta_{n,0}^{(2)} (-)^{\frac{n}{2}} e^{\frac{i\pi(1-2c)}{4}}
\ \nu^{-ip}~.
\end{align}
We observe that it agrees exactly with the singular part of 
eqn.~(\ref{crosscapO1mbasisallsec}) provided we make
the choice
\begin{equation}
\mathcal{N}_k ' \, e^{i\varphi(b,c)} = \sqrt{2} e^{\frac{i\pi(1-2c)}{4}}~.
\end{equation}
Actually, in order to obtain an orientifold with real tension in the string theory context, 
one can use the phase ambiguity in the definition of the crosscaps in the 
\textsc{ns-ns} sector that was mentioned in footnote~\ref{hosofoot}. This done, 
the exact crosscap wave-function of the ${\rm O}1$-plane in the \slc super-coset is
\begin{equation}
\Psi_{{\rm  O}_A}\left(p,n; \left[ {b \atop c} \right]\right)=
\sqrt{2} e^{\frac{i\pi(1-b-c)}{2}} \delta_{n,0}^{(2)}
\nu^{-ip} \cosh \frac{\pi p}{k}\,
\frac{\Gamma \left(1-\frac{2ip}{k} \right) \Gamma(-2ip)}{\Gamma(\frac{1}{2}-ip +\frac{n+b}{2})
\Gamma(\frac{1}{2}-ip-\frac{n+b}{2})}\, .
\label{crosscapO1final}
\end{equation}
As with the D1-brane wave-function there is no coupling  
to the states of discrete representations of \slc.

The case of the D1-brane with $\hat{\phi}=\frac{\pi}{2}$ can be treated along 
the same lines. First, one obtains in the open string channel a M\"obius strip amplitude 
similar to eqn.~(\ref{moebintwinding}) but with half-integral windings only 
({\it i.e.} in (\ref{moebintwinding}) $w$ should be replaced with $w+\hl$). In the
dual channel, one gets eqn.~(\ref{mobO1dual}) with an extra phase $e^{i\pi n/2}$. This phase
is precisely canceled by the phase $e^{-in \hat \phi}$ of the D1-brane wave-function
(see eqn.~(\ref{appbae})) since $\hat \phi = \pi/2$. This is a nice check of consistency,
because the crosscap wave-function~(\ref{crosscapO1final}) cannot
depend on the specific brane that we use in the derivation.

\subsection*{Regularized Klein bottle amplitude}
Having at our disposal the exact crosscap state for the ${\rm O}1$-plane 
we can compute the Klein bottle amplitude beyond the asymptotic 
region~(\ref{asymptklein}) and thus determine also the non-trivial regularized 
density of states. We start with the transverse channel amplitude
\begin{align}
\tilde{\mathcal{K}}_{{\rm  O}1} (t) &= 
\sum_{c \in \mathbb{Z}_2}
\int_0^{\infty}\!\!\! \di p'  \sum_{n \in \mathbb{Z}}
\Psi_{{\rm  O}_A}\left(-p' ,-n; \left[ {0 \atop -c} \right]\right)
\Psi_{{\rm  O}_A}\left(p' ,n; \left[ {0 \atop c} \right]\right)
ch_c \left( p',\frac{n}{2};-\frac{1}{2it}\right) \oao{0}{c}\nonumber\\
&= -\frac{1}{2k} 
\sum_{c \in \mathbb{Z}_2}  \int_0^{\infty}\!\!\!
\frac{\di p'}{\tanh \pi p' \, \tanh \frac{\pi p'}{k}}
\sum_{N \in \mathbb{Z}} ch_c \left(p',N;-\frac{1}{2it}\right) \oao{0}{c}
~.
\label{kleincrosscaps}
\end{align}
We observe that the integral over $p'$ has an \textsc{ir} divergence, 
which corresponds to the infinite volume of the cigar manifold.  
Taking the leading, divergent piece --~proportional to
$\delta (p')$~-- one recovers the direct channel Klein bottle amplitude 
(\ref{asymptklein}). 
The finite part of the Klein bottle amplitude in the direct  channel reads
\begin{equation}
\mathcal{K}_{{\rm  O}_A}= -  
\sum_{c\in \mathbb{Z}_2} \sum_{w \in \mathbb{Z}} 
\int_{0}^\infty \!\!\! \di P \ \rho_K (P) 
\  ch_c \left( P,\frac{kw}{2};2it \right) \oao{c}{0}\, ,
\end{equation}
with regularized density of states 
\begin{equation}
\rho_K (P) = \frac{1}{2i\pi} \frac{\di}{\di P} \ 
i \int_0^\infty \frac{\di y}{y} \left[ \frac{\sin \frac{4Py}{k}}{2\tanh y \tanh \frac{y}{k}}
-\frac{2P}{y}\right]\, .
\label{densklein}
\end{equation}
Such a density of states is not related to the closed string reflection 
amplitude as in the torus case 
\cite{Hanany:2002ev,Eguchi:2004yi,Israel:2004ir}. 
Indeed, the density of states that appears in the latter is
\begin{equation}
\rho_\mathcal{T} (P,n)=\frac{1}{2i\pi} \frac{\di}{\di P} \log \frac{\Gamma(\hl-iP+n)}{
\Gamma(\hl+iP+n)}~.
\end{equation}
and is naturally related to the closed string reflection amplitude 
given by eqn.~(\ref{reflampl}). 
In particular, this expression does not depend on the level 
$k$ in contrast with~(\ref{densklein}). This non-trivial (and perhaps
counter-intuitive) action of orientifolds on density of states is a general feature
(for related remarks in the case of the O2-plane see sect.~\ref{O2section}).

%%%%%%%%%%%%%%%%%%%%%%%%%%%%%%%%%%%%%%%%%%%%

\section*{Acknowledgments}

We would like to thank Yasuaki Hikida, Shinji Hirano, Volker Schomerus and Jan Troost
for interesting discussions and helpful correspondence. VN
acknowledges partial financial support by the EU under the contracts
MEXT-CT-2003-509661, MRTN-CT-2004-005104 and MRTN-CT-2004-503369. 
DI acknowledges partial financial support by the EU under the 
contract MEIF-CT-2005-024072.

\begin{appendix}

%%%%%%%%%%%%%%%%%%%%%%%%%%%%%%%%%%%%%%%%%%%%

\section{Conventions and useful material}
\label{appchar}

%%%%%%%%%%%%%%%%%%%%%%%%%

\subsection*{Free fermions}
Let us consider first the theory of a free Dirac fermion. We define the usual 
theta-functions as
$$\vartheta \oao{a}{b} (\tau,\nu )
= \sum_{n \in \Z} q^{\frac{1}{2} (n+\frac{a}{2})^2}
e^{2i \pi (n+\frac{a}{2})(\nu+\frac{b}{2})},$$
where $q=e^{ 2 \pi i \tau}$.
The fermionic characters are written as $\vartheta \oao{a}{b} (\tau;\nu)/\eta (\tau)$.
The \textsc{ns} sector (resp. \textsc{r} sector) is given by $a=0$ (resp. $a=1$), 
while characters with $b=1$ have a phase $e^{i\pi F}$ inserted in the trace. 
Their modular transformations read
\begin{subequations}
\begin{align}
\frac{\vartheta \oao{a}{b} (-\nicefrac{1}{\tau};-\nicefrac{\nu}{\tau})}{\eta (-\nicefrac{1}{\tau})} & =
e^{i\pi (\nicefrac{\nu^2}{\tau}-\nicefrac{ab}{2})} \frac{\vartheta \oao{-b}{a} (\tau;\nu)}{\eta (\tau)}
\\
\frac{\vartheta \oao{a}{b} (\tau+1;\nu)}{\eta (\tau+1)} & = e^{-i\frac{\pi}{4} a(a-2)}
\frac{\vartheta \oao{a}{a+b-1} (\tau;\nu)}{\eta (\tau)}
\end{align}
\end{subequations}

%%%%%%%%%%%%%%%%%%%%%%%%%

\subsection*{Characters of the non-minimal $\mathcal{N} =2$ superconformal algebra}
The characters of the \slc super-coset at level $k$ are 
characters of the $\mathcal{N} =2$ superconformal algebra 
with $c= 3+6/k$. They come in different classes corresponding to 
irreducible representations of the \slr algebra in the parent theory. 
In all cases the quadratic Casimir of the representations is $c_2=-j(j-1)$.
Here we summarize the basic representations.

First we consider the \emph{continuous representations} with $j = 1/2 + ip$,
$p \in \mathbb{R}^+$. The corresponding characters are denoted by
$ch_c (p,m) \oao{a}{b}$, where the $N=2$ superconformal $U(1)_R$ charge 
of the primary is $Q=2m/k$, $m \in \mathbb{R}$.\footnote{The spectrum of 
R-charges is not necessarily continuous and depends on the
model considered. For instance in the cigar \textsc{cft} one has $m=(n+kw)/2$ 
with $n,w \in \Z$.} The explicit form of the characters is
\begin{equation}
ch_c (p,m;\tau,\nu) \oao{a}{b} =
q^{\frac{p^2+m^2}{k}}e^{4i\pi\nu \frac{m}{k}} \frac{\vartheta \oao{a}{b} 
(\tau, \nu)}{\eta^3 (\tau)}\, .
\end{equation}
These are also the characters that appear in a free $\mathcal{N}=2$ linear 
dilaton theory. Their $S$-modular transformation is straightforward
\begin{multline}
\label{p2ae}
ch_c\left(P,m;-\nicefrac{1}{\tau}\right)\left[ {a\atop b} \right] =
\\
\frac{4}{k} e^{-\frac{i\pi ab}{2}}
\int_0^\infty \di P' \int_{-\infty}^\infty \di m' \  e^{-\frac{4\pi i  mm'}{k}}
\cos \left(\frac{4\pi PP'}{k}\right)
ch_c\left(P',m';\tau\right) \left[{-b \atop a}\right] \, .
\end{multline}

Another important class of representations comprises of
\emph{discrete representations} in the range $\nicefrac{1}{2} < j < \nicefrac{k+1}{2}$.
The corresponding characters are usually denoted as $ch_d (j,r) \oao{a}{b}$, 
and have $N=2$ $U(1)_R$ charge $Q= (2j+2r+a)/k$ with 
$r\in \Z $.\footnote{Only states with $r \geqslant 0$ are primaries
of \slr, however states with $r < 0$ are also primaries in the coset.} 
In the cigar, $j$ is quantized but not in the
non-compact $\NN=2$ Liouville theory. The explicit form of the 
discrete characters is
\begin{equation}
ch_d (j,r;\tau,\nu) \oao{a}{b} = q^{\frac{-(j-1/2)^2+(j+r+a/2)^2}{k}}
e^{2i\pi\nu \frac{2j+2r+a}{k}} \frac{1}{1+(-)^b \,
e^{2i\pi \nu} q^{1/2+r+a/2}} \frac{\vartheta \oao{a}{b} (\tau, \nu)}{\eta^3 (\tau)} .
\end{equation}
The discrete primary states are
\begin{eqnarray*}
|j,m=j+r\rangle&=&  |0\rangle_\textsc{ns} \otimes |j,m=j+r\rangle_\textsc{bos} \quad r \geqslant 0 \\
|j,m=j+r\rangle &=& \psi^{-}_{-\frac{1}{2}}|0\rangle_\textsc{ns} \otimes (J^{-}_{-1} )^{-r-1}
|j,j\rangle_\textsc{bos} \quad r<0 
\end{eqnarray*}

While the closed string spectrum in \slc contains only discrete and
continuous representations, the spectrum of open strings attached to
localized D-branes is built on the {\it identity representation}.
We denote the character of the identity representation
by $ch_\mathbb{I} (r) \oao{a}{b}$. It has the form
\begin{equation}
ch_\mathbb{I} (r;\tau,\nu) \oao{a}{b} =  \frac{(1-q)\
  q^{\frac{-1/4+(r+a/2)^2}{k}}
e^{2i\pi\nu \frac{2r+a}{k}}}{\left( 1+(-)^b \,
e^{2i\pi \nu} q^{1/2+r+a/2} \right)\left( 1+(-)^b \, e^{-2i\pi \nu}
q^{1/2-r-a/2}\right)} \frac{\vartheta \oao{a}{b} (\tau, \nu)}{\eta^3 (\tau)}.
\label{idchar}
\end{equation}
The identity primary states in the NS sector are 
the identity operator $|j=0,r=0\rangle \otimes
| 0\rangle_\textsc{ns}$ and the primary states 
\begin{eqnarray*}
|r\rangle = \psi^{+}_{-\frac{1}{2}} |0\rangle_\textsc{ns}
\otimes (J^{+}_{-1} )^{r-1} |0,0\rangle_\textsc{bos} \quad \text{for}\ r>0
\quad \text{with}\
&  L_0 & =  \frac{r^2}{k} +r - \frac{1}{2}~, \\
|r\rangle = \psi^{-}_{-\frac{1}{2}}|0\rangle_\textsc{ns}
\otimes (J^{+}_{-1} )^{-r-1} |0,0\rangle_\textsc{bos}
\quad \text{for}\ r<0
\quad \text{with}\
&L_0& = \frac{r^2}{k} -r - \frac{1}{2}\, .
\end{eqnarray*}

\paragraph{Reflection amplitude}
Among the known structure functions of the \slc conformal field theory, the 
two-point function, or reflection amplitude, plays a special role. It reads~\cite{Giveon:1999px} 
(with $\nu = \Gamma (1-\nicefrac{1}{k})/\Gamma (1+\nicefrac{1}{k})$):
\begin{align}
&\langle V^{1-j}_{-n\, -w} (z,\bar z) V^{j}_{n\, w} (0, 0)\rangle = 
|z|^{-4\Delta_{jnw}} R\left(j,\frac{n+kw}{2},\frac{n-kw}{2}\right)= \nonumber\\
&=
\frac{\nu^{1-2j} }{|z|^{4\Delta_{jnw}}}
\frac{\Gamma (1-2j )\Gamma (1+\frac{1-2j}{k})}{\Gamma (2j-1 )\Gamma (1+\frac{2j-1}{k})}
\frac{\Gamma (j+\frac{n+kw}{2}) \Gamma(j+\frac{n-kw}{2})}{
\Gamma (1-j+\frac{n+kw}{2}) \Gamma(1-j+\frac{n-kw}{2})}\, .
\label{reflampl}
\end{align}
As the name suggests, the reflection amplitude is related to the symmetry 
of the theory
\begin{equation}
V_{1-j,-m -\bar m} =  R\left(j,m,\bar m\right) V_{j,m \bar m}
~.
\label{reflproper}
\end{equation}

\paragraph{Extended characters}
When $k$ is rational it is often convenient to define 
\emph{extended characters} \cite{Eguchi:2003ik}.  Writing
$k=N/K$ with $K,N \in \mathbb{Z}_{>0}$, the extended characters are defined by 
partially summing over $N$ units of spectral flow. 
The resulting characters are characters of an extended chiral algebra 
similar to the extended chiral algebra of a 
$U(1)$ boson at rational radius squared. Explicitly, extended characters
(denoted by capital letters) are defined as
\begin{equation}
Ch_\star (\star,\star)\oao{a}{b} (\tau;\nu) = \sum_{\ell \in \Z}
ch_c \left( \star,\star \right)_\star \oao{a}{b} (\tau; \nu + N \ell \tau)
~.
\end{equation}
For example, the extended characters of the continuous representations are
for $k$ integer
\begin{equation}
Ch_c (P,m)\oao{a}{b} (\tau;\nu) = 
\frac{q^{\frac{P^2}{k}}}{\eta^3 (\tau )} \ \Theta_{2m,k} (\tau; \nicefrac{2\nu}{k} )
\ \vartheta \oao{a}{b} (\tau ; \nu)
~
\end{equation}
with $2 m \in \Z_{2k}$.

%%%%%%%%%%%%%%%%%%%%%%%%%%%%%%%%%%%%%%%%%%

\section{D-brane wave-functions}
\label{appwave-functions}

For the convenience of the reader we list here the one-point functions
of primary fields on the disc with boundary conditions corresponding to
the D0-, D1- and D2-branes of the supersymmetric cigar. As far as D2-branes
are concerned, for the purposes of the main text we will focus on the D2-branes of 
refs.~\cite{Fotopoulos:2004ut,Hosomichi:2004ph} that are based on the continuous
representations. For a more complete list of the D-branes of $\NN=2$ Liouville
theory we refer the reader to the excellent presentation of~\cite{Hosomichi:2004ph}.

The one-point function on the disc of a primary field
$\VV_{j,n,w}^{\left[{a \atop b}\right]}$
with quantum number $j=\frac{1}{2}+ip$, momentum $n$ and winding $w$ in
the $\left[ {a \atop b}\right]$-$\left[ {a \atop b}\right]$
sector\footnote{As usual
$\left[ {0 \atop 0}\right]=$NS, $\left[ {0 \atop 1} \right]=\tilde {\rm NS}$,
 $\left[ {1 \atop 0} \right]={\rm R}$,  $\left[ {1 \atop 1}
\right]={\rm \tilde R}$.} is
\beq
 \label{appbaa}
 \left \langle \VV_{p,n,w}^{\left[{a \atop b}\right]} \right\rangle_{D_\star}=
 \frac{\Phi_{D_\star}(j,n,w;\left[ {a \atop b}\right])}{|z-\bar
z|^{\Delta_{p,n,w;a}}}
\, ,
\eeq
where $\Delta_{j,n,w;a}$ is the scaling dimension of the primary field and
$D_\star$ the boundary condition of interest.

\paragraph{D0-branes} On the supersymmetric cigar there are (including
the fermionic contribution) four D0-branes with boundary states
$| D0;\left[ {a \atop b} \right] \rangle$ that obey B-type boundary conditions.
The one-point functions of primary fields on the disc with D0-boundary
conditions are\footnote{D0-branes couple both to continuous and discrete primary
fields. We present here only the one-point functions of continuous
states, which by definition
have $j=\frac{1}{2}+ip$, $p\in \R$.}
\begin{multline}
\label{appbab}
\Phi_{D0;\left[ {a \atop b} \right]}\left(p,n,w; \left[ {a' \atop
b'}\right]\right)= \\
= \, \delta_{n,0} \delta^{(2)}_{a',b} \delta^{(2)}_{b',a} ~ k^{-\frac{1}{2}}
(-)^{w(a-1)} e^{-\frac{i \pi ab}{2}} \nu^{-ip}\ 
\frac{\Gamma(\frac{1}{2}+ip+\frac{kw}{2}+\frac{b}{2})
\Gamma(\frac{1}{2}+ip-\frac{kw}{2}-\frac{b}{2})}
{\Gamma(2ip)\Gamma(1+\frac{2ip}{k})}\, ,
\end{multline}
where $\nu=\nicefrac{\Gamma(1-\frac{1}{k})}{\Gamma(1+\frac{1}{k})}$.

\paragraph{D2-branes} There are similarly four fermionic types of D2-branes
with boundary states $|D2;P,M;\left[ {a \atop b} \right]\rangle$ that obey
B-type boundary conditions. They are labeled by a continuous parameter
$P \in \R_{\geq 0}$ and a half-integer $M$ and exhibit the one-point functions
\bea
\label{appbad}
\Phi_{D2;P,M;\left[ {a \atop b}\right]}\left(p,n,w;\left[ {a' \atop
b'} \right] \right)&=&
\delta_{n,0} \delta^{(2)}_{a',b} \delta^{(2)}_{b',a} ~
\sqrt{\frac{2}{k}} (-)^w e^{-2\pi i M w} e^{-\frac{i\pi ab}{2}}\nu^{-ip}
\cos\left(\frac{4\pi p P}{k}\right)\ \times
\nonumber\\
&&\times \ \frac{\Gamma(-2ip)\Gamma(1-\frac{2ip}{k})}
{\Gamma(\frac{1}{2}-ip+\frac{kw}{2}+\frac{b}{2})
\Gamma(\frac{1}{2}-ip-\frac{kw}{2}-\frac{b}{2})}
~.
\eea
There are no couplings to discrete states in this case.

\paragraph{D1-branes} Finally, there are four fermionic types of D1-branes
with boundary states $|D1;\hat \rho,\hat\phi;\left[ {a \atop b} \right]\rangle$
($\hat\rho \in \R_{\geq 0}$, $\hat \phi \in [0,2\pi)$) that obey A-type boundary conditions.
The corresponding one-point functions on the disc are
\bea
\label{appbae}
\Phi_{D1;\hat \rho,\hat \phi;\left[ {a \atop b}\right]}\left(p,n,w;\left[ {a' \atop
b'} \right] \right)&=&
\delta_{w,0} \delta^{(2)}_{a',b} \delta^{(2)}_{b',a} ~
\frac{1}{\sqrt 2} e^{in \hat\phi} \nu^{-ip}
\left(e^{2i\hat\rho p}+(-)^n e^{-2i\hat \rho p}\right)\ \times
\nonumber\\
&&\times\ \frac{\Gamma(-2ip)\Gamma(1-\frac{2ip}{k})}
{\Gamma(\frac{1}{2}-ip+\frac{n}{2}+\frac{b}{2})\Gamma(\frac{1}{2}-ip-\frac{n}{2}-\frac{b}{2})}
~.
\eea
There are no couplings to discrete states in this case either.

%%%%%%%%%%%%%%%%%%%%%%%%%%%%%%%%%%%%%%%%%%

\section{The $\PP$-matrix of unextended and extended characters}
\label{Pmatapp}
In this appendix we determine the modular transformation properties
of the hatted identity representation characters under the 
$\PP$ modular transformation
\begin{equation*}
\tau = -\frac{1}{4it} + \frac12 \ \longrightarrow \ \tilde{\tau} = it + \frac12
~.
\end{equation*}
The results presented here are instrumental in the 
modular bootstrap approach of sec.\ \ref{O2section}.
 
We start with the hatted unextended character
that appears in the open string spectrum of the
D0-brane. Explicitly this character reads
\begin{equation}
\widehat{ch_\mathbb{I}} (r;\tau) \oao{a}{b}  = e^{-i\pi (\Delta-\nicefrac{c}{24})}
ch_\mathbb{I} (r;\tau+\hl) \oao{a}{b} ~.
\end{equation}
Using (\ref{idchar}) one finds
\begin{multline}
\widehat{ch_\mathbb{I}} (r;\tau) \oao{a}{b} 
=  e^{i\pi\left(r+\frac{1}{8} +\frac{1-a}{2}-\frac{a^2}{8}\right) } \ 
\frac{\vartheta \oao{a}{b} (\tau + \hl)}{\eta^3 (\tau + \hl)} \ 
q^{\frac{(r+\frac{a}{2})^2-\frac{1}{4}}{k}} \ \times \\ \times \ 
\left[ \frac{1}{1+e^{i\pi(b+r+\frac{a+1}{2})} q^{r+\frac{a+1}{2}}} -
\frac{1}{1+e^{i\pi(b+r+\frac{a-1}{2})} q^{r+\frac{a-1}{2}}}\right]
%\frac{\vartheta \oao{a}{b} (\tau + \hl)}{\eta^3 (\tau + \hl)}
~.
\end{multline}
Next inspired by~\cite{Miki:1989ri} we introduce "Miki's function"
\begin{equation}
\mathcal{I} \oao{a}{b} (N,\alpha,\beta;\tau)\equiv \sum_{s \in \Z+\frac{a+1}{2}}
e^{2\pi i s \alpha} \frac{q^{\beta s+\frac{N}{2}s^2}}{1+(-)^b e^{i\pi s} q^s}
~.
\end{equation}
One can rewrite the hatted identity character in terms of Miki's function as
\begin{multline}
\widehat{ch}_\mathbb{I} (r;\tau,0) \oao{a}{b} (r;\tau) =
e^{i\pi\left(r+\frac{1-a^2}{8} +\frac{1-a}{2}\right) } \
\frac{\vartheta \oao{a}{b} (\tau + \hl)}{\eta^3 (\tau + \hl)} \ \times \\ \times \ 
\ \int_0^1 \di \alpha \ e^{-2i\pi \alpha r}
\left[
e^{-i\pi \alpha (1+a)}\, \mathcal{I} \oao{a}{b}
\left(\frac{2}{k},\alpha,-\frac{1}{k};\tau\right) - e^{i\pi \alpha (1-a)}\, \mathcal{I}
\oao{a}{b} \left(\frac{2}{k},\alpha,\frac{1}{k};\tau\right) \right]
~.
\end{multline}

The computation of the $\mathcal{P}$-matrix elements of the identity character
proceeds as follows. First, let us take care of the bosonic and fermionic oscillators\footnote{For convenience 
we write the phase in front of the \textsc{rhs} 
in a simple way that is correct for $a,b \in \{0,1 \}$ 
but does not respect the periodicities of the $\vartheta$-functions.}
\begin{equation}
\frac{\vartheta \oao{a}{b} \left( -\frac{1}{4it} + \frac{1}{2}\right)}{
\eta^3 \left(-\frac{1}{4it} + \frac{1}{2}\right)}
= \frac{1}{2t}\, e^{\frac{i\pi}{2} (a-b+1+\frac{a-1}{2})} 
\frac{\vartheta \oao{a}{a-b+1} \left( it + \frac{1}{2}\right)}{
\eta^3 \left(it + \frac{1}{2}\right)}~.
\end{equation}
Then it will be useful to compute the quantity
$\II \oao{a}{b}(N,\alpha,\beta;-\frac{1}{4it})$. Explicitly, we have
\begin{equation}
\label{appeak}
\II \oao{a}{b} \left(N,\alpha,\beta;-\frac{1}{4it} \right)=
\sum_{s\in \Z+\frac{a+1}{2}}
e^{2\pi i s \alpha} \frac{e^{-\frac{\pi}{2t}(\beta s +\frac{N}{2}s^2)}}
{1+e^{i\pi (s +b)} e^{-\frac{\pi s}{2t}}} \, ,
\end{equation}
which we find useful to re-express as
\begin{equation}
\label{appeal}
\II \oao{a}{b} \left(N,\alpha ,\beta;-\frac{1}{4it}\right)=\II_+
\oao{a}{b} \left(N,\alpha ,\beta;-\frac{1}{4it}\right)\ +\
\II_-\oao{a}{b} \left(N,\alpha ,\beta;-\frac{1}{4it}\right)
\end{equation}
with
\begin{equation}
\label{appeam}
\II_\pm \oao{a}{b} \left(N,\alpha ,\beta;-\frac{1}{4it}\right) =\, 
\sum_{s\in 2\Z+\frac{a\pm 1}{2}}
e^{2\pi i s \alpha} \frac{e^{-\frac{\pi}{2t}(\beta s +\frac{N}{2}s^2)}}
{1+e^{i\pi (\frac{a+1}{2}+b)} e^{-\frac{\pi s}{2t}}}\, ,
\end{equation}
Then we observe that we can recast $\II_{\pm}$ as
\begin{equation}
\label{appeao}
\II_{\pm} \oao{a}{b} \left(N,\alpha,\beta;-\frac{1}{4it}\right)= 2t
\left[ \int_{-\infty-i\epsilon}^{\infty-i\epsilon}\!\!\!\!\!\! -
\int_{-\infty+i\epsilon}^{\infty +i\epsilon} \right]
\frac{ e^{-2\pi \beta x+8\pi i t x(\alpha+\frac{1}{4})-4N \pi t x^2}\, \di x }
{(e^{2\pi i t x} \mp  e^{-2\pi i (t x-\frac{a+1}{4})})
\left(1\pm  e^{i\pi (\frac{a+1}{2}+b)} e^{-2\pi x}\right)}
~.
\end{equation}
To evaluate the \textsc{rhs} of \eqref{appeao} in a different way we
make use of the expansions
\begin{subequations}
\begin{align}
\label{appeap}
\frac{1}{e^{2\pi i t x} \pm e^{-2\pi i (t x-\frac{a+1}{4})}}&=\sum_{n=0}^\infty e^{\frac{i\pi n}{2} (a \mp 1)}
e^{-2\pi i t x(2n+1)}~, ~ ~ \Im x<0
\, , \\
\label{appeaq}
\frac{1}{e^{2\pi i t x} \pm e^{-2\pi i (t x-\frac{a+1}{4})}}&=\ e^{-\frac{i\pi}{2} (a \pm 1)}
\sum_{n=0}^\infty e^{-\frac{i\pi n}{2} (a \mp 1)}
e^{2\pi i t x(2n+1)}~, ~ ~ \Im x>0 \, .
\end{align}
\end{subequations}
After a few steps of algebra, defining $q=e^{-2\pi t}$, we can show that
\beq
\label{appeas}
\II_{\pm} \oao{a}{b} \left(N,\alpha,\beta;-\frac{1}{4it}\right) =
2t\int_{-\infty}^\infty \di x\,  \sum_{n\in \Z} e^{\frac{i\pi}{2}(a \pm 1)n }
\frac{e^{-2\pi \beta x}}{1\pm (-)^b  e^{\frac{i \pi}{2} (a+1)} e^{-2\pi x}}
q^{2Nx^2-4ix(\alpha-\frac{n}{2})}
\, .
\eeq
Finally, shifting the contour of integration from the real line $\R$ to
$\R+i(\alpha-\frac{n}{2})/N$ we obtain
\beq
\label{appeat}
\frac{1}{2t} \II_{\pm} \oao{a}{b} \left(N,\alpha,\beta;-\frac{1}{4it}\right) =
\JJ_{cont,\pm}+\JJ_{disc,\pm}
\eeq
with continuous contribution
\beq
\label{appeau}
\JJ_{cont,\pm}=
\int_{-\infty}^\infty dx \sum_{n\in \Z} e^{\frac{i\pi}{2}(a \pm 1)n }
\frac{e^{-2\pi \beta (x+i \frac{2\alpha-n}{2N})}}
{1 \pm (-)^b  e^{\frac{i \pi}{2} (a+1)} e^{-2\pi (x+i\frac{2\alpha-n}{2N})}}
q^{2Nx^2+\frac{2}{N}(\alpha-\frac{n}{2})^2}
\eeq
and discrete contribution $\JJ_{disc,\pm}$
that arises when we pick up the appropriate poles
(this will be computed later). We are now ready to assemble the data, 
dealing only with the continuous pieces for the moment. By straightforward
algebra we find
\begin{multline}
\widehat{ch}_\mathbb{I}(r;-\frac{1}{4it}) \oao{a}{b}  =
e^{i\pi (r+\frac{7}{8}+\frac{3a}{8}-\frac{b}{2})}
\int_{-\infty}^{\infty} \di P  \int_0^1 \di \alpha\, e^{-2i\pi  \alpha (r+\frac{a}{2})}\ \times
 \\ \times \  \sum_{n \in \mathbb{Z}} e^{\frac{i \pi a n}{2}}
\left[ \frac{e^{\frac{i \pi n}{2}}}{
1 + (-)^b  e^{\frac{i \pi}{2} (a+1)} e^{-\pi (P+i\frac{k}{2}(2\alpha-n))}}
+\frac{e^{-\frac{i \pi n}{2}}}{
1 - (-)^b  e^{\frac{i \pi}{2} (a+1)} e^{-\pi (P+i\frac{k}{2}(2\alpha-n))}}\right]\ \times \\ \times \
\sinh \pi \left( \frac{P}{k}-\frac{i n}{2} \right) \
q^{\frac{P^2}{k}+k \left(\alpha - \frac{n}{2}\right)^2} 
\frac{\vartheta \oao{a}{a-b+1} \left( it + \frac{1}{2}\right)}
{\eta^3 \left(it + \frac{1}{2}\right)} \ + \ \text{discrete}\, .
\label{intermPmat}
\end{multline}
Let us write $n = -2\ell -\delta$ with $\delta=0,1$. We consider first 
the term with $\delta=0$. One can trade the summation over $\alpha$ and 
the sum over $\ell$ for an integral over $m = k(\ell + \alpha)$ and fold 
the integral over the $P$-axis:
\begin{multline}
\frac{1}{k}
e^{i\pi (r+\frac{7}{8}+\frac{3a}{8}-\frac{b}{2})}
\int_{0}^{\infty}\!\! \di P  \int_{-\infty}^{+\infty} \di m\
\frac{q^{\frac{P^2+m^2}{k}}}{
\eta^3 \left(it + \frac{1}{2}\right)}\ \vartheta \oao{a}{a-b+1} \left( it + \frac{1}{2}\right)\ \times
 \\ \times \  \
\frac{e^{-\frac{2i\pi}{k} m (r+\frac{a}{2})}\,
\sinh 2 \pi P \sinh \frac{\pi P}{k}}{\cosh \pi \left[P + i (m-\frac{a}{2})\right]
\cosh \pi \left[P - i (m-\frac{a}{2})\right]}
\end{multline}
Consider now the term with $\delta=1$. We define also 
$m = k(\ell + \hl+ \alpha)$ and obtain the result
\begin{multline}
2 \, e^{i\pi (b+r-\hl)}\ \frac{1}{k}
e^{i\pi (r+\frac{7}{8}+\frac{3a}{8}-\frac{b}{2})}
\int_{0}^{\infty}\!\! \di P  \int_{-\infty}^{+\infty} \di m\
\frac{q^{\frac{P^2+m^2}{k}}}{
\eta^3 \left(it + \frac{1}{2}\right)}\ \vartheta \oao{a}{a-b+1} \left( it + \frac{1}{2}\right)\ \times
 \\ \times \  \
\frac{e^{-\frac{2i\pi}{k} m (r+\frac{a}{2})}\,
\cosh  \pi P  \cos \pi \left(m-\frac{a}{2}\right)
\cosh \frac{\pi P}{k}}{\cosh \pi \left[P + i (m-\frac{a}{2})\right]
\cosh \pi \left[P - i (m-\frac{a}{2})\right]}
\end{multline}
The full result can be recast as follows
\begin{multline}
\widehat{ch}_\mathbb{I} \oao{a}{b} (r;-\frac{1}{4it}) =
\frac{2}{k}  \, e^{i \pi \left(\frac{2a+1}{4}+\frac{b}{2}\right)}
\int_{0}^{\infty}\!\!\!\! \di P   \int_{-\infty}^{\infty} \!\!\!\! \di m \
\ \times
\\ \times \  \ \frac{e^{-\frac{2i\pi}{k} m (r+\frac{a}{2})}\,
\cosh  \pi P  \left[\cos \pi \left(m-\frac{a}{2}\right)
\cosh \frac{\pi P}{k}
+i (-)^{b+r}\sinh  \pi P \sinh \frac{\pi P}{k}
 \right]}{\cosh \pi \left[P + i (m-\frac{a}{2})\right]
\cosh \pi \left[P - i (m-\frac{a}{2})\right]}\ \times  \\ \times \
e^{\frac{i \pi}{8}( 1-a^2)} \frac{q^{\frac{P^2+m^2}{k}}\, 
\vartheta \oao{a}{a-b+1} \left( it + \frac{1}{2}\right)}{
\eta^3 \left(it + \frac{1}{2}\right)} + \text{discrete}\, .
\end{multline}
In the last line we recognize an $\Omega$-twisted character 
for continuous representations.
It allows to obtain the $\mathcal{P}$-matrix elements of the identity
to the continuous representations as
{\large \begin{equation}
\begin{array}{|l|}
\hline
\mathcal{P}_{\mathbb{I} ;\, r \oao{a}{b}}^{c ;\, (P,m) \oao{a'}{b'}} =
\frac{2}{k}  \, e^{i \pi \left( \frac{2a+1}{4}+\frac{b}{2} \right)}
\delta_{a,a'}^{(2)} \delta_{a-b+1,b'}^{(2)}
\ \times
\\\phantom{aaaaaaaaaaaa}\times \
\frac{e^{-\frac{2i\pi}{k} m (r+\frac{a}{2})}\,
\cosh  \pi P  \left[\cos \pi \left(m-\frac{a}{2}\right)
\cosh \frac{\pi P}{k}
+i (-)^{b+r}\sinh  \pi P \sinh \frac{\pi P}{k}
 \right]}{\cosh \pi \left[P + i (m-\frac{a}{2})\right]
\cosh \pi \left[P - i (m-\frac{a}{2})\right]}
\\
\hline
\end{array}
\label{contPmat}
\end{equation}
}

\subsection*{Discrete representations}
While shifting the contour of integration over $x$ from 
$\R$ to $\R+i(\alpha-\frac{n}{2})/N$ in eqn.~(\ref{appeas}), 
one picks the residues of poles corresponding to the discrete 
representations of the coset \slc that appear in the closed string spectrum. 
To make the identification of discrete characters easier, 
we rewrite the integral~(\ref{intermPmat}) as (with $\tilde P = 2x$):
\begin{multline}
\frac{2}{k}
e^{i \pi (r+\frac{7+3a}{8}-\frac{b}{2})}
\frac{\vartheta \oao{a}{a-b+1} \left( it + \frac{1}{2}\right)}{
\eta^3 \left(it + \frac{1}{2}\right)}
\int \di \tilde P ~\di m \, e^{-\frac{2i\pi m (r+\nicefrac{a}{2})}{k}}\ \times \\ \times \
\left[ \sinh \frac{\pi (\tilde P-im)}{k} +e^{i \pi (b+r-\hl)} \cosh \frac{\pi (\tilde P-im)}{k}
e^{-\pi(\tilde P - \nicefrac{ia}{2})} \right] \frac{q^{\frac{\tilde P(\tilde P-2im)}{k}}}{1+e^{-2\pi(\tilde P
-\nicefrac{ia}{2})}}~.
\end{multline}
Assuming first $m>0$, the poles occur for
\begin{equation}
\tilde P = i \left(\upsilon + \frac{1+a}{2} \right)\ , 
\quad 0\leqslant \upsilon+\frac{1+a}{2} \leqslant m \ ,  \quad
\upsilon \in \mathbb{Z}
~.
\end{equation}
The sum over the residues reads
\begin{multline}
-\frac{2}{k}e^{i \pi (r+\frac{7+3a}{8}-\frac{b}{2})}
\frac{\vartheta \oao{a}{a-b+1} \left( it + \frac{1}{2}\right)}{
\eta^3 \left(it + \frac{1}{2}\right)}
 \sum_{\upsilon > -\frac{1+a}{2} } \int_{\upsilon+\frac{1+a}{2}}^{\infty} \di m
\, e^{-\frac{2i\pi m (r+\nicefrac{a}{2})}{k}}\ \times \\ \times \
\left[  \sin \frac{\pi (\upsilon-m+\frac{1+a}{2})}{k}
+ i e^{i \pi (b+r+\upsilon)} \cos \frac{\pi (\upsilon-m+\frac{1+a}{2})}{k}
\right] q^{\frac{(\upsilon + \frac{1+a}{2})(2m-\upsilon - \frac{1+a}{2})}{k}}
~.
\label{discrinterm}
\end{multline}
To proceed further we make the slicing
\begin{equation}
m=j+\upsilon+\frac{a}{2} +\frac{k \ell}{2}  \ , 
\quad 
 \frac{1}{2} \leqslant j \leqslant \frac{k+1}{2} \ \text{and}
\quad \ell=0,1,\ldots
\end{equation}
in terms of which we rewrite eqn~(\ref{discrinterm}) as
\begin{multline}
-\frac{2}{k} e^{i \pi (r+\frac{7+3a}{8}-\frac{b}{2})}
\frac{\vartheta \oao{a}{a-b+1} \left( it + \frac{1}{2}\right)}{
\eta^3 \left(it + \frac{1}{2}\right)}
 \sum_{\upsilon > -\frac{1+a}{2} } \int_{\frac{1}{2}}^{\frac{k+1}{2}} \di j \
\sum_{\ell=0}^{\infty}
\, e^{-\frac{2i\pi}{k} (j+\upsilon+\frac{a}{2})(r+\nicefrac{a}{2})} \ \times \\ \times \
e^{-i\pi \ell (r+\nicefrac{a}{2})} q^{\ell (\upsilon+\frac{a+1}{2})}
\left[  \sin \left( \frac{\pi(\hl-j)}{k}-\frac{\pi \ell}{2} \right)
+ i e^{i \pi (b+r+\upsilon)} \cos  \left( \frac{\pi(\hl-j)}{k}-\frac{\pi \ell}{2} \right)
\right] \ \times \\ \times \  q^{\frac{-(j-\hl)^2+(j+\upsilon+\nicefrac{a}{2})^2}{k}}
~.
\label{discintbis}
\end{multline}
The case $m<0$ will give a similar contribution with $\upsilon  <-\frac{1+a}{2}$.
We can now perform the sum over $\ell$ (for which we need to consider separately
the cases $\ell$ odd and $\ell$ even) obtaining
\begin{multline}
\frac{2  }{k}
e^{i \pi (r+\frac{2a-1}{4}-\frac{b}{2})}
  \int_{\frac{1}{2}}^{\frac{k+1}{2}} \di j \ \sum_{\upsilon\in \mathbb{Z} }
\, e^{-\frac{2i\pi}{k} (j+\upsilon+\frac{a}{2})(r+\nicefrac{a}{2})}\ \times  \\ \times \
\left[  \sin  \frac{\pi(\hl-j)}{k}
+  e^{i \pi (b+r+\upsilon+\hl)} \cos  \frac{\pi(\hl-j)}{k}
\right] \ \times  \\ \times \  e^{\frac{i\pi (1-a^2)}{8}}
\frac{q^{\frac{-(j-\hl)^2+(j+\upsilon+\nicefrac{a}{2})^2}{k}}}{1+(-)^b
e^{i\pi (\upsilon + \frac{a+1}{2})} q^{\upsilon+\frac{a+1}{2}}}
\frac{\vartheta \oao{a}{a-b+1} \left( it + \frac{1}{2}\right)}{
\eta^3 \left(it + \frac{1}{2}\right)}
~.
\end{multline}
In the last line we recognize the expression of the $\Omega$-inserted
discrete character $\widehat{ch}_d (j,\upsilon) \oao{a}{b}$. 
Note that {\it both} terms between square brackets
in~(\ref{discintbis}) are necessary in order to reconstruct characters 
for the discrete representations by summing
over $\ell$. We get finally the $\mathcal{P}$-matrix elements for the discrete representations as:
{\large \begin{equation}
\begin{array}{|l|}
\hline
\mathcal{P}_{\mathbb{I} ;\, r \oao{a}{b}}^{d ;\, (j,\upsilon) \oao{a'}{b'}} =
\frac{2}{k} \
e^{i \pi (r+\frac{2a-1}{4}-\frac{b}{2})}\ \delta_{a,a'}^{(2)}
 \delta_{a-b+1,b'}^{(2)}\ \times \\ \phantom{AAAAA}\times \
 e^{-\frac{2i\pi}{k} (j+\upsilon+\frac{a}{2})(r+\frac{a}{2})}
\left[  \sin  \frac{\pi(\frac12 -j)}{k}
+  e^{i \pi (b+r+\upsilon+\frac{1}{2})} \cos  \frac{\pi(\frac12 -j)}{k}
\right]
\\
\hline
\end{array}
\label{discPmat}
\end{equation}
}

\subsection*{Results relevant for the ${\rm O}_B$- 
and ${\rm \tilde O}_B$-planes of the cigar} 
The character $\sum_{r \in \mathbb{Z}} \widehat{ch}_\mathbb{I}
(r;-\frac{1}{4it}) \oao{a}{b}$ appears in the M\"obius strip amplitude of D0-branes 
on the cigar for the parity $\PP_2$. 
Using the $\PP$-matrix elements computed above 
we can easily obtain the modular transformation
\begin{multline}
\sum_{r \in \mathbb{Z}} \widehat{ch}_\mathbb{I}(r;-\frac{1}{4it})  \oao{a}{b} =
2  \, e^{i \pi \left( \frac{a+1}{4}+\frac{b}{2} \right)}
\int_{0}^{\infty}\!\!\!\! \di P   \
\sum_{w \in \mathbb{Z}}\, (-)^{aw}
\cosh  \pi P \ \times \\ \times\
\left\{\frac{ \cos \pi \left(kw-\frac{a}{2}\right)
\cosh \frac{\pi P}{k}}
{\cosh \pi \left[P + i (kw-\frac{a}{2})\right]
\cosh \pi \left[P - i (kw-\frac{a}{2})\right]}\widehat{ch}_c (P,kw;it)\oao{a}{a-b+1}\ +\right. \\
\left. +
\frac{\ e^{i \pi (b + \frac{1-a}{2})}\, \sinh  \pi P \sinh \frac{\pi P}{k}
}{\cosh \pi \left[P + i k(w+\frac12 )-\frac{ia}{2}\right]
\cosh \pi \left[P - i k(w+\frac12 )+\frac{ia}{2})
\right]} \widehat{ch}_c (P,k(w+\hl );it)\oao{a}{a-b+1}
\right\}
\\ +  \ 2
e^{i \pi (\frac{a+1}{4}+\frac{b}{2})}
  \int_{\frac{1}{2}}^{\frac{k+1}{2}} \di j \ \sum_{w,\upsilon\in \mathbb{Z} } (-)^{aw}
 \left[(-)^\upsilon \cos  \frac{\pi(j-\hl)}{k}\,
\delta\left(j+\upsilon+\frac{a}{2}-kw\right)
\ + \right. \\
\left.
+e^{i\pi(b+\frac{1-a}{2})}
  \sin  \frac{\pi(j-\hl)}{k}\, \delta\left(j+\upsilon+\frac{a}{2}-k(w+\hl)\right)\
\right] \  \widehat{ch}_d (j,\upsilon;it)\oao{a}{a-b+1} ~.
\label{Pmatcigfull}
\end{multline}
We observe the absence of boundary 
terms at $j=\hl$ or $j=\nicefrac{(k+1)}{2}$.\footnote{This is obvious from 
\eqref{Pmatcigfull} for generic non-integer $k$. One can 
check that this is also true for integer $k$.} Such terms 
would jeopardize the modular bootstrap results since they are not present
in the closed string spectrum~\cite{Eguchi:2003ik}. 

Similarly, the character
$\sum_{r \in \mathbb{Z}} (-)^r\widehat{ch}_\mathbb{I} (r;-\frac{1}{4it})\oao{a}{b}$
appears in the M\"obius strip amplitude of D0-branes for the parity $\tilde \PP_2$.
One can easily deduce from the above results the modular transformation
\begin{multline}
\sum_{r \in \mathbb{Z}} (-)^r\widehat{ch}_\mathbb{I} (r;-\frac{1}{4it})\oao{a}{b}=
2  \, e^{i \pi (\frac{3a+2}{8}+\frac{b}{2})}
\int_{0}^{\infty}\!\!\!\! \di P   \
\sum_{w \in \mathbb{Z}}\, (-)^{aw} \cosh  \pi P~ \times
\\ 
\left\{\frac{ e^{-\frac{i \pi a}{2}}\cos \pi \left(k(w+\hl)-\frac{a}{2}\right)
\cosh \frac{\pi P}{k} \, }{\cosh \pi \left[P + i (k(w+\hl)-\frac{a}{2})\right]
\cosh \pi \left[P - i (k(w+\hl)-\frac{a}{2})\right]}\ 
\widehat{ch}_c (P,k(w+\hl );it)\oao{a}{a-b+1}-
\right. \\ \phantom{aaaa}
\left. +i(-)^b
\frac{\sinh  \pi P \sinh \frac{\pi P}{k}\, }
{\cosh \pi \left[P + i (kw-\frac{a}{2})\right]\cosh \pi \left[P - i (kw-\frac{a}{2})\right]}
\widehat{ch}_c (P,kw;it)\oao{a}{a-b+1}
\right\}
\\ + \text{discrete}\, .
\end{multline}

\subsection*{$\PP$-matrix for continuous representations}
The computation of the $\PP$-modular transformation for the continuous representations is far less tedious. 
The result is: 
\begin{multline}
\label{obootab}
\widehat{ch_c}(p,m;-\nicefrac{1}{4\tau}) \left[{a \atop b}\right] =
\\ = \frac{2}{k} e^{\frac{i\pi }{4}(1-a-2b)}
\int_0^\infty \di p'\, \int_{-\infty}^\infty \di m' \,
e^{-2\pi i mm'/k} \cos\left(\frac{2\pi pp'}{k}\right)
\widehat{ch_c}(p',m';\tau)\left[ {a \atop a-b+1}\right]
\end{multline}

\end{appendix}

%\bibliography{orientlast}

\begin{thebibliography}{10}

\bibitem{Dabholkar:1997zd}
A.~Dabholkar, {\it Lectures on orientifolds and duality},
 \href{http://xxx.lanl.gov/abs/hep-th/9804208}{{\tt hep-th/9804208}}.

\bibitem{Giddings:2001yu}
S.~B. Giddings, S.~Kachru, and J.~Polchinski, {\it {Hierarchies from fluxes in
 string compactifications}},  {\em Phys. Rev.} {\bf D66} (2002) 106006,
 [\href{http://xxx.lanl.gov/abs/hep-th/0105097}{{\tt hep-th/0105097}}].

\bibitem{Govindarajan:2003vp}
S.~Govindarajan and J.~Majumder, {\it {Crosscaps in Gepner models and type IIA
 orientifolds}},  {\em JHEP} {\bf 02} (2004) 026,
 [\href{http://xxx.lanl.gov/abs/hep-th/0306257}{{\tt hep-th/0306257}}].

\bibitem{Aldazabal:2003ub}
G.~Aldazabal, E.~C. Andres, M.~Leston, and C.~Nunez, {\it {Type IIB
 orientifolds on Gepner points}},  {\em JHEP} {\bf 09} (2003) 067,
 [\href{http://xxx.lanl.gov/abs/hep-th/0307183}{{\tt hep-th/0307183}}].

\bibitem{Blumenhagen:2003su}
R.~Blumenhagen, {\it {Supersymmetric orientifolds of Gepner models}},  {\em
 JHEP} {\bf 11} (2003) 055,
 [\href{http://xxx.lanl.gov/abs/hep-th/0310244}{{\tt hep-th/0310244}}].

\bibitem{Brunner:2004zd}
I.~Brunner, K.~Hori, K.~Hosomichi, and J.~Walcher, {\it {Orientifolds of Gepner
 models}},  \href{http://xxx.lanl.gov/abs/hep-th/0401137}{{\tt
 hep-th/0401137}}.

\bibitem{Brunner:2006yi}
I.~Brunner and V.~Mitev, {\it {Permutation orientifolds}},
 \href{http://xxx.lanl.gov/abs/hep-th/0612108}{{\tt hep-th/0612108}}.

\bibitem{Hosomichi:2006yj}
K.~Hosomichi, {\it {Permutation orientifolds of Gepner models}},  {\em JHEP}
 {\bf 01} (2007) 081, [\href{http://xxx.lanl.gov/abs/hep-th/0612109}{{\tt
 hep-th/0612109}}].

\bibitem{fzz}
V.~Fateev, A.~B. Zamolodchikov, and A.~B. Zamolodchikov unpublished notes.

\bibitem{Kazakov:2000pm}
V.~Kazakov, I.~K. Kostov, and D.~Kutasov, {\it {A matrix model for the
 two-dimensional black hole}},  {\em Nucl. Phys.} {\bf B622} (2002) 141--188,
 [\href{http://xxx.lanl.gov/abs/hep-th/0101011}{{\tt hep-th/0101011}}].

\bibitem{Hori:2001ax}
K.~Hori and A.~Kapustin, {\it {Duality of the fermionic 2d black hole and N = 2
 Liouville theory as mirror symmetry}},  {\em JHEP} {\bf 08} (2001) 045,
 [\href{http://xxx.lanl.gov/abs/hep-th/0104202}{{\tt hep-th/0104202}}].

\bibitem{Ooguri:1995wj}
H.~Ooguri and C.~Vafa, {\it {Two-Dimensional Black Hole and Singularities of CY
 Manifolds}},  {\em Nucl. Phys.} {\bf B463} (1996) 55--72,
 [\href{http://xxx.lanl.gov/abs/hep-th/9511164}{{\tt hep-th/9511164}}].

\bibitem{Giveon:1999zm}
A.~Giveon, D.~Kutasov, and O.~Pelc, {\it {Holography for non-critical
 superstrings}},  {\em JHEP} {\bf 10} (1999) 035,
 [\href{http://xxx.lanl.gov/abs/hep-th/9907178}{{\tt hep-th/9907178}}].


\bibitem{Sfetsos:1998xd}
K.~Sfetsos, {\it Branes for Higgs phases and exact conformal field theories},
 {\em JHEP} {\bf 01} (1999) 015,
 [\href{http://xxx.lanl.gov/abs/hep-th/9811167}{{\tt hep-th/9811167}}].


\bibitem{Giveon:1999px}
A.~Giveon and D.~Kutasov, {\it {Little string theory in a double scaling
 limit}},  {\em JHEP} {\bf 10} (1999) 034,
 [\href{http://xxx.lanl.gov/abs/hep-th/9909110}{{\tt hep-th/9909110}}].

\bibitem{Seiberg:1997zk}
N.~Seiberg, {\it {New theories in six dimensions and matrix description of M-
 theory on T**5 and T**5/Z(2)}},  {\em Phys. Lett.} {\bf B408} (1997) 98--104,
 [\href{http://xxx.lanl.gov/abs/hep-th/9705221}{{\tt hep-th/9705221}}].

\bibitem{Aharony:1999ks}
O.~Aharony, {\it A brief review of 'little string theories'},  {\em Class.
 Quant. Grav.} {\bf 17} (2000) 929--938,
 [\href{http://xxx.lanl.gov/abs/hep-th/9911147}{{\tt hep-th/9911147}}].

\bibitem{Kutasov:2001uf}
D.~Kutasov, {\it Introduction to little string theory}, . Prepared for ICTP
 Spring School on Superstrings and Related Matters, Trieste, Italy, 2-10 Apr
 2001.

\bibitem{Hanany:1996ie}
A.~Hanany and E.~Witten, {\it {Type IIB superstrings, BPS monopoles, and
 three-dimensional gauge dynamics}},  {\em Nucl. Phys.} {\bf B492} (1997)
 152--190, [\href{http://xxx.lanl.gov/abs/hep-th/9611230}{{\tt
 hep-th/9611230}}].

\bibitem{Witten:1997sc}
E.~Witten, {\it {Solutions of four-dimensional field theories via M-theory}},
 {\em Nucl. Phys.} {\bf B500} (1997) 3--42,
 [\href{http://xxx.lanl.gov/abs/hep-th/9703166}{{\tt hep-th/9703166}}].

\bibitem{Giveon:1998sr}
A.~Giveon and D.~Kutasov, {\it {Brane dynamics and gauge theory}},  {\em Rev.
 Mod. Phys.} {\bf 71} (1999) 983--1084,
 [\href{http://xxx.lanl.gov/abs/hep-th/9802067}{{\tt hep-th/9802067}}].

\bibitem{Ribault:2003ss}
S.~Ribault and V.~Schomerus, {\it {Branes in the 2-D black hole}},  {\em JHEP}
 {\bf 02} (2004) 019, [\href{http://xxx.lanl.gov/abs/hep-th/0310024}{{\tt
 hep-th/0310024}}].

\bibitem{Israel:2004jt}
D.~Israel, A.~Pakman, and J.~Troost, {\it {D-branes in N = 2 Liouville theory
 and its mirror}},  {\em Nucl. Phys.} {\bf B710} (2005) 529--576,
 [\href{http://xxx.lanl.gov/abs/hep-th/0405259}{{\tt hep-th/0405259}}].

\bibitem{Eguchi:2003ik}
T.~Eguchi and Y.~Sugawara, {\it {Modular bootstrap for boundary N = 2 Liouville
 theory}},  {\em JHEP} {\bf 01} (2004) 025,
 [\href{http://xxx.lanl.gov/abs/hep-th/0311141}{{\tt hep-th/0311141}}].

\bibitem{Ahn:2003tt}
C.~Ahn, M.~Stanishkov, and M.~Yamamoto, {\it {One-point functions of N = 2
 super-Liouville theory with boundary}},  {\em Nucl. Phys.} {\bf B683} (2004)
 177--195, [\href{http://xxx.lanl.gov/abs/hep-th/0311169}{{\tt
 hep-th/0311169}}].

\bibitem{Fotopoulos:2004ut}
A.~Fotopoulos, V.~Niarchos, and N.~Prezas, {\it {D-branes and extended
 characters in SL(2,R)/U(1)}},  {\em Nucl. Phys.} {\bf B710} (2005) 309--370,
 [\href{http://xxx.lanl.gov/abs/hep-th/0406017}{{\tt hep-th/0406017}}].

\bibitem{Hosomichi:2004ph}
K.~Hosomichi, {\it {N = 2 Liouville theory with boundary}},  {\em JHEP} {\bf
 12} (2006) 061, [\href{http://xxx.lanl.gov/abs/hep-th/0408172}{{\tt
 hep-th/0408172}}].

\bibitem{Fotopoulos:2005cn}
A.~Fotopoulos, V.~Niarchos, and N.~Prezas, {\it {D-branes and SQCD in
 non-critical superstring theory}},  {\em JHEP} {\bf 10} (2005) 081,
 [\href{http://xxx.lanl.gov/abs/hep-th/0504010}{{\tt hep-th/0504010}}].

\bibitem{Ashok:2005py}
S.~K. Ashok, S.~Murthy, and J.~Troost, {\it {D-branes in non-critical
 superstrings and minimal super Yang-Mills in various dimensions}},  {\em
 Nucl. Phys.} {\bf B749} (2006) 172--205,
 [\href{http://xxx.lanl.gov/abs/hep-th/0504079}{{\tt hep-th/0504079}}].

\bibitem{Israel:2005zp}
D.~Israel, {\it {Non-critical string duals of N = 1 quiver theories}},  {\em
 JHEP} {\bf 04} (2006) 029,
 [\href{http://xxx.lanl.gov/abs/hep-th/0512166}{{\tt hep-th/0512166}}].

\bibitem{Murthy:2006xt}
S.~Murthy and J.~Troost, {\it {D-branes in non-critical superstrings and
 duality in N = 1 gauge theories with flavor}},  {\em JHEP} {\bf 10} (2006)
 019, [\href{http://xxx.lanl.gov/abs/hep-th/0606203}{{\tt hep-th/0606203}}].

\bibitem{Hikida:2002bt}
Y.~Hikida, {\it {Liouville field theory on a unoriented surface}},  {\em JHEP}
 {\bf 05} (2003) 002, [\href{http://xxx.lanl.gov/abs/hep-th/0210305}{{\tt
 hep-th/0210305}}].

\bibitem{Hikida:2002fh}
Y.~Hikida, {\it {Crosscap states for orientifolds of euclidean AdS(3)}},  {\em
 JHEP} {\bf 05} (2002) 021,
 [\href{http://xxx.lanl.gov/abs/hep-th/0203030}{{\tt hep-th/0203030}}].

\bibitem{Hikida:2002ws}
Y.~Hikida, {\it {Orientifolds of SU(2)/U(1) WZW models}},  {\em JHEP} {\bf 11}
 (2002) 035, [\href{http://xxx.lanl.gov/abs/hep-th/0201175}{{\tt
 hep-th/0201175}}].

\bibitem{Brunner:2002em}
I.~Brunner and K.~Hori, {\it {Notes on orientifolds of rational conformal field
 theories}},  {\em JHEP} {\bf 07} (2004) 023,
 [\href{http://xxx.lanl.gov/abs/hep-th/0208141}{{\tt hep-th/0208141}}].

\bibitem{Brunner:2003zm}
I.~Brunner and K.~Hori, {\it {Orientifolds and mirror symmetry}},  {\em JHEP}
 {\bf 11} (2004) 005, [\href{http://xxx.lanl.gov/abs/hep-th/0303135}{{\tt
 hep-th/0303135}}].

\bibitem{Nakayama:2004at}
Y.~Nakayama, {\it {Crosscap states in N = 2 Liouville theory}},  {\em Nucl.
 Phys.} {\bf B708} (2005) 345--380,
 [\href{http://xxx.lanl.gov/abs/hep-th/0409039}{{\tt hep-th/0409039}}].

\bibitem{noncritzeroprimeB}
  D.~Israel and V.~Niarchos,
  ``Tree-level stability without spacetime fermions: Novel examples in string
  theory,''
  JHEP {\bf 0707} (2007) 065
  [arXiv:0705.2140 [hep-th]].
  %%CITATION = JHEPA,0707,065;%% 

\bibitem{Evans:1997hk}
N.~J. Evans, C.~V. Johnson, and A.~D. Shapere, {\it {Orientifolds, branes, and
 duality of 4D gauge theories}},  {\em Nucl. Phys.} {\bf B505} (1997)
 251--271, [\href{http://xxx.lanl.gov/abs/hep-th/9703210}{{\tt
 hep-th/9703210}}].

\bibitem{Balog:1988jb}
J.~Balog, L.~O'Raifeartaigh, P.~Forgacs, and A.~Wipf, {\it Consistency of
 string propagation on curved space-times: An su(1,1) based counterexample},
 {\em Nucl. Phys.} {\bf B325} (1989) 225.

\bibitem{Henningson:1991jc}
M.~Henningson, S.~Hwang, P.~Roberts, and B.~Sundborg, {\it {Modular invariance
 of SU(1,1) strings}},  {\em Phys. Lett.} {\bf B267} (1991) 350--355.

\bibitem{Maldacena:2000hw}
J.~M. Maldacena and H.~Ooguri, {\it {Strings in AdS(3) and SL(2,R) WZW model.
 I}},  {\em J. Math. Phys.} {\bf 42} (2001) 2929--2960,
 [\href{http://xxx.lanl.gov/abs/hep-th/0001053}{{\tt hep-th/0001053}}].

\bibitem{Elitzur:1991cb}
S.~Elitzur, A.~Forge, and E.~Rabinovici, {\it {Some global aspects of string
 compactifications}},  {\em Nucl. Phys.} {\bf B359} (1991) 581--610.

\bibitem{Mandal:1991tz}
G.~Mandal, A.~M. Sengupta, and S.~R. Wadia, {\it {Classical solutions of
 two-dimensional string theory}},  {\em Mod. Phys. Lett.} {\bf A6} (1991)
 1685--1692.

\bibitem{Witten:1991yr}
E.~Witten, {\it {On string theory and black holes}},  {\em Phys. Rev.} {\bf
 D44} (1991) 314--324.

\bibitem{Dijkgraaf:1991ba}
R.~Dijkgraaf, H.~L. Verlinde, and E.~P. Verlinde, {\it {String propagation in a
 black hole geometry}},  {\em Nucl. Phys.} {\bf B371} (1992) 269--314.

\bibitem{Teschner:1997ft}
J.~Teschner, {\it {On structure constants and fusion rules in the SL(2,C)/SU(2)
 WZNW model}},  {\em Nucl. Phys.} {\bf B546} (1999) 390--422,
 [\href{http://xxx.lanl.gov/abs/hep-th/9712256}{{\tt hep-th/9712256}}].

\bibitem{Teschner:1999ug}
J.~Teschner, {\it {Operator product expansion and factorization in the H-3+
 WZNW model}},  {\em Nucl. Phys.} {\bf B571} (2000) 555--582,
 [\href{http://xxx.lanl.gov/abs/hep-th/9906215}{{\tt hep-th/9906215}}].

\bibitem{Bars:1992sr}
I.~Bars and K.~Sfetsos, {\it {Conformally exact metric and dilaton in string
 theory on curved space-time}},  {\em Phys. Rev.} {\bf D46} (1992) 4510--4519,
 [\href{http://xxx.lanl.gov/abs/hep-th/9206006}{{\tt hep-th/9206006}}].

\bibitem{Tseytlin:1993my}
A.~A. Tseytlin, {\it {Conformal sigma models corresponding to gauged
 Wess-Zumino- Witten theories}},  {\em Nucl. Phys.} {\bf B411} (1994)
 509--558, [\href{http://xxx.lanl.gov/abs/hep-th/9302083}{{\tt
 hep-th/9302083}}].

\bibitem{Giveon:2001up}
A.~Giveon and D.~Kutasov, {\it {Notes on AdS(3)}},  {\em Nucl. Phys.} {\bf
 B621} (2002) 303--336, [\href{http://xxx.lanl.gov/abs/hep-th/0106004}{{\tt
 hep-th/0106004}}].

\bibitem{Tong:2003ik}
D.~Tong, {\it {Mirror mirror on the wall: On two-dimensional black holes and
 Liouville theory}},  {\em JHEP} {\bf 04} (2003) 031,
 [\href{http://xxx.lanl.gov/abs/hep-th/0303151}{{\tt hep-th/0303151}}].

\bibitem{Israel:2003ry}
D.~Israel, C.~Kounnas, and M.~P. Petropoulos, {\it {Superstrings on NS5
 backgrounds, deformed AdS(3) and holography}},  {\em JHEP} {\bf 10} (2003)
 028, [\href{http://xxx.lanl.gov/abs/hep-th/0306053}{{\tt hep-th/0306053}}].

\bibitem{Lee:2001gh}
P.~Lee, H.~Ooguri, and J.-w. Park, {\it {Boundary states for AdS(2) branes in
 AdS(3)}},  {\em Nucl. Phys.} {\bf B632} (2002) 283--302,
 [\href{http://xxx.lanl.gov/abs/hep-th/0112188}{{\tt hep-th/0112188}}].

\bibitem{Ponsot:2001gt}
B.~Ponsot, V.~Schomerus, and J.~Teschner, {\it {Branes in the Euclidean
 AdS(3)}},  {\em JHEP} {\bf 02} (2002) 016,
 [\href{http://xxx.lanl.gov/abs/hep-th/0112198}{{\tt hep-th/0112198}}].

\bibitem{Israel:2005ek}
D.~Israel, {\it {D-branes in Lorentzian AdS(3)}},  {\em JHEP} {\bf 06} (2005)
 008, [\href{http://xxx.lanl.gov/abs/hep-th/0502159}{{\tt hep-th/0502159}}].

\bibitem{Bachas:2000fr}
C.~Bachas and M.~Petropoulos, {\it {Anti-de-Sitter D-branes}},  {\em JHEP} {\bf
 02} (2001) 025, [\href{http://xxx.lanl.gov/abs/hep-th/0012234}{{\tt
 hep-th/0012234}}].

\bibitem{Maldacena:2001ky}
J.~M. Maldacena, G.~W. Moore, and N.~Seiberg, {\it {Geometrical interpretation
 of D-branes in gauged WZW models}},  {\em JHEP} {\bf 07} (2001) 046,
 [\href{http://xxx.lanl.gov/abs/hep-th/0105038}{{\tt hep-th/0105038}}].

\bibitem{Israel:2005fn}
D.~Israel, A.~Pakman, and J.~Troost, {\it {D-branes in little string theory}},
 {\em Nucl. Phys.} {\bf B722} (2005) 3--64,
 [\href{http://xxx.lanl.gov/abs/hep-th/0502073}{{\tt hep-th/0502073}}].

\bibitem{Fotopoulos:2003vc}
A.~Fotopoulos, {\it {Semiclassical description of D-branes in SL(2)/U(1) gauged
 WZW model}},  {\em Class. Quant. Grav.} {\bf 20} (2003) S465--S472,
 [\href{http://xxx.lanl.gov/abs/hep-th/0304015}{{\tt hep-th/0304015}}].

\bibitem{Ribault:2005pq}
S.~Ribault, {\it {Discrete D-branes in AdS(3) and in the 2d black hole}},  {\em
 JHEP} {\bf 08} (2006) 015,
 [\href{http://xxx.lanl.gov/abs/hep-th/0512238}{{\tt hep-th/0512238}}].

\bibitem{Adorf:2007ah}
H.~Adorf and M.~Flohr, {\it On the vario{us types of D-branes in the boundary
 H(3)+ model}},  \href{http://xxx.lanl.gov/abs/hep-th/0702158}{{\tt
 hep-th/0702158}}.

\bibitem{Gomis:2003vi}
J.~Gomis and A.~Kapustin, {\it {Two-dimensional unoriented strings and matrix
 models}},  {\em JHEP} {\bf 06} (2004) 002,
 [\href{http://xxx.lanl.gov/abs/hep-th/0310195}{{\tt hep-th/0310195}}].

\bibitem{Bergman:2003yp}
O.~Bergman and S.~Hirano, {\it {The cap in the hat: Unoriented 2D strings and
 matrix(- vector) models}},  {\em JHEP} {\bf 01} (2004) 043,
 [\href{http://xxx.lanl.gov/abs/hep-th/0311068}{{\tt hep-th/0311068}}].

\bibitem{Bachas:2001id}
C.~Bachas, N.~Couchoud, and P.~Windey, {\it {Orientifolds of the 3-sphere}},
 {\em JHEP} {\bf 12} (2001) 003,
 [\href{http://xxx.lanl.gov/abs/hep-th/0111002}{{\tt hep-th/0111002}}].

\bibitem{Hanany:2002ev}
A.~Hanany, N.~Prezas, and J.~Troost, {\it {The partition function of the
 two-dimensional black hole conformal field theory}},  {\em JHEP} {\bf 04}
 (2002) 014, [\href{http://xxx.lanl.gov/abs/hep-th/0202129}{{\tt
 hep-th/0202129}}].

\bibitem{Eguchi:2004yi}
T.~Eguchi and Y.~Sugawara, {\it {SL(2,R)/U(1) supercoset and elliptic genera of
 non-compact Calabi-Yau manifolds}},  {\em JHEP} {\bf 05} (2004) 014,
 [\href{http://xxx.lanl.gov/abs/hep-th/0403193}{{\tt hep-th/0403193}}].

\bibitem{Israel:2004ir}
D.~Israel, C.~Kounnas, A.~Pakman, and J.~Troost, {\it The partition function of
 the supersymmetric two- dimensional black hole and little string theory},
 {\em JHEP} {\bf 06} (2004) 033,
 [\href{http://xxx.lanl.gov/abs/hep-th/0403237}{{\tt hep-th/0403237}}].

\bibitem{Angelantonj:2002ct}
C.~Angelantonj and A.~Sagnotti, {\it {Open strings}},  {\em Phys. Rept.} {\bf
 371} (2002) 1--150, [\href{http://xxx.lanl.gov/abs/hep-th/0204089}{{\tt
 hep-th/0204089}}].

\bibitem{Nakayama:2003ep}
Y.~Nakayama, {\it {Tadpole cancellation in unoriented Liouville theory}},  {\em
 JHEP} {\bf 11} (2003) 017,
 [\href{http://xxx.lanl.gov/abs/hep-th/0309063}{{\tt hep-th/0309063}}].

\bibitem{Nakayama:2004vk}
Y.~Nakayama, {\it {Liouville field theory: A decade after the revolution}},
 {\em Int. J. Mod. Phys.} {\bf A19} (2004) 2771--2930,
 [\href{http://xxx.lanl.gov/abs/hep-th/0402009}{{\tt hep-th/0402009}}].


\bibitem{Murthy:2003es}
 S.~Murthy, {\it Notes on non-critical superstrings in various dimensions},
 {\em JHEP} {\bf 11} (2003) 056, [\href{http://xxx.lanl.gov/abs/hep-th/0305197}{{\tt
hep-th/0305197}}].


\bibitem{Miki:1989ri}
K.~Miki, {\it The representation theory of the so(3) invariant superconformal
 algebra},  {\em Int. J. Mod. Phys.} {\bf A5} (1990) 1293.

\end{thebibliography}

\providecommand{\href}[2]{#2}\begingroup\raggedright\endgroup

\end{document}